\documentclass[a4paper,twocolumn,11pt,unpublished]{quantumarticle}
\pdfoutput=1
\input{glyphtounicode}\pdfgentounicode=1
\usepackage[utf8]{inputenc}
\usepackage[english]{babel}
\usepackage[T1]{fontenc}
\usepackage{amsmath,amssymb}
\usepackage{graphicx}
\usepackage{booktabs}
\usepackage{tabularx}
\usepackage{adjustbox}
\newcolumntype{L}{>{\raggedright\arraybackslash}X}
\usepackage{placeins}
\usepackage{float}
\usepackage[numbers,sort&compress]{natbib}
\usepackage{hyperref}
\AtBeginDocument{\normalfont}
\newcommand{\ket}[1]{|#1\rangle}

\newcommand{\figdir}{figures}
\graphicspath{{figures/}}
\makeatletter
\expandafter\def\csname qv@A1.dropped\endcsname{\ensuremath{143\text{--}1{,}083}}
\expandafter\def\csname qv@A1.npairs\endcsname{12}
\expandafter\def\csname qv@A1.phi\endcsname{\ensuremath{0.008\text{--}0.020}}
\expandafter\def\csname qv@A1.t.R\endcsname{\ensuremath{2.79\text{--}13.32}}
\expandafter\def\csname qv@A1.t.holm\endcsname{\ensuremath{6.0\times10^{-4}}}
\expandafter\def\csname qv@A1.t.holmperm\endcsname{\ensuremath{1.2\times10^{-3}}}
\expandafter\def\csname qv@A1.t.lomin\endcsname{\ensuremath{2.38}}
\expandafter\def\csname qv@A1.u.R\endcsname{\ensuremath{3.08\text{--}16.27}}
\expandafter\def\csname qv@A1.u.holm\endcsname{\ensuremath{6.0\times10^{-4}}}
\expandafter\def\csname qv@A1.u.holmperm\endcsname{\ensuremath{1.2\times10^{-3}}}
\expandafter\def\csname qv@A1.u.lomin\endcsname{\ensuremath{2.67}}
\expandafter\def\csname qv@A11.chi\endcsname{\ensuremath{32\text{--}60}}
\expandafter\def\csname qv@A11.n\endcsname{1,000}
\expandafter\def\csname qv@A12.x.Dcf.A\endcsname{\ensuremath{0.521\ [0.361,\,0.641]}}
\expandafter\def\csname qv@A12.x.Dcf.B\endcsname{\ensuremath{0.042\ [0.025,\,0.079]}}
\expandafter\def\csname qv@A12.x.Dclip.A\endcsname{\ensuremath{0.341\ [0.196,\,0.341]}}
\expandafter\def\csname qv@A12.x.Dclip.B\endcsname{\ensuremath{0.016\ [0.002,\,0.019]}}
\expandafter\def\csname qv@A12.z.Dcf.A\endcsname{\ensuremath{0.433\ [0.301,\,0.564]}}
\expandafter\def\csname qv@A12.z.Dcf.B\endcsname{\ensuremath{0.038\ [0.034,\,0.075]}}
\expandafter\def\csname qv@A12.z.Dclip.A\endcsname{\ensuremath{0.273\ [0.152,\,0.276]}}
\expandafter\def\csname qv@A12.z.Dclip.B\endcsname{\ensuremath{0.015\ [0.002,\,0.021]}}
\expandafter\def\csname qv@A2.x.diff\endcsname{\ensuremath{-0.052\ [-0.428,\,0.304]}}
\expandafter\def\csname qv@A2.x.p\endcsname{\ensuremath{0.591}}
\expandafter\def\csname qv@A2.z.diff\endcsname{\ensuremath{-0.279\ [-0.699,\,0.096]}}
\expandafter\def\csname qv@A2.z.p\endcsname{\ensuremath{0.922}}
\expandafter\def\csname qv@A3.rcf.A\endcsname{\ensuremath{2.4\text{--}2.9}}
\expandafter\def\csname qv@A3.x.cf.A\endcsname{\ensuremath{0.369\ [0.232,\,0.553]}}
\expandafter\def\csname qv@A3.x.cf.B\endcsname{\ensuremath{0.450\ [0.404,\,0.478]}}
\expandafter\def\csname qv@A3.x.rcf.A\endcsname{\ensuremath{2.41}}
\expandafter\def\csname qv@A3.x.rcf.B\endcsname{\ensuremath{1.09}}
\expandafter\def\csname qv@A3.z.cf.A\endcsname{\ensuremath{0.225\ [0.087,\,0.374]}}
\expandafter\def\csname qv@A3.z.cf.B\endcsname{\ensuremath{0.386\ [0.337,\,0.402]}}
\expandafter\def\csname qv@A3.z.rcf.A\endcsname{\ensuremath{2.92}}
\expandafter\def\csname qv@A3.z.rcf.B\endcsname{\ensuremath{1.10}}
\expandafter\def\csname qv@A4.differ.A\endcsname{2}
\expandafter\def\csname qv@A4.differ.B\endcsname{1}
\expandafter\def\csname qv@A4.n.A\endcsname{1,200}
\expandafter\def\csname qv@A4.n.B\endcsname{1,189}
\expandafter\def\csname qv@A4.need.x\endcsname{22.7\%}
\expandafter\def\csname qv@A4.need.z\endcsname{17.9\%}
\expandafter\def\csname qv@A4.recl.A\endcsname{1}
\expandafter\def\csname qv@A4.recl.B\endcsname{0}
\expandafter\def\csname qv@A4.upB\endcsname{0.25\%}
\expandafter\def\csname qv@A5.x.R1\endcsname{\ensuremath{[-0.246,\,-0.007]}}
\expandafter\def\csname qv@A5.x.R2\endcsname{\ensuremath{[-0.149,\,0.069]}}
\expandafter\def\csname qv@A5.x.R3\endcsname{\ensuremath{[-0.00152,\,0.00098]}}
\expandafter\def\csname qv@A5.z.R1\endcsname{\ensuremath{[-0.226,\,-0.018]}}
\expandafter\def\csname qv@A5.z.R2\endcsname{\ensuremath{[-0.135,\,0.055]}}
\expandafter\def\csname qv@A5.z.R3\endcsname{\ensuremath{[-0.00318,\,-0.00004]}}
\expandafter\def\csname qv@A6.lo\endcsname{\ensuremath{0.99962}}
\expandafter\def\csname qv@A6.shots\endcsname{10,000}
\expandafter\def\csname qv@A8.Lplus\endcsname{\ensuremath{0.988}}
\expandafter\def\csname qv@A8.gamma\endcsname{\ensuremath{0.0128, 0.0128, 0.0101}}
\expandafter\def\csname qv@A8.rotated.diffuse\endcsname{32\%}
\expandafter\def\csname qv@A8.rotated.ge2\endcsname{66\%}
\expandafter\def\csname qv@A8.rotated.mean\endcsname{\ensuremath{1.709}}
\expandafter\def\csname qv@A8.rotated.ratio\endcsname{\ensuremath{1.73}}
\expandafter\def\csname qv@A8.standard.diffuse\endcsname{1\%}
\expandafter\def\csname qv@A8.standard.ge2\endcsname{0\%}
\expandafter\def\csname qv@A8.standard.mean\endcsname{\ensuremath{0.988}}
\expandafter\def\csname qv@A8.standard.ratio\endcsname{\ensuremath{1.00}}
\expandafter\def\csname qv@A9.runs\endcsname{\ensuremath{1{,}072\text{--}2{,}296}}
\expandafter\def\csname qv@DG.cells\endcsname{36}
\expandafter\def\csname qv@DG.cz\endcsname{\ensuremath{1.04}}
\expandafter\def\csname qv@DG.cz.iqr\endcsname{\ensuremath{1.00\text{--}1.08}}
\expandafter\def\csname qv@DG.cz.range\endcsname{\ensuremath{0.88\text{--}1.22}}
\expandafter\def\csname qv@DG.cz.sep\endcsname{11}
\expandafter\def\csname qv@DG.h\endcsname{\ensuremath{1.06}}
\expandafter\def\csname qv@DG.h.iqr\endcsname{\ensuremath{1.01\text{--}1.13}}
\expandafter\def\csname qv@DG.h.range\endcsname{\ensuremath{0.81\text{--}1.73}}
\expandafter\def\csname qv@DG.h.sep\endcsname{0}
\expandafter\def\csname qv@DG.id\endcsname{\ensuremath{1.03}}
\expandafter\def\csname qv@DG.id.iqr\endcsname{\ensuremath{0.98\text{--}1.08}}
\expandafter\def\csname qv@DG.id.range\endcsname{\ensuremath{0.95\text{--}1.18}}
\expandafter\def\csname qv@DG.id.sep\endcsname{9}
\expandafter\def\csname qv@E4.allundo\endcsname{104,000}
\expandafter\def\csname qv@E4.cells\endcsname{36}
\expandafter\def\csname qv@E4.rho\endcsname{\ensuremath{0.0005\text{--}0.0110}}
\expandafter\def\csname qv@E7.maxrel\endcsname{\ensuremath{0.46\%}}
\expandafter\def\csname qv@H1.lo\endcsname{\ensuremath{0.99985}}
\expandafter\def\csname qv@H1.p\endcsname{\ensuremath{4.2\times10^{-113}}}
\expandafter\def\csname qv@H1.pairs\endcsname{25,746}
\expandafter\def\csname qv@H1.runs\endcsname{471,614}
\expandafter\def\csname qv@H2a.R\endcsname{\ensuremath{2.8\text{--}13.4}}
\expandafter\def\csname qv@H2a.f0\endcsname{\ensuremath{1.4\text{--}9.9\%}}
\expandafter\def\csname qv@H2a.f05\endcsname{\ensuremath{18.1\text{--}27.3\%}}
\expandafter\def\csname qv@H4p.f0\endcsname{\ensuremath{0.33\text{--}0.82}}
\expandafter\def\csname qv@H4p.f05\endcsname{\ensuremath{0.11\text{--}0.19}}
\expandafter\def\csname qv@H4p.npairs\endcsname{6}
\expandafter\def\csname qv@H4p.nreg\endcsname{5}
\expandafter\def\csname qv@H5.x.Aonly\endcsname{1,258}
\expandafter\def\csname qv@H5.x.Bonly\endcsname{1,271}
\expandafter\def\csname qv@H5.x.N\endcsname{136,232}
\expandafter\def\csname qv@H5.x.both\endcsname{742}
\expandafter\def\csname qv@H5.x.neither\endcsname{132,961}
\expandafter\def\csname qv@H5.x.p\endcsname{\ensuremath{1.4\times10^{-111}}}
\expandafter\def\csname qv@H5.x.ratio\endcsname{\ensuremath{0.994\ [0.953,\,1.035]}}
\expandafter\def\csname qv@H5.x.shared\endcsname{\ensuremath{37\%}}
\expandafter\def\csname qv@H5.x.tailA\endcsname{\ensuremath{25.0\%}}
\expandafter\def\csname qv@H5.x.tailA.n\endcsname{499}
\expandafter\def\csname qv@H5.x.tailB\endcsname{\ensuremath{2.2\%}}
\expandafter\def\csname qv@H5.x.tailB.n\endcsname{45}
\expandafter\def\csname qv@H5.x.tailboth\endcsname{3}
\expandafter\def\csname qv@H5.z.Aonly\endcsname{832}
\expandafter\def\csname qv@H5.z.Bonly\endcsname{869}
\expandafter\def\csname qv@H5.z.N\endcsname{108,019}
\expandafter\def\csname qv@H5.z.both\endcsname{1,168}
\expandafter\def\csname qv@H5.z.neither\endcsname{105,150}
\expandafter\def\csname qv@H5.z.p\endcsname{\ensuremath{1.6\times10^{-81}}}
\expandafter\def\csname qv@H5.z.ratio\endcsname{\ensuremath{0.982\ [0.949,\,1.015]}}
\expandafter\def\csname qv@H5.z.shared\endcsname{\ensuremath{58\%}}
\expandafter\def\csname qv@H5.z.tailA\endcsname{\ensuremath{20.2\%}}
\expandafter\def\csname qv@H5.z.tailA.n\endcsname{404}
\expandafter\def\csname qv@H5.z.tailB\endcsname{\ensuremath{2.3\%}}
\expandafter\def\csname qv@H5.z.tailB.n\endcsname{47}
\expandafter\def\csname qv@H5.z.tailboth\endcsname{13}
\expandafter\def\csname qv@H6.Alow\endcsname{\ensuremath{-0.12}}
\expandafter\def\csname qv@H6.n\endcsname{200,000}
\expandafter\def\csname qv@H6.x.A\endcsname{\ensuremath{-0.039\ [-0.097,\,0.028]}}
\expandafter\def\csname qv@H6.x.A.p\endcsname{\ensuremath{0.01445}}
\expandafter\def\csname qv@H6.x.A.round\endcsname{\ensuremath{-0.037}}
\expandafter\def\csname qv@H6.x.B\endcsname{\ensuremath{-0.039\ [-0.053,\,-0.024]}}
\expandafter\def\csname qv@H6.x.B.p\endcsname{\ensuremath{0.01506}}
\expandafter\def\csname qv@H6.x.B.round\endcsname{\ensuremath{-0.037}}
\expandafter\def\csname qv@H6.x.holm\endcsname{\ensuremath{0.61}}
\expandafter\def\csname qv@H6.x.zzratio\endcsname{\ensuremath{0.96}}
\expandafter\def\csname qv@H6.z.A\endcsname{\ensuremath{-0.075\ [-0.124,\,-0.017]}}
\expandafter\def\csname qv@H6.z.A.p\endcsname{\ensuremath{0.01829}}
\expandafter\def\csname qv@H6.z.A.round\endcsname{\ensuremath{-0.041}}
\expandafter\def\csname qv@H6.z.B\endcsname{\ensuremath{-0.060\ [-0.074,\,-0.045]}}
\expandafter\def\csname qv@H6.z.B.p\endcsname{\ensuremath{0.01831}}
\expandafter\def\csname qv@H6.z.B.round\endcsname{\ensuremath{-0.036}}
\expandafter\def\csname qv@H6.z.holm\endcsname{\ensuremath{0.61}}
\expandafter\def\csname qv@H6.z.zzratio\endcsname{\ensuremath{0.95}}
\expandafter\def\csname qv@H7.Aup\endcsname{\ensuremath{1.45}}
\expandafter\def\csname qv@H7.nF\endcsname{2,000}
\expandafter\def\csname qv@H7.nS\endcsname{10,000}
\expandafter\def\csname qv@H7.x.A\endcsname{\ensuremath{1.09\ [0.70,\,1.45]}}
\expandafter\def\csname qv@H7.x.B\endcsname{\ensuremath{1.15\ [1.12,\,1.18]}}
\expandafter\def\csname qv@H7.x.holm\endcsname{\ensuremath{1.00}}
\expandafter\def\csname qv@H7.x.kF.A\endcsname{\ensuremath{0.89}}
\expandafter\def\csname qv@H7.x.kF.B\endcsname{\ensuremath{0.49}}
\expandafter\def\csname qv@H7.x.kS.A\endcsname{61/10,000}
\expandafter\def\csname qv@H7.x.kS.B\endcsname{4/10,000}
\expandafter\def\csname qv@H7.x.kappa.A\endcsname{\ensuremath{0.47}}
\expandafter\def\csname qv@H7.x.kappa.B\endcsname{\ensuremath{0.05}}
\expandafter\def\csname qv@H7.x.p\endcsname{\ensuremath{0.60}}
\expandafter\def\csname qv@H7.z.A\endcsname{\ensuremath{0.79\ [0.38,\,1.15]}}
\expandafter\def\csname qv@H7.z.B\endcsname{\ensuremath{1.07\ [1.01,\,1.11]}}
\expandafter\def\csname qv@H7.z.holm\endcsname{\ensuremath{1.00}}
\expandafter\def\csname qv@H7.z.kF.A\endcsname{\ensuremath{0.66}}
\expandafter\def\csname qv@H7.z.kF.B\endcsname{\ensuremath{0.42}}
\expandafter\def\csname qv@H7.z.kS.A\endcsname{69/10,000}
\expandafter\def\csname qv@H7.z.kS.B\endcsname{4/10,000}
\expandafter\def\csname qv@H7.z.kappa.A\endcsname{\ensuremath{0.57}}
\expandafter\def\csname qv@H7.z.kappa.B\endcsname{\ensuremath{0.05}}
\expandafter\def\csname qv@H7.z.p\endcsname{\ensuremath{0.93}}
\expandafter\def\csname qv@M1.s0\endcsname{\ensuremath{9.56\ [7.16,\,13.55]}}
\expandafter\def\csname qv@M1.s0.tw\endcsname{\ensuremath{9.46}}
\expandafter\def\csname qv@M1.sr\endcsname{\ensuremath{8.77\ [6.68,\,12.14]}}
\expandafter\def\csname qv@M1.sr.tw\endcsname{\ensuremath{8.55}}
\expandafter\def\csname qv@M2.rank\endcsname{21/32}
\expandafter\def\csname qv@M2.sign\endcsname{32/32}
\expandafter\def\csname qv@M2B.rank\endcsname{30/32}
\expandafter\def\csname qv@M2B.sign\endcsname{32/32}
\expandafter\def\csname qv@N1.nA\endcsname{1,000}
\expandafter\def\csname qv@N1.x.D.R0\endcsname{\ensuremath{20.5\ [17.7,\,23.2]}}
\expandafter\def\csname qv@N1.x.D.R4\endcsname{\ensuremath{24.5\ [21.6,\,27.4]}}
\expandafter\def\csname qv@N1.x.T.R0\endcsname{22.9\%}
\expandafter\def\csname qv@N1.x.T.R4\endcsname{26.9\%}
\expandafter\def\csname qv@N1.x.TB\endcsname{2.4\%}
\expandafter\def\csname qv@N1.x.dE\endcsname{\ensuremath{0.001\ [-0.094,\,0.094]}}
\expandafter\def\csname qv@N1.x.dT\endcsname{\ensuremath{4.0\ [1.4,\,6.6]}}
\expandafter\def\csname qv@N1.x.dZZ\endcsname{\ensuremath{0.003\ [-0.090,\,0.094]}}
\expandafter\def\csname qv@N1.x.gg\endcsname{200}
\expandafter\def\csname qv@N1.x.ghp\endcsname{\ensuremath{0.019}}
\expandafter\def\csname qv@N1.x.nB\endcsname{1,070}
\expandafter\def\csname qv@N1.z.D.R0\endcsname{\ensuremath{18.7\ [16.2,\,21.3]}}
\expandafter\def\csname qv@N1.z.D.R4\endcsname{\ensuremath{20.5\ [17.9,\,23.2]}}
\expandafter\def\csname qv@N1.z.T.R0\endcsname{21.1\%}
\expandafter\def\csname qv@N1.z.T.R4\endcsname{22.9\%}
\expandafter\def\csname qv@N1.z.TB\endcsname{2.4\%}
\expandafter\def\csname qv@N1.z.dE\endcsname{\ensuremath{-0.066\ [-0.163,\,0.021]}}
\expandafter\def\csname qv@N1.z.dT\endcsname{\ensuremath{1.8\ [-0.7,\,4.3]}}
\expandafter\def\csname qv@N1.z.dZZ\endcsname{\ensuremath{-0.057\ [-0.140,\,0.021]}}
\expandafter\def\csname qv@N1.z.gg\endcsname{200}
\expandafter\def\csname qv@N1.z.ghp\endcsname{\ensuremath{0.0023}}
\expandafter\def\csname qv@N1.z.nB\endcsname{1,046}
\expandafter\def\csname qv@N2.D\endcsname{\ensuremath{17.0\ [15.3,\,18.9]}}
\expandafter\def\csname qv@N2.Dfix\endcsname{\ensuremath{17.9}}
\expandafter\def\csname qv@N2.Dhalf\endcsname{\ensuremath{8.9}}
\expandafter\def\csname qv@N2.Gi\endcsname{400}
\expandafter\def\csname qv@N2.Gj\endcsname{50}
\expandafter\def\csname qv@N2.Lcf.A\endcsname{\ensuremath{0.222}}
\expandafter\def\csname qv@N2.Lcf.B\endcsname{\ensuremath{0.370}}
\expandafter\def\csname qv@N2.Lclip.A\endcsname{\ensuremath{0.395}}
\expandafter\def\csname qv@N2.Lclip.B\endcsname{\ensuremath{0.403}}
\expandafter\def\csname qv@N2.TA\endcsname{19.3\%}
\expandafter\def\csname qv@N2.TB\endcsname{2.3\%}
\expandafter\def\csname qv@N2.kF.A\endcsname{\ensuremath{0.694}}
\expandafter\def\csname qv@N2.kF.B\endcsname{\ensuremath{0.433}}
\expandafter\def\csname qv@N2.kappa.A\endcsname{\ensuremath{0.53}}
\expandafter\def\csname qv@N2.kappa.B\endcsname{\ensuremath{0.08}}
\expandafter\def\csname qv@N2.mult.A\endcsname{\ensuremath{2.89}}
\expandafter\def\csname qv@N2.mult.B\endcsname{\ensuremath{1.14}}
\expandafter\def\csname qv@N2.nA\endcsname{2,000}
\expandafter\def\csname qv@N2.nB\endcsname{1,945}
\expandafter\def\csname qv@N2.nS\endcsname{10,000}
\expandafter\def\csname qv@N2.ratio\endcsname{\ensuremath{1.028\ [0.993,\,1.064]}}
\expandafter\def\csname qv@N2.rcf.A\endcsname{\ensuremath{3.12}}
\expandafter\def\csname qv@N2.rcf.B\endcsname{\ensuremath{1.17}}
\expandafter\def\csname qv@N2.rclip.A\endcsname{\ensuremath{1.75}}
\expandafter\def\csname qv@N2.rclip.B\endcsname{\ensuremath{1.07}}
\expandafter\def\csname qv@N2.seed\endcsname{20261110}
\expandafter\def\csname qv@N3.n\endcsname{2,772}
\expandafter\def\csname qv@N3.x.A.F.acc\endcsname{441}
\expandafter\def\csname qv@N3.x.A.F.accpct\endcsname{94\%}
\expandafter\def\csname qv@N3.x.A.F.margin\endcsname{\ensuremath{0.0014}}
\expandafter\def\csname qv@N3.x.A.F.mix\endcsname{25}
\expandafter\def\csname qv@N3.x.A.F.multi\endcsname{467}
\expandafter\def\csname qv@N3.x.A.F.par\endcsname{1}
\expandafter\def\csname qv@N3.x.A.F.same\endcsname{365}
\expandafter\def\csname qv@N3.x.A.F.shift\endcsname{\ensuremath{0.0043}}
\expandafter\def\csname qv@N3.x.A.S.acc\endcsname{23}
\expandafter\def\csname qv@N3.x.A.S.accpct\endcsname{100\%}
\expandafter\def\csname qv@N3.x.A.S.margin\endcsname{\ensuremath{0.0053}}
\expandafter\def\csname qv@N3.x.A.S.mix\endcsname{0}
\expandafter\def\csname qv@N3.x.A.S.multi\endcsname{23}
\expandafter\def\csname qv@N3.x.A.S.par\endcsname{0}
\expandafter\def\csname qv@N3.x.A.S.same\endcsname{22}
\expandafter\def\csname qv@N3.x.A.S.shift\endcsname{\ensuremath{0.0071}}
\expandafter\def\csname qv@N3.x.B.F.acc\endcsname{45}
\expandafter\def\csname qv@N3.x.B.F.accpct\endcsname{38\%}
\expandafter\def\csname qv@N3.x.B.F.margin\endcsname{\ensuremath{0.49}}
\expandafter\def\csname qv@N3.x.B.F.mix\endcsname{2}
\expandafter\def\csname qv@N3.x.B.F.multi\endcsname{117}
\expandafter\def\csname qv@N3.x.B.F.par\endcsname{70}
\expandafter\def\csname qv@N3.x.B.F.same\endcsname{7}
\expandafter\def\csname qv@N3.x.B.F.shift\endcsname{\ensuremath{1.00}}
\expandafter\def\csname qv@N3.z.A.F.acc\endcsname{349}
\expandafter\def\csname qv@N3.z.A.F.accpct\endcsname{95\%}
\expandafter\def\csname qv@N3.z.A.F.margin\endcsname{\ensuremath{0.0016}}
\expandafter\def\csname qv@N3.z.A.F.mix\endcsname{18}
\expandafter\def\csname qv@N3.z.A.F.multi\endcsname{368}
\expandafter\def\csname qv@N3.z.A.F.par\endcsname{1}
\expandafter\def\csname qv@N3.z.A.F.same\endcsname{293}
\expandafter\def\csname qv@N3.z.A.F.shift\endcsname{\ensuremath{0.0043}}
\expandafter\def\csname qv@N3.z.A.S.acc\endcsname{23}
\expandafter\def\csname qv@N3.z.A.S.accpct\endcsname{100\%}
\expandafter\def\csname qv@N3.z.A.S.margin\endcsname{\ensuremath{0.03}}
\expandafter\def\csname qv@N3.z.A.S.mix\endcsname{0}
\expandafter\def\csname qv@N3.z.A.S.multi\endcsname{23}
\expandafter\def\csname qv@N3.z.A.S.par\endcsname{0}
\expandafter\def\csname qv@N3.z.A.S.same\endcsname{23}
\expandafter\def\csname qv@N3.z.A.S.shift\endcsname{\ensuremath{0.16}}
\expandafter\def\csname qv@N3.z.B.F.acc\endcsname{33}
\expandafter\def\csname qv@N3.z.B.F.accpct\endcsname{39\%}
\expandafter\def\csname qv@N3.z.B.F.margin\endcsname{\ensuremath{0.51}}
\expandafter\def\csname qv@N3.z.B.F.mix\endcsname{0}
\expandafter\def\csname qv@N3.z.B.F.multi\endcsname{85}
\expandafter\def\csname qv@N3.z.B.F.par\endcsname{52}
\expandafter\def\csname qv@N3.z.B.F.same\endcsname{3}
\expandafter\def\csname qv@N3.z.B.F.shift\endcsname{\ensuremath{1.00}}
\expandafter\def\csname qv@S10.f05.min\endcsname{\ensuremath{97.1\%}}
\expandafter\def\csname qv@S13.Gb.n\endcsname{48,000}
\expandafter\def\csname qv@S13.MB\endcsname{\ensuremath{1.5\times10^{-14}}}
\expandafter\def\csname qv@S13.cont.nsites\endcsname{8}
\expandafter\def\csname qv@S13.diag.n\endcsname{100,000}
\expandafter\def\csname qv@S13.diag.x.T2\endcsname{\ensuremath{0.00060\ [-0.00028,\,0.00148]}}
\expandafter\def\csname qv@S13.diag.x.pred\endcsname{\ensuremath{0.0033}}
\expandafter\def\csname qv@S13.diag.z.T2\endcsname{\ensuremath{0.00035\ [-0.00030,\,0.00100]}}
\expandafter\def\csname qv@S13.diag.z.pred\endcsname{\ensuremath{0.0028}}
\expandafter\def\csname qv@S13.eps\endcsname{\ensuremath{10^{-12}}}
\expandafter\def\csname qv@S13.moved.A\endcsname{\ensuremath{38\text{--}40\%}}
\expandafter\def\csname qv@S13.moved.n\endcsname{59,040}
\expandafter\def\csname qv@S13.nF\endcsname{2,000}
\expandafter\def\csname qv@S13.nS\endcsname{10,000}
\expandafter\def\csname qv@S13.r12.max\endcsname{\ensuremath{80}}
\expandafter\def\csname qv@S13.r12.min\endcsname{\ensuremath{57}}
\expandafter\def\csname qv@S13.toy.C.R0\endcsname{\ensuremath{0.12\text{--}0.14}}
\expandafter\def\csname qv@S13.toy.C.R1\endcsname{\ensuremath{1.00\text{--}1.03}}
\expandafter\def\csname qv@S13.toy.C.R2\endcsname{\ensuremath{0.96\text{--}1.00}}
\expandafter\def\csname qv@S13.toy.C.twirl\endcsname{\ensuremath{0.19}}
\expandafter\def\csname qv@S13.toy.cells\endcsname{120}
\expandafter\def\csname qv@S13.toy.ok\endcsname{120}
\expandafter\def\csname qv@S13.x.C.R0\endcsname{\ensuremath{0.117\ [0.073,\,0.171]}}
\expandafter\def\csname qv@S13.x.C.R1\endcsname{\ensuremath{0.977\ [0.971,\,0.982]}}
\expandafter\def\csname qv@S13.x.C.R3\endcsname{\ensuremath{0.117\ [0.074,\,0.170]}}
\expandafter\def\csname qv@S13.x.D.R3\endcsname{\ensuremath{-0.005\ [-0.020,\,0.007]}}
\expandafter\def\csname qv@S13.x.Drel.R3\endcsname{\ensuremath{0.6\%}}
\expandafter\def\csname qv@S13.x.Lplus.A\endcsname{\ensuremath{0.55}}
\expandafter\def\csname qv@S13.x.Lplus.B\endcsname{\ensuremath{0.48}}
\expandafter\def\csname qv@S13.x.kF.R0.A\endcsname{\ensuremath{0.89\ [0.81,\,0.98]}}
\expandafter\def\csname qv@S13.x.kF.R0.B\endcsname{\ensuremath{0.49\ [0.47,\,0.52]}}
\expandafter\def\csname qv@S13.x.kF.R1.A\endcsname{\ensuremath{44.1\ [42.7,\,45.5]}}
\expandafter\def\csname qv@S13.x.kF.R1.B\endcsname{\ensuremath{0.49\ [0.47,\,0.52]}}
\expandafter\def\csname qv@S13.x.kF.R2.A\endcsname{\ensuremath{38.1\ [36.6,\,39.6]}}
\expandafter\def\csname qv@S13.x.kF.R2.B\endcsname{\ensuremath{0.49\ [0.47,\,0.52]}}
\expandafter\def\csname qv@S13.x.kF.R3.A\endcsname{\ensuremath{0.89\ [0.80,\,0.97]}}
\expandafter\def\csname qv@S13.x.kF.R3.B\endcsname{\ensuremath{0.49\ [0.47,\,0.52]}}
\expandafter\def\csname qv@S13.x.r.R0.A\endcsname{\ensuremath{1.62}}
\expandafter\def\csname qv@S13.x.r.R0.B\endcsname{\ensuremath{1.03}}
\expandafter\def\csname qv@S13.x.r.R1.A\endcsname{\ensuremath{80.3}}
\expandafter\def\csname qv@S13.x.r.R1.B\endcsname{\ensuremath{1.03}}
\expandafter\def\csname qv@S13.x.r.R2.A\endcsname{\ensuremath{69.3}}
\expandafter\def\csname qv@S13.x.r.R2.B\endcsname{\ensuremath{1.03}}
\expandafter\def\csname qv@S13.x.r.R3.A\endcsname{\ensuremath{1.61}}
\expandafter\def\csname qv@S13.x.r.R3.B\endcsname{\ensuremath{1.03}}
\expandafter\def\csname qv@S13.x.rrange\endcsname{\ensuremath{[1.610,\,1.620]}}
\expandafter\def\csname qv@S13.x.tail.R0.A\endcsname{\ensuremath{11.8\%}}
\expandafter\def\csname qv@S13.x.tail.R0.B\endcsname{\ensuremath{0.1\%}}
\expandafter\def\csname qv@S13.x.tail.R1.A\endcsname{\ensuremath{75.3\%}}
\expandafter\def\csname qv@S13.x.tail.R1.B\endcsname{\ensuremath{0.1\%}}
\expandafter\def\csname qv@S13.x.tail.R2.A\endcsname{\ensuremath{65.8\%}}
\expandafter\def\csname qv@S13.x.tail.R2.B\endcsname{\ensuremath{0.1\%}}
\expandafter\def\csname qv@S13.x.tail.R3.A\endcsname{\ensuremath{11.6\%}}
\expandafter\def\csname qv@S13.x.tail.R3.B\endcsname{\ensuremath{0.1\%}}
\expandafter\def\csname qv@S13.z.C.R0\endcsname{\ensuremath{0.074\ [0.029,\,0.124]}}
\expandafter\def\csname qv@S13.z.C.R1\endcsname{\ensuremath{0.962\ [0.953,\,0.971]}}
\expandafter\def\csname qv@S13.z.C.R3\endcsname{\ensuremath{0.075\ [0.029,\,0.127]}}
\expandafter\def\csname qv@S13.z.D.R3\endcsname{\ensuremath{-0.015\ [-0.026,\,-0.006]}}
\expandafter\def\csname qv@S13.z.Drel.R3\endcsname{\ensuremath{2.4\%}}
\expandafter\def\csname qv@S13.z.Lplus.A\endcsname{\ensuremath{0.39}}
\expandafter\def\csname qv@S13.z.Lplus.B\endcsname{\ensuremath{0.41}}
\expandafter\def\csname qv@S13.z.kF.R0.A\endcsname{\ensuremath{0.66\ [0.60,\,0.72]}}
\expandafter\def\csname qv@S13.z.kF.R0.B\endcsname{\ensuremath{0.42\ [0.40,\,0.45]}}
\expandafter\def\csname qv@S13.z.kF.R1.A\endcsname{\ensuremath{23.1\ [21.8,\,24.4]}}
\expandafter\def\csname qv@S13.z.kF.R1.B\endcsname{\ensuremath{0.42\ [0.40,\,0.45]}}
\expandafter\def\csname qv@S13.z.kF.R2.A\endcsname{\ensuremath{22.0\ [20.7,\,23.3]}}
\expandafter\def\csname qv@S13.z.kF.R2.B\endcsname{\ensuremath{0.42\ [0.40,\,0.45]}}
\expandafter\def\csname qv@S13.z.kF.R3.A\endcsname{\ensuremath{0.64\ [0.58,\,0.71]}}
\expandafter\def\csname qv@S13.z.kF.R3.B\endcsname{\ensuremath{0.42\ [0.40,\,0.45]}}
\expandafter\def\csname qv@S13.z.r.R0.A\endcsname{\ensuremath{1.71}}
\expandafter\def\csname qv@S13.z.r.R0.B\endcsname{\ensuremath{1.04}}
\expandafter\def\csname qv@S13.z.r.R1.A\endcsname{\ensuremath{60.0}}
\expandafter\def\csname qv@S13.z.r.R1.B\endcsname{\ensuremath{1.04}}
\expandafter\def\csname qv@S13.z.r.R2.A\endcsname{\ensuremath{57.0}}
\expandafter\def\csname qv@S13.z.r.R2.B\endcsname{\ensuremath{1.04}}
\expandafter\def\csname qv@S13.z.r.R3.A\endcsname{\ensuremath{1.67}}
\expandafter\def\csname qv@S13.z.r.R3.B\endcsname{\ensuremath{1.04}}
\expandafter\def\csname qv@S13.z.rrange\endcsname{\ensuremath{[1.667,\,1.707]}}
\expandafter\def\csname qv@S13.z.tail.R0.A\endcsname{\ensuremath{8.9\%}}
\expandafter\def\csname qv@S13.z.tail.R0.B\endcsname{\ensuremath{0.0\%}}
\expandafter\def\csname qv@S13.z.tail.R1.A\endcsname{\ensuremath{43.5\%}}
\expandafter\def\csname qv@S13.z.tail.R1.B\endcsname{\ensuremath{0.0\%}}
\expandafter\def\csname qv@S13.z.tail.R2.A\endcsname{\ensuremath{41.7\%}}
\expandafter\def\csname qv@S13.z.tail.R2.B\endcsname{\ensuremath{0.0\%}}
\expandafter\def\csname qv@S13.z.tail.R3.A\endcsname{\ensuremath{8.6\%}}
\expandafter\def\csname qv@S13.z.tail.R3.B\endcsname{\ensuremath{0.0\%}}
\expandafter\def\csname qv@S13c.H8.n\endcsname{200,000}
\expandafter\def\csname qv@S13c.H8.x.dC\endcsname{\ensuremath{8.00\times10^{-4}}}
\expandafter\def\csname qv@S13c.H8.x.dN\endcsname{\ensuremath{8.85\times10^{-4}}}
\expandafter\def\csname qv@S13c.H8.x.holm\endcsname{\ensuremath{0.405}}
\expandafter\def\csname qv@S13c.H8.x.p\endcsname{\ensuremath{0.203}}
\expandafter\def\csname qv@S13c.H8.z.dC\endcsname{\ensuremath{8.20\times10^{-4}}}
\expandafter\def\csname qv@S13c.H8.z.dN\endcsname{\ensuremath{7.85\times10^{-4}}}
\expandafter\def\csname qv@S13c.H8.z.holm\endcsname{\ensuremath{0.677}}
\expandafter\def\csname qv@S13c.H8.z.p\endcsname{\ensuremath{0.677}}
\expandafter\def\csname qv@S13c.x.ov8.A\endcsname{1/8}
\expandafter\def\csname qv@S13c.x.ov8.B\endcsname{8/8}
\expandafter\def\csname qv@S13c.x.tau.A\endcsname{\ensuremath{0.38\ [0.35,\,0.62]}}
\expandafter\def\csname qv@S13c.x.tau.B\endcsname{\ensuremath{0.94\ [0.92,\,0.99]}}
\expandafter\def\csname qv@S13c.z.ov8.A\endcsname{4/8}
\expandafter\def\csname qv@S13c.z.ov8.B\endcsname{7/8}
\expandafter\def\csname qv@S13c.z.tau.A\endcsname{\ensuremath{0.41\ [0.33,\,0.62]}}
\expandafter\def\csname qv@S13c.z.tau.B\endcsname{\ensuremath{0.94\ [0.91,\,0.99]}}
\expandafter\def\csname qv@T71.x.L.A\endcsname{\ensuremath{0.47\ [0.30,\,0.63]}}
\expandafter\def\csname qv@T71.x.L.B\endcsname{\ensuremath{0.47\ [0.43,\,0.50]}}
\expandafter\def\csname qv@T71.x.dL\endcsname{\ensuremath{0.00\ [-0.17,\,0.17]}}
\expandafter\def\csname qv@T71.x.kappa.A\endcsname{\ensuremath{0.47\ [0.32,\,0.66]}}
\expandafter\def\csname qv@T71.x.kappa.B\endcsname{\ensuremath{0.05\ [0.01,\,0.12]}}
\expandafter\def\csname qv@T71.x.ratio.A\endcsname{\ensuremath{1.90\ [1.47,\,2.91]}}
\expandafter\def\csname qv@T71.x.ratio.B\endcsname{\ensuremath{1.06\ [1.01,\,1.13]}}
\expandafter\def\csname qv@T71.z.L.A\endcsname{\ensuremath{0.28\ [0.12,\,0.43]}}
\expandafter\def\csname qv@T71.z.L.B\endcsname{\ensuremath{0.40\ [0.37,\,0.43]}}
\expandafter\def\csname qv@T71.z.dL\endcsname{\ensuremath{-0.12\ [-0.28,\,0.03]}}
\expandafter\def\csname qv@T71.z.kappa.A\endcsname{\ensuremath{0.57\ [0.37,\,0.81]}}
\expandafter\def\csname qv@T71.z.kappa.B\endcsname{\ensuremath{0.05\ [0.01,\,0.10]}}
\expandafter\def\csname qv@T71.z.ratio.A\endcsname{\ensuremath{2.32\ [1.59,\,5.21]}}
\expandafter\def\csname qv@T71.z.ratio.B\endcsname{\ensuremath{1.05\ [1.01,\,1.12]}}
\expandafter\def\csname qv@T72.ci\endcsname{\ensuremath{0.93\text{--}1.09}}
\expandafter\def\csname qv@T72.nsel\endcsname{0}
\expandafter\def\csname qv@T72.pfail\endcsname{\ensuremath{0.5\text{--}13\%}}
\expandafter\def\csname qv@T72.ratio\endcsname{\ensuremath{0.99\text{--}1.03}}
\expandafter\def\csname qv@T73.differ\endcsname{2}
\expandafter\def\csname qv@T73.extra\endcsname{0}
\expandafter\def\csname qv@T73.missed\endcsname{3}
\expandafter\def\csname qv@T73.n\endcsname{1,219}
\expandafter\def\csname qv@T73.same\endcsname{1,217}
\expandafter\def\csname qv@T73.sites.x\endcsname{1,182}
\expandafter\def\csname qv@T73.sites.z\endcsname{1,128}
\expandafter\def\csname qv@T73.tailchg\endcsname{0}
\expandafter\def\csname qv@T73.up\endcsname{\ensuremath{0.5\%}}
\expandafter\def\csname qv@T74.aligned.exact.L\endcsname{\ensuremath{1.64}}
\expandafter\def\csname qv@T74.aligned.exact.Lplus\endcsname{\ensuremath{1.64}}
\expandafter\def\csname qv@T74.aligned.exact.antithetic\endcsname{\ensuremath{3.00}}
\expandafter\def\csname qv@T74.aligned.exact.independent\endcsname{\ensuremath{2.88}}
\expandafter\def\csname qv@T74.aligned.exact.kappa\endcsname{\ensuremath{0.00}}
\expandafter\def\csname qv@T74.aligned.exact.over\endcsname{\ensuremath{1.00}}
\expandafter\def\csname qv@T74.aligned.exact.p\endcsname{\ensuremath{0.0873}}
\expandafter\def\csname qv@T74.aligned.exact.shared\endcsname{\ensuremath{1.64}}
\expandafter\def\csname qv@T74.aligned.theta\endcsname{\ensuremath{0.20, 0.20, 0.20}}
\expandafter\def\csname qv@T74.aligned.twirl.L\endcsname{\ensuremath{0.98}}
\expandafter\def\csname qv@T74.aligned.twirl.Lplus\endcsname{\ensuremath{0.98}}
\expandafter\def\csname qv@T74.aligned.twirl.antithetic\endcsname{\ensuremath{1.00}}
\expandafter\def\csname qv@T74.aligned.twirl.independent\endcsname{\ensuremath{1.00}}
\expandafter\def\csname qv@T74.aligned.twirl.kappa\endcsname{\ensuremath{0.02}}
\expandafter\def\csname qv@T74.aligned.twirl.over\endcsname{\ensuremath{1.02}}
\expandafter\def\csname qv@T74.aligned.twirl.p\endcsname{\ensuremath{0.0293}}
\expandafter\def\csname qv@T74.aligned.twirl.shared\endcsname{\ensuremath{1.00}}
\expandafter\def\csname qv@T74.opposed.exact.L\endcsname{\ensuremath{1.01}}
\expandafter\def\csname qv@T74.opposed.exact.Lplus\endcsname{\ensuremath{1.78}}
\expandafter\def\csname qv@T74.opposed.exact.antithetic\endcsname{\ensuremath{3.00}}
\expandafter\def\csname qv@T74.opposed.exact.independent\endcsname{\ensuremath{2.96}}
\expandafter\def\csname qv@T74.opposed.exact.kappa\endcsname{\ensuremath{0.43}}
\expandafter\def\csname qv@T74.opposed.exact.over\endcsname{\ensuremath{1.00}}
\expandafter\def\csname qv@T74.opposed.exact.p\endcsname{\ensuremath{0.0223}}
\expandafter\def\csname qv@T74.opposed.exact.shared\endcsname{\ensuremath{1.78}}
\expandafter\def\csname qv@T74.opposed.theta\endcsname{\ensuremath{0.20, 0.20, -0.10}}
\expandafter\def\csname qv@T74.opposed.twirl.L\endcsname{\ensuremath{0.99}}
\expandafter\def\csname qv@T74.opposed.twirl.Lplus\endcsname{\ensuremath{0.99}}
\expandafter\def\csname qv@T74.opposed.twirl.antithetic\endcsname{\ensuremath{1.00}}
\expandafter\def\csname qv@T74.opposed.twirl.independent\endcsname{\ensuremath{1.00}}
\expandafter\def\csname qv@T74.opposed.twirl.kappa\endcsname{\ensuremath{0.01}}
\expandafter\def\csname qv@T74.opposed.twirl.over\endcsname{\ensuremath{1.01}}
\expandafter\def\csname qv@T74.opposed.twirl.p\endcsname{\ensuremath{0.0221}}
\expandafter\def\csname qv@T74.opposed.twirl.shared\endcsname{\ensuremath{1.00}}
\expandafter\def\csname qv@T75.nsites\endcsname{96}
\expandafter\def\csname qv@T75.x.Lplus.A\endcsname{\ensuremath{0.55\ [0.52,\,0.72]}}
\expandafter\def\csname qv@T75.x.Lplus.B\endcsname{\ensuremath{0.48\ [0.46,\,0.51]}}
\expandafter\def\csname qv@T75.x.negmass.A\endcsname{\ensuremath{0.08}}
\expandafter\def\csname qv@T75.x.negmass.B\endcsname{\ensuremath{0.01}}
\expandafter\def\csname qv@T75.x.nneg.A\endcsname{0}
\expandafter\def\csname qv@T75.x.nneg.B\endcsname{0}
\expandafter\def\csname qv@T75.x.npt.A\endcsname{15}
\expandafter\def\csname qv@T75.x.npt.B\endcsname{4}
\expandafter\def\csname qv@T75.x.over.A\endcsname{\ensuremath{1.62}}
\expandafter\def\csname qv@T75.x.over.B\endcsname{\ensuremath{1.03}}
\expandafter\def\csname qv@T75.z.Lplus.A\endcsname{\ensuremath{0.39\ [0.37,\,0.53]}}
\expandafter\def\csname qv@T75.z.Lplus.B\endcsname{\ensuremath{0.41\ [0.39,\,0.44]}}
\expandafter\def\csname qv@T75.z.negmass.A\endcsname{\ensuremath{0.10}}
\expandafter\def\csname qv@T75.z.negmass.B\endcsname{\ensuremath{0.01}}
\expandafter\def\csname qv@T75.z.nneg.A\endcsname{0}
\expandafter\def\csname qv@T75.z.nneg.B\endcsname{0}
\expandafter\def\csname qv@T75.z.npt.A\endcsname{20}
\expandafter\def\csname qv@T75.z.npt.B\endcsname{2}
\expandafter\def\csname qv@T75.z.over.A\endcsname{\ensuremath{1.71}}
\expandafter\def\csname qv@T75.z.over.B\endcsname{\ensuremath{1.04}}
\expandafter\def\csname qv@T77.aligned.exact.L\endcsname{\ensuremath{1.51}}
\expandafter\def\csname qv@T77.aligned.exact.Lplus\endcsname{\ensuremath{1.51}}
\expandafter\def\csname qv@T77.aligned.exact.antithetic\endcsname{\ensuremath{2.75}}
\expandafter\def\csname qv@T77.aligned.exact.independent\endcsname{\ensuremath{2.66}}
\expandafter\def\csname qv@T77.aligned.exact.kappa\endcsname{\ensuremath{0.06}}
\expandafter\def\csname qv@T77.aligned.exact.over\endcsname{\ensuremath{1.06}}
\expandafter\def\csname qv@T77.aligned.exact.p\endcsname{\ensuremath{0.1199}}
\expandafter\def\csname qv@T77.aligned.exact.shared\endcsname{\ensuremath{1.60}}
\expandafter\def\csname qv@T77.aligned.theta\endcsname{\ensuremath{0.20, 0.20, 0.20; 0.20, 0.20}}
\expandafter\def\csname qv@T77.aligned.twirl.L\endcsname{\ensuremath{0.96}}
\expandafter\def\csname qv@T77.aligned.twirl.Lplus\endcsname{\ensuremath{0.96}}
\expandafter\def\csname qv@T77.aligned.twirl.antithetic\endcsname{\ensuremath{1.00}}
\expandafter\def\csname qv@T77.aligned.twirl.independent\endcsname{\ensuremath{1.00}}
\expandafter\def\csname qv@T77.aligned.twirl.kappa\endcsname{\ensuremath{0.04}}
\expandafter\def\csname qv@T77.aligned.twirl.over\endcsname{\ensuremath{1.04}}
\expandafter\def\csname qv@T77.aligned.twirl.p\endcsname{\ensuremath{0.0479}}
\expandafter\def\csname qv@T77.aligned.twirl.shared\endcsname{\ensuremath{1.00}}
\expandafter\def\csname qv@T77.opposed.exact.L\endcsname{\ensuremath{1.27}}
\expandafter\def\csname qv@T77.opposed.exact.Lplus\endcsname{\ensuremath{1.53}}
\expandafter\def\csname qv@T77.opposed.exact.antithetic\endcsname{\ensuremath{2.39}}
\expandafter\def\csname qv@T77.opposed.exact.independent\endcsname{\ensuremath{2.36}}
\expandafter\def\csname qv@T77.opposed.exact.kappa\endcsname{\ensuremath{0.20}}
\expandafter\def\csname qv@T77.opposed.exact.over\endcsname{\ensuremath{1.04}}
\expandafter\def\csname qv@T77.opposed.exact.p\endcsname{\ensuremath{0.0600}}
\expandafter\def\csname qv@T77.opposed.exact.shared\endcsname{\ensuremath{1.59}}
\expandafter\def\csname qv@T77.opposed.theta\endcsname{\ensuremath{0.20, 0.20, -0.10; 0.25, 0.15}}
\expandafter\def\csname qv@T77.opposed.twirl.L\endcsname{\ensuremath{0.97}}
\expandafter\def\csname qv@T77.opposed.twirl.Lplus\endcsname{\ensuremath{0.97}}
\expandafter\def\csname qv@T77.opposed.twirl.antithetic\endcsname{\ensuremath{1.00}}
\expandafter\def\csname qv@T77.opposed.twirl.independent\endcsname{\ensuremath{1.00}}
\expandafter\def\csname qv@T77.opposed.twirl.kappa\endcsname{\ensuremath{0.03}}
\expandafter\def\csname qv@T77.opposed.twirl.over\endcsname{\ensuremath{1.03}}
\expandafter\def\csname qv@T77.opposed.twirl.p\endcsname{\ensuremath{0.0422}}
\expandafter\def\csname qv@T77.opposed.twirl.shared\endcsname{\ensuremath{1.00}}
\expandafter\def\csname qv@T79.B\endcsname{10,000}
\expandafter\def\csname qv@T79.Bp1\endcsname{10{,}001}
\expandafter\def\csname qv@T79.bias.A\endcsname{\ensuremath{0.062\text{--}0.068}}
\expandafter\def\csname qv@T79.bias.B\endcsname{\ensuremath{0.005\text{--}0.006}}
\expandafter\def\csname qv@T79.cov.A\endcsname{\ensuremath{1\%}}
\expandafter\def\csname qv@T79.cov.B\endcsname{\ensuremath{93\%}}
\expandafter\def\csname qv@T79.nullmax\endcsname{\ensuremath{3.7}}
\expandafter\def\csname qv@T79.permp\endcsname{\ensuremath{1.0\times10^{-4}}}
\expandafter\def\csname qv@T79.upper\endcsname{\ensuremath{96}}
\expandafter\def\csname qv@T79.x.Lplus.A\endcsname{\ensuremath{0.55}}
\expandafter\def\csname qv@T79.x.Lplus.B\endcsname{\ensuremath{0.48}}
\expandafter\def\csname qv@T79.x.diff\endcsname{\ensuremath{22.7\ [20.8,\,24.7]}}
\expandafter\def\csname qv@T79.x.indep.A\endcsname{\ensuremath{94.6}}
\expandafter\def\csname qv@T79.x.indep.B\endcsname{\ensuremath{94.6}}
\expandafter\def\csname qv@T79.x.lo.A\endcsname{\ensuremath{1.29}}
\expandafter\def\csname qv@T79.x.lo.B\endcsname{\ensuremath{1.00}}
\expandafter\def\csname qv@T79.x.ratio.A\endcsname{\ensuremath{1.62}}
\expandafter\def\csname qv@T79.x.ratio.B\endcsname{\ensuremath{1.03}}
\expandafter\def\csname qv@T79.x.rci\endcsname{\ensuremath{[0.954,\,1.036]}}
\expandafter\def\csname qv@T79.z.Lplus.A\endcsname{\ensuremath{0.39}}
\expandafter\def\csname qv@T79.z.Lplus.B\endcsname{\ensuremath{0.41}}
\expandafter\def\csname qv@T79.z.diff\endcsname{\ensuremath{17.9\ [16.1,\,19.7]}}
\expandafter\def\csname qv@T79.z.indep.A\endcsname{\ensuremath{94.3}}
\expandafter\def\csname qv@T79.z.indep.B\endcsname{\ensuremath{94.3}}
\expandafter\def\csname qv@T79.z.lo.A\endcsname{\ensuremath{1.29}}
\expandafter\def\csname qv@T79.z.lo.B\endcsname{\ensuremath{1.00}}
\expandafter\def\csname qv@T79.z.ratio.A\endcsname{\ensuremath{1.71}}
\expandafter\def\csname qv@T79.z.ratio.B\endcsname{\ensuremath{1.04}}
\expandafter\def\csname qv@T79.z.rci\endcsname{\ensuremath{[0.950,\,1.015]}}
\expandafter\def\csname qv@T80.R\endcsname{400}
\expandafter\def\csname qv@T80.cov.A\endcsname{\ensuremath{100\%}}
\expandafter\def\csname qv@T80.cov.B\endcsname{\ensuremath{100\%}}
\expandafter\def\csname qv@T80.nscen\endcsname{4}
\expandafter\def\csname qv@T80.s.boundary.cov\endcsname{\ensuremath{100\% / 100\%}}
\expandafter\def\csname qv@T80.s.boundary.under\endcsname{\ensuremath{0.23\text{--}0.29 / 0.01\text{--}0.02}}
\expandafter\def\csname qv@T80.s.half.cov\endcsname{\ensuremath{100\% / 100\%}}
\expandafter\def\csname qv@T80.s.half.under\endcsname{\ensuremath{0.28\text{--}0.36 / 0.02}}
\expandafter\def\csname qv@T80.s.measured.cov\endcsname{\ensuremath{100\% / 100\%}}
\expandafter\def\csname qv@T80.s.measured.under\endcsname{\ensuremath{0.18\text{--}0.22 / 0.01}}
\expandafter\def\csname qv@T80.s.sparse.cov\endcsname{\ensuremath{100\% / 100\%}}
\expandafter\def\csname qv@T80.s.sparse.under\endcsname{\ensuremath{0.06\text{--}0.09 / 0.01}}
\expandafter\def\csname qv@T80.under.A\endcsname{\ensuremath{0.06\text{--}0.36}}
\expandafter\def\csname qv@T80.under.B\endcsname{\ensuremath{0.01\text{--}0.02}}
\expandafter\def\csname qv@T81.Bcov\endcsname{400}
\expandafter\def\csname qv@T81.cov.A\endcsname{\ensuremath{100\%}}
\expandafter\def\csname qv@T81.cov.B\endcsname{\ensuremath{100\%}}
\expandafter\def\csname qv@T81.covlo\endcsname{\ensuremath{99.3\%}}
\expandafter\def\csname qv@T81.s.rows.cov\endcsname{\ensuremath{100\% / 100\%}}
\expandafter\def\csname qv@T81.s.rows.under\endcsname{\ensuremath{0.18\text{--}0.23 / 0.01}}
\expandafter\def\csname qv@T81.s.rows_half.cov\endcsname{\ensuremath{100\% / 100\%}}
\expandafter\def\csname qv@T81.s.rows_half.under\endcsname{\ensuremath{0.27\text{--}0.35 / 0.02}}
\expandafter\def\csname qv@T81.under.A\endcsname{\ensuremath{0.18\text{--}0.35}}
\expandafter\def\csname qv@T81.under.B\endcsname{\ensuremath{0.01\text{--}0.02}}
\expandafter\def\csname qv@T81.x.dci\endcsname{\ensuremath{[20.8,\,24.8]}}
\expandafter\def\csname qv@T81.z.dci\endcsname{\ensuremath{[16.1,\,19.7]}}
\expandafter\def\csname qv@T87.x.clmed\endcsname{\ensuremath{36}}
\expandafter\def\csname qv@T87.x.conditional\endcsname{\ensuremath{0.10}}
\expandafter\def\csname qv@T87.x.factual\endcsname{\ensuremath{0.84}}
\expandafter\def\csname qv@T87.x.ncl\endcsname{819}
\expandafter\def\csname qv@T87.x.nv\endcsname{29,009}
\expandafter\def\csname qv@T87.x.one\endcsname{\ensuremath{0.35}}
\expandafter\def\csname qv@T87.z.clmed\endcsname{\ensuremath{52}}
\expandafter\def\csname qv@T87.z.conditional\endcsname{\ensuremath{0.53}}
\expandafter\def\csname qv@T87.z.factual\endcsname{\ensuremath{0.62}}
\expandafter\def\csname qv@T87.z.ncl\endcsname{804}
\expandafter\def\csname qv@T87.z.nv\endcsname{44,501}
\expandafter\def\csname qv@T87.z.one\endcsname{\ensuremath{0.74}}
\expandafter\def\csname qv@ZZ.Bmore\endcsname{24/24}
\expandafter\def\csname qv@ZZ.cells\endcsname{24}
\expandafter\def\csname qv@ZZ.perA\endcsname{\ensuremath{2.6\text{--}3.4}}
\expandafter\def\csname qv@ZZ.perB\endcsname{\ensuremath{1.05\text{--}1.14}}
\expandafter\def\csname qv@ZZ.x4.any.A\endcsname{\ensuremath{30\%}}
\expandafter\def\csname qv@ZZ.x4.any.B\endcsname{\ensuremath{42\%}}
\expandafter\def\csname qv@ZZ.x4.noZZ.A\endcsname{\ensuremath{2.1\%}}
\expandafter\def\csname qv@ZZ.x4.noZZ.B\endcsname{\ensuremath{1.0\%}}
\expandafter\def\csname qv@ZZ.z4.any.A\endcsname{\ensuremath{23\%}}
\expandafter\def\csname qv@ZZ.z4.any.B\endcsname{\ensuremath{39\%}}
\expandafter\def\csname qv@ZZ.z4.noZZ.A\endcsname{\ensuremath{1.5\%}}
\expandafter\def\csname qv@ZZ.z4.noZZ.B\endcsname{\ensuremath{0.9\%}}
\expandafter\def\csname qv@audit.differ\endcsname{3}
\expandafter\def\csname qv@audit.extra\endcsname{0}
\expandafter\def\csname qv@audit.missed\endcsname{4}
\expandafter\def\csname qv@audit.n\endcsname{2,389}
\expandafter\def\csname qv@audit.ncells\endcsname{4}
\expandafter\def\csname qv@audit.same\endcsname{2,386}
\expandafter\def\csname qv@audit.sites.max\endcsname{2,256}
\expandafter\def\csname qv@audit.sites.min\endcsname{1,032}
\expandafter\def\csname qv@audit.tailchg\endcsname{1}
\expandafter\def\csname qv@audit.tailup\endcsname{\ensuremath{0.20\%}}
\expandafter\def\csname qv@audit.up\endcsname{\ensuremath{0.3\%}}
\expandafter\def\csname qv@audit.x4f05.A.differ\endcsname{0}
\expandafter\def\csname qv@audit.x4f05.A.missed\endcsname{0}
\expandafter\def\csname qv@audit.x4f05.A.n\endcsname{300}
\expandafter\def\csname qv@audit.x4f05.A.tail\endcsname{75 / 75}
\expandafter\def\csname qv@audit.x4f05.B.differ\endcsname{1}
\expandafter\def\csname qv@audit.x4f05.B.missed\endcsname{1}
\expandafter\def\csname qv@audit.x4f05.B.n\endcsname{297}
\expandafter\def\csname qv@audit.x4f05.B.tail\endcsname{5 / 5}
\expandafter\def\csname qv@audit.z4f0.A.differ\endcsname{0}
\expandafter\def\csname qv@audit.z4f0.A.missed\endcsname{0}
\expandafter\def\csname qv@audit.z4f0.A.n\endcsname{300}
\expandafter\def\csname qv@audit.z4f0.A.tail\endcsname{5 / 5}
\expandafter\def\csname qv@audit.z4f0.B.differ\endcsname{0}
\expandafter\def\csname qv@audit.z4f0.B.missed\endcsname{0}
\expandafter\def\csname qv@audit.z4f0.B.n\endcsname{300}
\expandafter\def\csname qv@audit.z4f0.B.tail\endcsname{5 / 5}
\expandafter\def\csname qv@audit.z4f05.A.differ\endcsname{1}
\expandafter\def\csname qv@audit.z4f05.A.missed\endcsname{2}
\expandafter\def\csname qv@audit.z4f05.A.n\endcsname{300}
\expandafter\def\csname qv@audit.z4f05.A.tail\endcsname{64 / 64}
\expandafter\def\csname qv@audit.z4f05.B.differ\endcsname{0}
\expandafter\def\csname qv@audit.z4f05.B.missed\endcsname{0}
\expandafter\def\csname qv@audit.z4f05.B.n\endcsname{322}
\expandafter\def\csname qv@audit.z4f05.B.tail\endcsname{7 / 7}
\expandafter\def\csname qv@audit.z8f05.A.differ\endcsname{1}
\expandafter\def\csname qv@audit.z8f05.A.missed\endcsname{1}
\expandafter\def\csname qv@audit.z8f05.A.n\endcsname{300}
\expandafter\def\csname qv@audit.z8f05.A.tail\endcsname{68 / 69}
\expandafter\def\csname qv@audit.z8f05.B.differ\endcsname{0}
\expandafter\def\csname qv@audit.z8f05.B.missed\endcsname{0}
\expandafter\def\csname qv@audit.z8f05.B.n\endcsname{270}
\expandafter\def\csname qv@audit.z8f05.B.tail\endcsname{3 / 3}
\expandafter\def\csname qv@cell.sites.z4\endcsname{1,032}
\expandafter\def\csname qv@cell.sites.z4f05\endcsname{1,128}
\expandafter\def\csname qv@cell.zz.z4\endcsname{96}
\expandafter\def\csname qv@d5.f0.h\endcsname{\ensuremath{6.1}}
\expandafter\def\csname qv@d5.f05.s\endcsname{7,611}
\expandafter\def\csname qv@kappa.s0\endcsname{\ensuremath{0.26\ [0.00,\,0.68]}}
\expandafter\def\csname qv@kappa.sr\endcsname{\ensuremath{0.17\ [0.04,\,0.39]}}
\expandafter\def\csname qv@noise.nzz\endcsname{96}
\expandafter\def\csname qv@noise.theta\endcsname{\ensuremath{0.075}}
\expandafter\def\csname qv@null.n\endcsname{8,000}
\expandafter\def\csname qv@nz.1q.depol\endcsname{\ensuremath{2.9\times10^{-4}}}
\expandafter\def\csname qv@nz.1q.err\endcsname{\ensuremath{3.5\times10^{-4}}}
\expandafter\def\csname qv@nz.1q.gamma\endcsname{\ensuremath{3.7\times10^{-4}}}
\expandafter\def\csname qv@nz.1q.lambda\endcsname{\ensuremath{1.9\times10^{-4}}}
\expandafter\def\csname qv@nz.1q.ns\endcsname{\ensuremath{25}}
\expandafter\def\csname qv@nz.T1\endcsname{\ensuremath{67.2\,\mu\mathrm{s}}}
\expandafter\def\csname qv@nz.T2\endcsname{\ensuremath{88.9\,\mu\mathrm{s}}}
\expandafter\def\csname qv@nz.anc.gamma\endcsname{\ensuremath{1.0\times10^{-2}}}
\expandafter\def\csname qv@nz.anc.lambda\endcsname{\ensuremath{5.2\times10^{-3}}}
\expandafter\def\csname qv@nz.anc.ns\endcsname{\ensuremath{680}}
\expandafter\def\csname qv@nz.cz.depol.f0\endcsname{\ensuremath{2.85\times10^{-3}}}
\expandafter\def\csname qv@nz.cz.depol.f05\endcsname{\ensuremath{1.42\times10^{-3}}}
\expandafter\def\csname qv@nz.cz.err\endcsname{\ensuremath{2.8\times10^{-3}}}
\expandafter\def\csname qv@nz.cz.excess\endcsname{\ensuremath{2.28\times10^{-3}}}
\expandafter\def\csname qv@nz.cz.gamma\endcsname{\ensuremath{5.1\times10^{-4}}}
\expandafter\def\csname qv@nz.cz.lambda\endcsname{\ensuremath{2.6\times10^{-4}}}
\expandafter\def\csname qv@nz.cz.ns\endcsname{\ensuremath{34}}
\expandafter\def\csname qv@nz.cz.rcoh\endcsname{\ensuremath{1.1\times10^{-3}}}
\expandafter\def\csname qv@nz.idle.gamma\endcsname{\ensuremath{1.3\times10^{-2}}}
\expandafter\def\csname qv@nz.idle.lambda\endcsname{\ensuremath{6.5\times10^{-3}}}
\expandafter\def\csname qv@nz.idle.ns\endcsname{\ensuremath{864}}
\expandafter\def\csname qv@repo.commit\endcsname{\texttt{c4a458f3}}
\makeatother
\newcommand{\val}[1]{\ifcsname qv@#1\endcsname\csname qv@#1\endcsname\else\textsf{[\detokenize{#1}]}\fi}
\newcommand{\tabin}[1]{\IfFileExists{\figdir/tables/#1}{\adjustbox{max width=\linewidth}{\input{\figdir/tables/#1}}}{\fbox{table \detokenize{#1} pending}}}
\newcommand{\figin}[2]{\IfFileExists{\figdir/#2}{\includegraphics[width=#1]{#2}}{\fbox{figure \detokenize{#2} pending}}}
\newcommand{\fcoh}{f_{\mathrm{coh}}}
\newcommand{\pfail}{p_{\mathrm{fail}}}
\newcommand{\doI}[1]{\mathrm{do}(#1\leftarrow I)}
\makeatletter\let\switch@array\relax\makeatother

\makeatletter\renewcommand\section{\@startsection{section}{1}{\z@}{-3ex \@plus -1ex \@minus -.2ex}{2ex \@plus.2ex}{\normalfont\Large\sffamily}}
\renewcommand\subsection{\@startsection{subsection}{2}{\z@}{-2.75ex\@plus -1ex \@minus -.2ex}{1.3ex \@plus .2ex}{\normalfont\large\sffamily}}\makeatother

\makeatletter\renewcommand\paragraph{\@startsection{paragraph}{4}{\z@}{1.2ex \@plus.4ex \@minus.2ex}{-1em}{\normalfont\normalsize\bfseries}}\makeatother
\begin{document}

\title{Counterfactual attribution of surface-code failures under non-Pauli noise: dependence on trajectory pairing}
\author{Kisung Jin}
\orcid{0000-0002-1997-3019}
\email{ksjin@etri.re.kr}
\affiliation{Electronics and Telecommunications Research Institute (ETRI), Daejeon, Republic of Korea}
\author{Gyuil Cha}
\orcid{0009-0000-6168-8177}
\email{gicha@etri.re.kr}
\affiliation{Electronics and Telecommunications Research Institute (ETRI), Daejeon, Republic of Korea}
\maketitle

\begin{abstract}
Under Pauli noise in a stabilizer memory, attributing a quantum error-correction failure to the noise that caused it has a canonical answer; under non-Pauli noise the attribution depends on how trajectories are paired with their counterfactuals. We define a failing shot's cause set, the sites whose removal alone removes the failure, and separate it from pairing-independent net effects. The pairing enters wherever a fixed random draw meets a Born probability; making it maximal at every random step selects shared random numbers, while continuity does not. In a distance-three surface code, exact non-Pauli noise and its Pauli twirl fail at rates equivalent within a registered margin, yet under shared draws \val{H5.z.tailA} of exact-noise failures need three or more sites, against \val{H5.z.tailB} under the twirl. Other admissible pairings move the mean coherent cause count from \val{S13.x.r.R3.A} to \val{S13.r12.max} times the minimum the marginals allow, yet pairing-independent quantities barely separate the two models. Shot-level attributions should therefore state their pairing and report pairing-independent effects alongside.
\end{abstract}

\section{Introduction}\label{sec:intro}

A logical failure in a quantum error-correction experiment is the end of a chain: physical noise produces a syndrome, the decoder chooses a correction, and the correction combined with the residual error flips the logical observable. When the noise is Pauli, the chain can be read backwards for each failing shot. A Pauli-frame simulator samples discrete events, and the failure can be traced to the subset of events that, together with the decoder, produced it~\cite{Gidney2021Stim}. This is how fault configurations are enumerated and how rare failures are found and their rates estimated~\cite{BravyiVargo2013,Beverland2025}.

Superconducting processors are not subject to Pauli noise alone. Energy relaxation, dephasing and residual coherent couplings act on every qubit at every step~\cite{Ghosh2012,Willow2025}, and simulations of surface and color codes under such noise report logical error rates that differ from those of the Pauli-twirled model by amounts that depend on the code, schedule and decoder~\cite{Darmawan2017,Bravyi2018,Tuloup2026,Barone2026,Hines2026,Garner2025,HouRavi2026}. Yet tracing a failure back to sampled Pauli-frame events fails for such noise: under a coherent rotation no branch is sampled at all, and under amplitude damping the no-jump branch changes the state even in shots where no jump occurs; a ledger of sampled branches therefore does not say which noise was responsible for a failure, and the question ``which error caused this failure'' has no answer until a counterfactual construction is stated.

\begin{table*}[t]
\centering\small
\begin{tabularx}{\linewidth}{@{}lL@{}}
\toprule
Status & Result \\
\midrule
Proved & Pauli lemma (Appendix~\ref{app:kraus}); Kraus-order invariance for two-branch and state-independent channels (Section~\ref{sec:m-layers}); uniqueness of the stepwise maximal pairing for each intervention, that it implies continuity, and that continuity does not fix the pairing, the detector-restoring pairing R3 being continuous as well (Appendix~\ref{app:canon}); admissible minimum in the parity toy (Section~\ref{sec:m-toy}) \\
Known result, applied & Marginal bounds and their sharpness~\cite{TianPearl2000}; stepwise maximal coupling as counterfactual stability at binary steps~\cite{OberstSontag2019} \\
Preregistered, measured & H1 engine against Pauli frame; H2a spread of cause sets with $\fcoh$; H5 tail difference at equivalent failure rates; four-pairing comparison and its prespecified reading; the fall of the reset-wait share with $\fcoh$ (H4$'$); continuity of the shared and antithetic pairings; criteria not met for H4 (class-level reset wait across idle states), H6 (interaction between halves), H7 (site-level redundancy) and H8 (removing frequent causes against large net effects); exploratory but registered before the run: a convention between sites that re-draws inside the factual outcome (N1), random signs of the coherent rotations (N2) and the measurements a cause flips (N3) (Appendix~\ref{app:hyp}) \\
Exploratory or open & Ratios of cause counts to the sharp bound $L^+$ (Eq.~(\ref{eq:sum}); Section~\ref{sec:r-free}) and their decomposition; paired re-analyses of the registered tests chosen after the data (Appendix~\ref{app:stat}); rank comparison of cause frequency and net effect; the measured continuity of the detector-restoring pairing R3 (Section~\ref{sec:r-coup}); the smallest cause count over all admissible pairings \\
\bottomrule
\end{tabularx}
\caption{Status of the results. H1--H8 are the registered hypotheses, defined in Appendix~\ref{app:hyp} (Table~\ref{tab:hyp}); $\fcoh$ is the share of the excess controlled-$Z$ error made coherent (Section~\ref{sec:r-setup}).}
\label{tab:status}
\end{table*}

Recent work attributes logical failures to circuit components at the population level, as sensitivities of the logical error rate~\cite{ShaoLiu2026,FuStaplesThompson2026}. Population sensitivities answer which components matter on average; they do not say what happened in a given shot. A shot-level answer needs an intervention: change one noise application in the history of the shot and see whether the failure remains. We make this precise with replayable quantum trajectories, in which every random draw of a shot can be held fixed while one noise site is replaced by the identity, and we call the sites whose replacement removes the failure the shot's cause set.

Counterfactuals defined by re-running with the same random numbers depend on that choice of pairing between the factual and the counterfactual run~\cite{Pearl2009}; the probability that a cause was necessary is not identified from interventional data, and the marginals only bound it~\cite{TianPearl2000}. For Pauli noise on a Clifford circuit there is nevertheless a canonical answer: every pairing that holds fixed the draws whose probabilities the removal leaves unchanged gives the decoded counterfactual of removing the sampled event (Section~\ref{sec:m-layers}, Appendix~\ref{app:kraus}). For non-Pauli noise that class of pairings leaves the measurement draws free: removing a small coherent rotation shifts later measurement probabilities, and whether a fixed random number lands on the other side of a threshold depends on how the two draws are paired. On the surface code this freedom moves the mean coherent-$ZZ$ cause count by more than an order of magnitude (Section~\ref{sec:r-coup}). Counterfactual stability~\cite{OberstSontag2019}, imposed at every random step, singles out the shared random draw for the channels used here and their unravelling (Appendix~\ref{app:canon}). We therefore separate attribution statements into two layers: coupling-free statements, differences of marginal probabilities such as the change in the failure rate when sites are removed, which do not depend on the pairing, and coupling-dependent ones, such as the cause set of one shot, which are defined only for a chosen pairing.

We apply the definition to a distance-three rotated surface-code memory under circuit-level noise in two arms that share all random draws: an exact arm with amplitude damping, dephasing and coherent $ZZ$ rotations, and the same circuit with every channel replaced by its Pauli twirl~\cite{Wallman2016}. The comparison is a natural test case for the two layers: on this circuit the twirl leaves the failure rate within a few percent while it changes the nature of the noise. The order-of-magnitude difference in cause sets that we find between the two models is therefore not by itself a difference between the noise models that the failure rate misses: it reflects the noise model and the pairing together, and Section~\ref{sec:r-coup} measures how far the pairing alone can move a cause set when the noise is coherent.

Our contributions are the following (Table~\ref{tab:status} gives the status of each result):
\begin{itemize}
\item \emph{where the pairing enters.} An operational definition of the cause set on replayable trajectories; a lemma that every admissible pairing gives the Pauli-frame counterfactual for Pauli noise in a stabilizer memory; the location of the dependence for non-Pauli noise in the comparison of fixed draws with Born probabilities and the reuse of later draws, not in the order of the Kraus operators; and, applying counterfactual stability, that the stepwise maximal pairing is unique for each intervention and is the shared draw, while continuity does not fix the pairing (Sections~\ref{sec:m-layers} and~\ref{sec:m-toy}, Appendices~\ref{app:kraus} and~\ref{app:canon});
\item \emph{how much it matters.} At failure rates equivalent within a registered margin the cause sets of the exact model and its twirl differ by up to an order of magnitude in their tail and differ in their mean coherent cause count as well (Section~\ref{sec:r-dep}). Over four admissible pairings the exact model's coherent cause count spans more than an order of magnitude while the twirl's does not move, mostly from two discontinuous pairings; the two continuous ones differ little (Section~\ref{sec:r-coup}). The tail difference survives a second convention between sites and random rotation signs (Sections~\ref{sec:r-dep} and~\ref{sec:r-mech});
\item \emph{what can be said without it.} Coupling-free quantities, the sharp marginal bound $L^+$ and the split of a net effect into failure-removing and failure-creating flips, with which the registered tests did not detect the multiplicity of the exact model's cause sets at the population level (Sections~\ref{sec:r-free} and~\ref{sec:r-coup}).
\end{itemize}
That counterfactuals depend on the coupling and are bounded by the marginals is known from causal inference~\cite{Pearl2009,TianPearl2000}, as are counterfactual stability and its stepwise use~\cite{OberstSontag2019,Tsirtsis2021}; what we add is where the dependence enters for quantum trajectories, why admissibility is weak under coherent noise, and how large the dependence is for a surface-code memory. Table~\ref{tab:status} separates what is proved from what is measured, and all registered tests are reported, including the four whose criteria were not met (Appendix~\ref{app:hyp}). In short: the pairing enters at measurements (Section~\ref{sec:method}); on the surface code it changes cause sets a great deal while the coupling-free quantities agree (Section~\ref{sec:results}); and Section~\ref{sec:discussion} says what to report. For work with non-Pauli noise this means that agreeing failure rates need not mean agreeing shot-level explanations, that tools deriving error labels from trajectories must state their pairing, and that the net effect, not the cause frequency, is a coupling-invariant target for choosing where to reduce noise, although we do not show that it leads to better choices (Section~\ref{sec:discussion}).

\section{Background}\label{sec:background}

\paragraph{Surface-code memory and decoding.} We use the rotated surface code of distance three with $d^2=9$ data and $d^2-1=8$ ancilla qubits and $t$ rounds of syndrome extraction followed by a data readout. Detectors are parities of consecutive syndrome outcomes; a detector error model (DEM) lists the fault mechanisms of the circuit with the detectors and the logical observable each flips, and a minimum-weight perfect-matching decoder chooses a correction from the fired detectors~\cite{Higgott2023}. A shot fails when the correction and the error together flip the logical observable.

\paragraph{Non-Pauli trajectories and twirling.} A channel with Kraus operators $\{K_j\}$ is simulated on a pure state $\psi$ by choosing branch $j$ with probability $\|K_j\psi\|^2$ and renormalizing~\cite{Dalibard1992}. For a unitary mixture (a Pauli channel) the branch probabilities do not depend on the state, and a branch other than the identity is an event. For amplitude damping with damping probability $\gamma$ the no-jump branch $K_0=\mathrm{diag}(1,\sqrt{1-\gamma})$ acts in every shot in which no jump occurs at that site, without an event, and a coherent rotation $\exp(-i\theta ZZ/2)$ by an angle $\theta$ has a single branch. The Pauli twirl of a channel keeps the diagonal of its Pauli transfer matrix and turns it into a Pauli channel~\cite{Wallman2016}; at the circuit level it is the standard approximation behind stabilizer simulation of realistic noise~\cite{Ghosh2012}. For coherent errors in surface codes, the twirled model can underestimate the logical error rate below threshold even with noiseless syndrome measurement~\cite{Bravyi2018}, and detector error models estimated from data acquire hyperedges that the twirled model lacks~\cite{TakouBrown2026}.

\paragraph{Attributing logical failures.} Three lines of work are closest. Population sensitivities estimate the derivative of the logical error rate with respect to the strength of each circuit component from the same Monte Carlo samples, and use it to choose where to reduce noise~\cite{ShaoLiu2026,FuStaplesThompson2026}; the score-function estimator of~\cite{ShaoLiu2026} assumes that the distribution of a component's events depends only on its own strength, which does not hold for branch probabilities that depend on the incoming state. Rare-event methods find failures by sampling or enumeration of low-weight fault configurations~\cite{BravyiVargo2013,Beverland2025}. Trajectory samplers that record which error branch each trajectory took exist for unitary-mixture noise~\cite{Patti2025}. None of these defines the cause of an individual failure under non-Pauli noise.

\paragraph{Counterfactuals and couplings.} A counterfactual of an individual unit requires a joint distribution of factual and counterfactual outcomes, which the marginal distributions do not determine~\cite{Pearl2009}. Re-running a stochastic process with the same random numbers fixes one such coupling. Replay of software agents attributes failures to individual steps by replaying the run up to an intervened step and resampling what follows, leaving common random numbers across branches open~\cite{Shah2026}. The probabilities of necessity and sufficiency of a cause are bounded by the interventional marginals and are identified under monotonicity~\cite{TianPearl2000}. Counterfactual stability constrains the coupling of categorical variables, is satisfied by the Gumbel-max mechanism, and reduces to monotonicity for binary ones~\cite{OberstSontag2019}; it has been applied transition by transition in sequential decision processes~\cite{Tsirtsis2021}, and bounds on counterfactuals over all mechanisms of a dynamic latent-state model, with stability or monotonicity as linear constraints, are computed by non-convex polynomial optimization~\cite{HaughSingal2023}. For two binary distributions the coupling that minimizes the probability of disagreement is the standard maximal coupling~\cite{LevinPeresWilmer}. Quantum causal models define causal structure through unitary dynamics, with interventions on quantum channels~\cite{Allen2017QCM,Barrett2019QCM}, and counterfactual queries in them have been given a semantics that generalizes Pearl's abduction, action and prediction~\cite{Suresh2023QCF}; these act on quantum systems and channels, whereas we ask about the classical record of a simulated trajectory, where a counterfactual needs a pairing of random draws. In quantum trajectory theory the analogous fact is that ensemble averages do not depend on the unravelling of a master equation while properties of individual trajectories can~\cite{WisemanMilburn2009}. We borrow the construction and ask which of its statements survive a change of coupling when the process is a quantum trajectory, and how far the cause sets of a surface-code memory are from the bounds.

\section{Method}\label{sec:method}

\subsection{Replayable trajectories}\label{sec:m-traj}

A shot is a trajectory of the circuit under noise. Every noise application --- a channel acting on one or two qubits after a gate, an idle period or a readout --- is a site $i$. The simulator consumes exactly one uniform random number per site, per measurement and per reset, in a fixed canonical order ($r$ at a site, $u$ at a measurement), and chooses the branch of a channel by comparing the draw with the cumulative branch probabilities. All draws of a shot are generated from its seed. Replaying a shot with the same seed reproduces it bit for bit, and records the branch taken at every site (the ledger).

\subsection{Interventions and the cause set}\label{sec:m-cause}

An intervention $\doI{S}$ replaces the channels at the sites in $S$ by the identity and re-runs the shot with the same seed. The identity channel still consumes its draw, so every other site and every measurement sees the same random number as in the factual run. For a failing shot $\tau$ we define the flip indicator and the cause set
\begin{equation}
\begin{split}\phi_i(\tau)&=\mathbf 1\big[Y^{\doI{i}}(\tau)\ne Y(\tau)\big],\\ E(\tau)&=\{\,i:\phi_i(\tau)=1\,\},\end{split}
\end{equation}
where $Y$ is the decoded logical outcome; in a memory experiment a failure is a decoded outcome different from the prepared one, so a flip of $Y$ in a failing shot removes the failure. Write $F$ for the failure event, $\neg F$ for its complement (a success), and $F_i$ for failure of the run with site $i$ removed; $\pfail=P(F)$ is the failure rate of the factual run and $\pfail^{\doI{S}}$ that of the same circuit under $\doI{S}$. Membership in $E$ is a but-for statement about one site at a time, a counterfactual cause-set membership: removing $i$ alone, with everything else as in the shot, removes the failure. It is not a statement about cooperating causes, and $|E|$ counts the sites to which the decoded outcome is individually sensitive; it is not the number of faults. A failure with $E=\emptyset$ while removing all sites removes the failure is diffuse, and a failure can need a set of sites none of which is individually necessary (Fig.~\ref{fig:cases}c). Statements about $E$ refer to the twirled decoder of Section~\ref{sec:r-setup} unless stated otherwise.

\paragraph{Site resolution.} What counts as one site is part of the definition: splitting a rotation into two halves leaves the circuit's channel unchanged but changes which interventions exist, so one cause can become two or none. We use one site per noise application (Section~\ref{sec:r-setup}).

\subsection{Two layers of attribution}\label{sec:m-layers}

The pairing of a factual and a counterfactual run through shared random numbers is a coupling of two probability distributions. Statements that depend only on the marginals are coupling-free (pairing-independent): the net effect of a site set $S$,
\begin{equation}
\Delta_S=\pfail-\pfail^{\doI{S}},
\end{equation}
is the same for every coupling, and so are the failure rate of an arm and the net effect of a gate class. Statements about the cause set of a shot, its size, membership and joint causes are coupling-dependent. For a single site $i$ write $\Delta_i=\pfail\,r^{\mathrm{fs}}_i-(1-\pfail)\,r^{\mathrm{sf}}_i$, where $r^{\mathrm{fs}}_i$ is the share of failing shots that removing $i$ turns into successes and $r^{\mathrm{sf}}_i$ the share of succeeding shots it turns into failures. Cause sets count only the first term. For a set of sites $G$ the ratio
\begin{equation}
\kappa_G=\frac{(1-\pfail)\sum_{i\in G}r^{\mathrm{sf}}_i}{\pfail\sum_{i\in G}r^{\mathrm{fs}}_i}
\end{equation}
is the share of the failure-to-success mass that success-to-failure flips cancel and that is therefore invisible at the population level; it is defined when the failure-to-success mass of $G$ is positive and lies in $[0,1]$ when the summed net effect of $G$ is not negative. At $\kappa_G=0$ no site in $G$ turns a success into a failure (monotonicity), and the whole failure-to-success mass appears in the net effect; as $\kappa_G$ approaches $1$, almost all of it cancels and little remains in the net effect. Both terms are coupling-dependent; only their difference is not.

Two properties locate the coupling dependence. First, for the channels in this paper the joint distribution of factual and counterfactual outcomes does not depend on the order in which a channel's Kraus operators are listed: reordering a two-branch channel maps the draw $r$ to $1-r$ in both runs, a state-independent channel applies one fixed measure-preserving map to both runs, and a coherent rotation has a single branch. For a channel with three or more branches whose probabilities depend on the state this fails (Appendix~\ref{app:kraus}). Second, what remains is the comparison of a fixed draw with a Born probability at each measurement, and the reuse of later draws after an intervention, made precise in the next paragraph. Invariance under relabelling is not invariance under a change of unravelling: rewriting the Kraus operators of a channel as unitary mixtures of one another gives different trajectories and, in general, different cause sets; for dephasing we use the unitary-mixture form in both arms (Section~\ref{sec:r-setup}). The unravelling is therefore a degree of freedom of the same rank as the pairing. For one qubit in $\ket1$ passing three amplitude-damping sites ($\gamma=\val{A8.gamma}$) and failing on outcome $0$, the Kraus operators $\{K_0,K_1\}$ give a mean cause count equal to the sharp bound, \val{A8.standard.mean}, while $\{(K_0\pm K_1)/\sqrt2\}$, the same channel with the same net effects, give \val{A8.rotated.ratio} times the bound and two or more causes in \val{A8.rotated.ge2} of failures (exact, exploratory). Coupling-free statements are invariant under both changes, as long as the channels and the site-level identity intervention stay the same.

\paragraph{Admissible couplings and the Pauli case.} A coupling here is a probabilistic pairing of factual and counterfactual runs, not a physical coupling between qubits. Call a coupling of the factual and a counterfactual run admissible if it holds fixed the draw of every random element --- a site's branch choice or a measurement outcome --- whose probability is the same in both runs. Each run's probabilities are taken conditional on its own history up to the draw, so an admissible coupling is built draw by draw; once the two runs have split, the comparison is between their two conditional distributions, with branches matched by name (Appendix~\ref{app:canon}). Admissibility is thus the minimal requirement that a choice whose probability the removal leaves unchanged stays unchanged. Under Pauli noise it alone fixes the counterfactual; under non-Pauli noise it leaves a large share of the measurement draws free (next paragraph), and one more principle is needed. Consider Pauli noise on a Clifford circuit with stabilizer-state preparation, Pauli measurements and resets, and detectors and a logical observable that are deterministic parities of the measurement record in the noiseless circuit. There the admissible counterfactual of a single-site removal is unique, and its decoded outcome is that of removing the sampled event. Every site's branch probabilities are fixed numbers, unchanged by the removal. A removed Pauli event $Q$ multiplies the state by $Q$, and the outcome probability of a Pauli measurement $M$ on a stabilizer state is $0$, $1$ or $1/2$: if it is $1/2$ it stays $1/2$, and if it is $0$ or $1$ it stays or flips deterministically according to whether $Q$ commutes with $M$. Every random element with a non-trivial distribution therefore keeps it, its draw is held fixed, and the counterfactual is fixed. It need not equal the Pauli-frame record shot by shot: when $Q$ anticommutes with a measurement whose outcome is random, removing $Q$ from the frame flips that outcome, while the admissible replay keeps it. The two counterfactual states then differ by a stabilizer of the state before the measurement, which changes no deterministic parity, so the detectors, the observable and the decoded outcome agree (Appendix~\ref{app:kraus}). The shared-seed coupling is admissible, and on the Pauli arm it reproduced the decoded Pauli-frame counterfactual on every tested pair (Section~\ref{sec:r-valid}).

\begin{table}[t]
\centering\footnotesize
\begin{tabularx}{\linewidth}{@{}>{\raggedright\arraybackslash}p{0.40\linewidth}L@{}}
\toprule
Statement & What it needs \\
\midrule
failure rate, net effects $\Delta_S$, bounds $L$, $L^+$ (Eq.~(\ref{eq:sum})) & the marginals only (coupling-free) \\
$\mathrm{PN}_i$, mean cause count $\sum_i\mathrm{PN}_i$, $\kappa$ & the joint law of the factual run and each single intervention, fixed by stepwise maximality \\
tail $P(|E|\ge3)$, joint membership, size distribution of $E$ & in addition, the coupling of the counterfactuals of different sites with one another (a convention) \\
\bottomrule
\end{tabularx}
\caption{Which coupling each statement needs. Kraus order does not enter for the channels used here; the unravelling enters every row but the first.}
\label{tab:levels}
\end{table}

\paragraph{Non-Pauli noise and the stepwise maximal coupling.} Under amplitude damping or a coherent rotation, removing a site moves later Born probabilities and state-dependent branch probabilities to values that differ between the runs, and admissibility leaves the coupling of those draws free; the shared draw is one choice. This freedom is large. In a check of the engine of Section~\ref{sec:r-coup} on 20 shots per cell of the exact arm from a stream not used elsewhere, every fourth noise site was removed in turn and the shot replayed under the rules R1--R3, and every measurement of each replay was checked, including those before the intervened site. A rule changes a draw only where its probability moved, and it changed at least \val{S13.moved.A} of the \val{S13.moved.n} draws checked per cell: these draws differ in probability between the two runs by more than a tolerance of \val{S13.eps}, so admissibility leaves them free. A stronger requirement is that the pairing be maximal at every random step: given the common history, the factual and counterfactual branches disagree with the smallest probability their two distributions allow, their total-variation distance. For a binary step this coupling is unique and is the shared threshold draw~\cite{LevinPeresWilmer}; for the channels used here the stepwise maximal pairing is therefore unique for each intervention and is the shared draw, and it is admissible and continuous in the strength of the intervention (Appendix~\ref{app:canon}). This fixes the joint law of the factual run and each single intervention, and with it the mean cause count $\sum_i\mathrm{PN}_i$ (defined below); the full law of the cause set, such as its tail, also depends on how the counterfactuals of different sites are coupled with one another, and we use the same seed for all of them. That is a further convention, not implied by stepwise maximality: the mean cause count and the bounds below do not depend on it, but the tail of the cause-set size and the registered tests on it (H2a, H5) do (Table~\ref{tab:levels}). Section~\ref{sec:r-dep} measures the tail under a second convention. It is the trajectory form of counterfactual stability~\cite{OberstSontag2019}, a principle of minimal disturbance and not an optimum: it does not minimize the cause count, and continuity alone does not single it out. The canonical counterfactual is not monotone even under Pauli noise: removing a Pauli event can create a failure as well as remove one, for example when the failure is a parity of two events. Throughout the paper, the shared draw, the shared-seed coupling, the (stepwise) maximal pairing and the rule R0 of Section~\ref{sec:r-coup} name this one pairing.

\paragraph{Bounds from the marginals.} Failure rates alone do not fix how many causes a failure has: the marginals bound the cause count, and the coupling of the two runs decides where in that range it falls. If removing a site lowers the failure rate from 10\,\% to 8\,\%, the removal may have rescued 2\,\% of the shots and harmed none, or it may have rescued every failure while turning successes amounting to 8\,\% of the shots into failures. The rates are the same, yet the site is in the cause set of one failure in five in the first case and of every failure in the second. Let $\mathrm{PN}_i=P(i\in E\mid F)$ be the probability that site $i$ is in the cause set of a failing shot, the probability of necessity of~\cite{TianPearl2000}, and $q_i=\pfail^{\doI{i}}$. It is a coupling-dependent quantity, but the marginals bound it: for an intervention that, as here, is applied independently of the shot,
\begin{equation}
\max\!\big(0,\Delta_i/\pfail\big)\le\mathrm{PN}_i\le\min\!\big(1,(1-q_i)/\pfail\big),
\end{equation}
and the upper bound is $1$ whenever $\pfail+q_i\le1$, as in every cell here~\cite{TianPearl2000}. For $\Delta_i\ge0$ the lower bound is attained by a coupling under which removing $i$ never turns a success into a failure (monotonicity); for $\Delta_i<0$, by one under which it never turns a failure into a success. Monotonicity is therefore a property of a noise model together with a coupling, not of the noise model alone, and the marginals identify $\mathrm{PN}_i$ only if one assumes it. Summing over a set of sites $G$, the conditional mean number of members of $G$ in a cause set satisfies
\begin{equation}
\begin{split}\mathbb E\big[\,|E\cap G|\;\big|\;F\big]&=\sum_{i\in G}\mathrm{PN}_i\\ &\ge L^+_G=\frac{1}{\pfail}\sum_{i\in G}\max(0,\Delta_i)\\ &\ge L_G=\frac{1}{\pfail}\sum_{i\in G}\Delta_i.\end{split}\label{eq:sum}
\end{equation}
With $a_i=P(F,\neg F_i)$ and $b_i=P(\neg F,F_i)$, the joint probabilities under the pairing that removing $i$ removes or creates a failure, so that $\Delta_i=a_i-b_i$ and $\mathrm{PN}_i=a_i/\pfail$, the gap to the middle bound $L^+_G$, which we call the sharp bound for the reason below, is $\sum_{i\in G}\min(a_i,b_i)/\pfail$ and the gap to the signed bound $L_G$ is $\sum_{i\in G}b_i/\pfail$. The signed bound is attained only if no site of $G$ turns a success into a failure, which is impossible when some $\Delta_i<0$. The sharp bound is attained when every site flips the outcome in one direction only, and it is sharp in the following sense: the coupling of the outcomes in which one uniform number $U$ decides all of them, $F=\mathbf 1[U<\pfail]$ and $F_i=\mathbf 1[U<q_i]$, attains it at every site at once, and it can be realized on trajectories by drawing each run conditional on its outcome. The marginals therefore cannot exclude a mean cause count as low as the sharp bound: they do not force a cause count above it. The bound is sharp over all couplings of the outcomes, not over admissible ones. The construction above need not hold the trajectory's draws fixed (in the one-qubit toy of Section~\ref{sec:m-toy} the shared draw realizes it), and an admissible coupling need not reach the bound; in the parity toy of Section~\ref{sec:m-toy} none does. How far admissible couplings can lower the cause count on the surface code is not determined here. Under the shared-seed coupling, for $\kappa_G<1$, $\mathbb E[|E\cap G|\,|\,F]/L_G=1/(1-\kappa_G)$: the distance of the cause sets from the signed bound and the cancelled share $\kappa_G$ are one quantity, not two. The other extreme is easy to write down. Under independent replays, $\mathrm{PN}_i=1-q_i$, close to $1$ for every site, and nearly every site would be in every cause set. The shared-seed coupling is one point in this range; the toy circuits below show how far a cause set moves across it.

\subsection{Toy circuits with a variable coupling}\label{sec:m-toy}

One qubit prepared in $\ket0$ passes three noise sites and is measured in the $Z$ basis; the shot fails on outcome $1$. In the exact model site $k$ applies $\exp(-i\theta_k X/2)$, with $\theta=(\val{T74.opposed.theta})$, so the only randomness is the measurement draw; in the twirled model site $k$ applies $X$ with probability $\sin^2(\theta_k/2)$, the site draws are held fixed and the measurement is deterministic. The angles are illustrative and small, the third has the opposite sign so that removing it raises the failure rate ($\Delta_3<0$), and $\theta_3=-\theta_1\theta_2/(\theta_1+\theta_2)$ makes the two models fail at nearly the same rate (\val{T74.opposed.exact.p} and \val{T74.opposed.twirl.p}); the qualitative statements below do not depend on the angles as long as they are not degenerate. We compare three couplings of the factual and counterfactual measurement draws with the same marginals: the shared draw, the antithetic draw $r\mapsto1-r$, and an independent draw. In the exact model the mean cause-set size is \val{T74.opposed.exact.shared}, \val{T74.opposed.exact.antithetic} and \val{T74.opposed.exact.independent} under the three; the antithetic draw puts every site in every cause set because $\pfail+q_i\le1$. The twirled model's cause set, the set of $X$ events, is \val{T74.opposed.twirl.shared} under all three and lies \val{T74.opposed.twirl.over} times above its sharp bound, because two $X$ events cancel and removing one can create a failure. Since the only random element of the exact model is the final measurement, the shared draw is the comonotone coupling of the outcome and attains the sharp bound of Eq.~(\ref{eq:sum}) exactly; $\kappa=\val{T74.opposed.exact.kappa}$ there comes from the third site's net effect, which the marginals force.

In this toy family and under the shared draw, the exact model leaves the sharp bound only if the outcome combines several measurements. In a second toy two qubits pass their own sites and the shot fails on the parity of the two outcomes; removing a site of the first qubit changes only its probability of outcome $1$, from $p_1$ to $p_1'<p_1$, while the second qubit, with probability $0<p_2<1$, keeps its draw. Whatever the second outcome, an admissible coupling then flips the first outcome with probability at least $p_1-p_1'$, so the removal turns failures into successes with probability $a_i\ge(p_1-p_1')(1-p_2)$ and successes into failures with probability $b_i\ge(p_1-p_1')\,p_2>0$, and symmetrically when $p_1'>p_1$. The shared draw attains both minima and gives the smallest cause sets of all admissible couplings, yet $b_i>0$ keeps every admissible coupling above the sharp bound: with $\theta=(\val{T77.opposed.theta})$ the shared draw lies \val{T77.opposed.exact.over} times above it. Values for all configurations are in the artifact. The toys show where the coupling enters --- the measurement draw, and outcomes composed of several measurements --- but not the size of the surface-code excess (Section~\ref{sec:r-free}), which may involve mid-circuit measurements, the reuse of later draws and a decoder.

\subsection{Screening the cause sets}\label{sec:m-est}

Testing every site individually costs one replay per site and failing shot, and the $Z$-basis circuit with $t=4$ has \val{cell.sites.z4} to \val{cell.sites.z4f05} sites. We therefore screen in three steps. (i) \emph{Group test.} The sites are split into 24 blocks; each block is removed as a whole, and sites are tested individually only inside the blocks whose removal flips the outcome. (ii) \emph{Coherent $ZZ$ sites.} The group test assumes that removing a block flips the outcome when one of its sites does. This can fail for coherent $ZZ$, where removing a block can leave the failure while removing one $ZZ$ site inside it removes it, as we observed in exploratory runs; every coherent-$ZZ$ site, and in arm B every twirled channel at a $ZZ$ location, is therefore always tested individually, and the $ZZ$ membership of a cause set is not affected by the screening. (iii) \emph{Checks.} A candidate diffuse failure is confirmed by testing every site individually, and removing all sites is checked to remove each failure. The same violation can in principle occur at other sites, because removing several Pauli faults together can restore the parity of a failure, so the screening is audited in both arms against an exhaustive single-site scan (Section~\ref{sec:r-valid}, Table~\ref{tab:audit}). The result is the \emph{screened cause set} $\hat E\subseteq E$, which can miss members but adds none; we write $E$ for $\hat E$ in the results, and every tail mass below is one of screened cause sets. The algorithm is described in Appendix~\ref{app:stat}.

The coupling-free counterpart of a cause set with several members is an interaction. Split a set of sites into groups $G_1,G_2$ and measure, on the same shots, the net effects $\Delta_1,\Delta_2$ of removing each and $\Delta_{12}$ of removing both. The normalized interaction
\begin{equation}
\eta=\frac{\Delta_{12}-\Delta_1-\Delta_2}{\Delta_1+\Delta_2}
\end{equation}
is $0$ when the two groups contribute additively and $-1/2$ when they are jointly necessary in an idealized sense: the failures they cause carry the same mass, removing either group removes all of it, and neither removal has any other effect. Being a combination of marginals, $\eta$ is coupling-free; shared random numbers leave its population value unchanged and change the variance of its estimate through the covariance of the terms. Because it is a net quantity, success-to-failure flips enter it with the opposite sign, so a test on $\eta$ excludes this idealized pattern rather than every form of joint necessity in individual shots.

\section{Results}\label{sec:results}

\subsection{Setup}\label{sec:r-setup}

The experiment compares two noise models built to fail at the same rate while differing in kind: an exact model with amplitude damping, dephasing and a coherent rotation, and its Pauli twirl. Both run on the same random numbers, so that a difference in attribution between them can come neither from different failure rates nor from different luck.

\paragraph{Circuit and noise.} We simulate the $d=3$ rotated surface-code memory with $t\in\{2,4,8\}$ rounds in the $Z$ and $X$ bases, with controlled-$Z$ syndrome extraction. Relaxation times, gate durations and average gate errors are the medians of a hand-authored device model built from representative published values of a superconducting processor~\cite{Willow2025}, not a per-qubit calibration (its controlled-$Z$ duration, \val{nz.cz.ns}~ns, is shorter than the 42~ns of that processor's error budget~\cite{Willow2025}, so relaxation during the gate is weaker and more of the fixed gate error is depolarizing or coherent). They set relaxation and dephasing at every gate and idle location, depolarizing noise for the remaining single-qubit error, two-qubit depolarizing noise for the excess controlled-$Z$ error, and an optional idle of the ancilla after reset (the reset wait; cells $\sigma_{\mathrm{reset}}$, without it $\sigma_0$). A fraction $\fcoh\in\{0,0.25,0.5\}$ of the controlled-$Z$ excess error is moved into a coherent rotation $\exp(-i\theta ZZ/2)$ of fixed sign at every gate, keeping the average gate error fixed to first order; at $\fcoh=0.5$, $\theta=\val{noise.theta}$, a share and sign set by this study (Appendix~\ref{app:noise}). Such a rotation, the channel $\mathcal U_\theta(\rho)=e^{-i\theta ZZ/2}\rho\,e^{i\theta ZZ/2}$, is at diamond distance $\tfrac12\|\mathcal U_\theta-\mathrm{id}\|_\diamond=\sin(\theta/2)$ from the identity but has average error $4\sin^2(\theta/2)/5$, so equal average error is not equal worst-case error; we compare the arms at failure rates shown to be equivalent within a registered margin. Readout is noiseless in all reported cells except the null-intervention check of Section~\ref{sec:r-valid} and the class-level test H4 of Section~\ref{sec:r-wait}. The $Z$-basis circuit with $t=4$ has \val{cell.sites.z4} noise sites at $\fcoh=0$ and \val{cell.sites.z4f05} at $\fcoh=0.5$, of which \val{cell.zz.z4} are coherent $ZZ$ sites. In arm A every channel is exact; in arm B every channel is replaced by its Pauli twirl, and both arms use the same seeds. Dephasing, a Pauli channel equal to its own twirl, is registered in both arms in the same unitary-mixture form, so the channels that differ are amplitude damping and the coherent rotation; the form changes how draws are paired, not the noise. The noiseless circuit is a fixed Clifford schedule with stabilizer preparation, $Z$- or $X$-basis measurements and resets and no feed-forward, and the decoder reads only the detectors, so arm B meets the assumptions of the lemma in Appendix~\ref{app:kraus}.

\paragraph{Simulation and decoding.} Shots are simulated with the matrix-product-state backend of an extension of the sparse state-vector emulator QPlayer~\cite{Jin2023QPlayer}, used before for logical-qubit synthesis~\cite{Jin2023Multilayered} and backend selection~\cite{Jin2026Affine}, with a bond-dimension cap of 256 and a singular-value cutoff of $10^{-12}$; the recorded truncation error is zero, so the simulation is exact up to floating-point arithmetic; the largest bond dimension reached in the \val{A11.n} recorded failing shots of each of four $Z$-basis $t=4$ cells is \val{A11.chi}. All runs are reproduced by the public release v6.0.1-Lynx of the emulator, whose source is included in the artifact~\cite{JinPaper2Artifact}; the pairings of Section~\ref{sec:r-coup}, the second convention of Section~\ref{sec:r-dep} and the runs of Section~\ref{sec:r-mech} use in addition research patches, included there as well, that record the measurement draws and re-pair them or replace the channels of single sites. The decoder is minimum-weight matching on the DEM of the twirled model (twirled decoder) and, as a check, on the same graph with uniform weights (uniform decoder). As a further check we use a decoder whose DEM edge weights are fitted to the two-point correlations of the detectors in the data.

\begin{figure*}[t]
\centering
\figin{\linewidth}{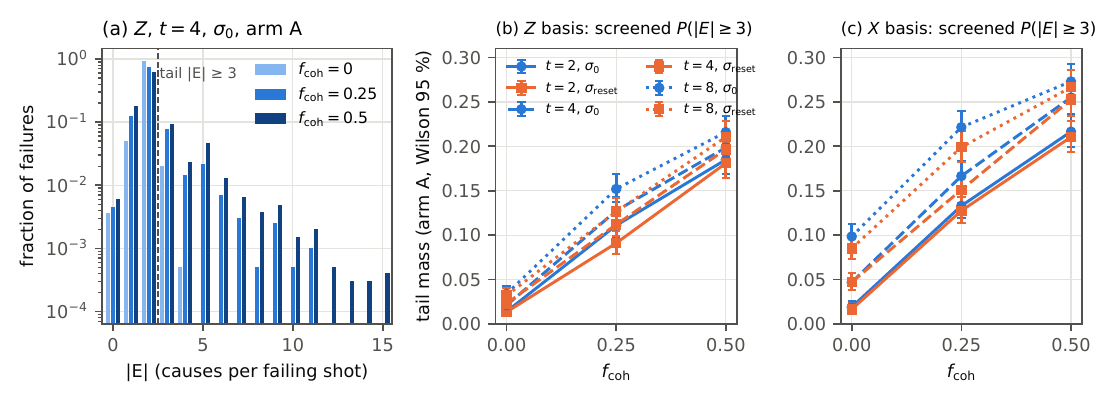}
\caption{Screened cause-set sizes (Section~\ref{sec:m-est}), main-grid stream. A failure is a decoded logical outcome $Y$ different from the prepared one (event $F$); $|E|$ counts the sites whose removal alone flips $Y$, not the faults. (a) Distribution of $|E|$ over failing shots of arm A for $\fcoh=0$, $0.25$ and $0.5$. (b, c) Tail mass $P(|E|\ge3)$ against $\fcoh$ for the twelve cell pairs, with Wilson 95\,\% intervals.}
\label{fig:sets}
\end{figure*}

\begin{table*}[t]
\centering\small
\begin{tabular}{lccrc}
\toprule
Cell pair & screened $P(|E|\ge3)$, $f_{\mathrm{coh}}=0\to0.25\to0.5$ & $R$ [95\,\%] & Holm $p$ & uniform decoder \\
\midrule
z t2 $\sigma_0$ & 1.4 $\to$ 11.1 $\to$ 18.5 & 13.2 [9.4, 20.4] & $9.4\times10^{-84}$ & 1.3 $\to$ 17.4 \\
z t2 $\sigma_{\mathrm{reset}}$ & 1.4 $\to$ 9.1 $\to$ 18.1 & 13.4 [9.5, 21.0] & $3.4\times10^{-82}$ & 1.3 $\to$ 19.5 \\
z t4 $\sigma_0$ & 2.1 $\to$ 12.8 $\to$ 19.9 & 9.5 [8.3, 10.9] & $<10^{-300}$ & 2.0 $\to$ 20.1 \\
z t4 $\sigma_{\mathrm{reset}}$ & 2.3 $\to$ 11.2 $\to$ 19.7 & 8.5 [7.5, 9.8] & $<10^{-300}$ & 1.9 $\to$ 20.2 \\
z t8 $\sigma_0$ & 3.4 $\to$ 15.2 $\to$ 21.6 & 6.4 [5.0, 8.3] & $1.0\times10^{-73}$ & 3.6 $\to$ 21.8 \\
z t8 $\sigma_{\mathrm{reset}}$ & 3.2 $\to$ 12.7 $\to$ 21.1 & 6.6 [5.2, 8.7] & $3.7\times10^{-73}$ & 2.6 $\to$ 22.5 \\
x t2 $\sigma_0$ & 1.9 $\to$ 13.3 $\to$ 21.7 & 11.4 [8.5, 16.5] & $6.3\times10^{-95}$ & 1.5 $\to$ 24.0 \\
x t2 $\sigma_{\mathrm{reset}}$ & 1.7 $\to$ 12.8 $\to$ 21.1 & 12.8 [9.3, 18.9] & $2.4\times10^{-95}$ & 1.7 $\to$ 22.8 \\
x t4 $\sigma_0$ & 4.7 $\to$ 16.7 $\to$ 25.5 & 5.4 [4.4, 6.8] & $1.1\times10^{-80}$ & 4.4 $\to$ 27.1 \\
x t4 $\sigma_{\mathrm{reset}}$ & 4.7 $\to$ 15.1 $\to$ 25.3 & 5.4 [4.4, 6.7] & $2.3\times10^{-79}$ & 4.2 $\to$ 26.5 \\
x t8 $\sigma_0$ & 9.9 $\to$ 22.2 $\to$ 27.3 & 2.8 [2.4, 3.2] & $3.5\times10^{-47}$ & 9.1 $\to$ 28.2 \\
x t8 $\sigma_{\mathrm{reset}}$ & 8.5 $\to$ 19.9 $\to$ 26.7 & 3.1 [2.7, 3.7] & $3.9\times10^{-53}$ & 8.1 $\to$ 26.9 \\
\bottomrule
\end{tabular}

\caption{Tail mass of the screened cause-set size (\%) by cell pair, main-grid stream, twirled decoder (uniform decoder in the last column). $R$ is the tail mass at $\fcoh=0.5$ over that at $0$, with a 95\,\% interval from independent binomials; each cell ran until arm A had its failure target. Holm $p$ is the registered one-sided Fisher test, which treats the two cells as independent; ``$<10^{-300}$'' is the floating-point limit. The paired analysis that respects the shared random numbers is in the text.}
\label{tab:sets}
\end{table*}

\paragraph{Registration.} Hypotheses, test statistics, stopping rules and random streams were approved and written down before each measurement in a version-controlled protocol, published in the artifact~\cite{JinPaper2Artifact}; Appendix~\ref{app:hyp} lists every hypothesis with its status. A timestamp shows this only for some versions: the first commit of the protocol version precedes the first shot for H5, the distance-five gate, the 32-cell part of H2a, the four pairings with continuity and H8, and the runs N1--N3, while the versions of H1, H4, H4$'$, H6, H7, the four-cell part of H2a and the diagnostic of Section~\ref{sec:r-coup} were first committed after their first shot; the artifact lists both times for every version. Confirmatory measurements used one seed base with disjoint offsets per experiment; exploratory runs used other seed bases or offsets that no confirmatory measurement used. The main grid covers \val{E4.cells} cells (two bases $\times$ $t\in\{2,4,8\}$ $\times$ with or without reset wait $\times$ three values of $\fcoh$), each run until arm A collected 10,000 failures (the four $Z$-basis $t=4$ cells with $\fcoh\in\{0,0.5\}$) or 2,000 failures (the other cells; Appendix~\ref{app:stat}). Cells that share the basis, $t$ and the reset wait and differ only in $\fcoh$ form a cell pair, and the registered tests compare $\fcoh=0.5$ with $\fcoh=0$ in each of the twelve cell pairs. The random streams depend on the basis and $t$ but not on the noise, so the two cells of a pair run on the same random numbers; the registered tests nevertheless treat them as independent samples. We therefore give, for each registered test, an analysis that respects the dependence, chosen after the registration and marked exploratory, and report the registered $p$ values as registered (Appendix~\ref{app:stat}, Table~\ref{tab:dep}). Analyses marked exploratory were chosen after the data were seen.

\begin{figure*}[t]
\centering
\figin{\linewidth}{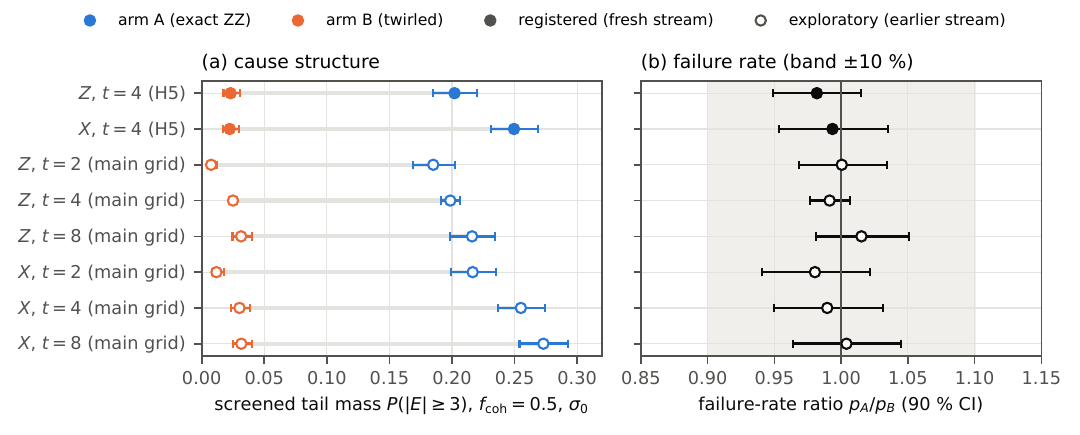}
\caption{Exact (A) against twirled (B) noise at $\fcoh=0.5$, $\sigma_0$; rows are cells labelled by basis and rounds $t$. (a) Share of failing shots whose screened cause set has three or more sites, $P(|E|\ge3)$, with Wilson 95\,\% intervals: in every cell arm A (blue) lies far above arm B (orange). This is coupling-dependent, a statement under the shared-seed pairing. (b) Ratio of the failure rates of the two arms, $p_A/p_B$, with 90\,\% intervals; the shaded band is the registered $\pm10\,\%$ equivalence margin, and the registered criterion of equivalence is a 90\,\% interval inside it, which every interval meets. This is coupling-free. Together the panels show failure rates that agree with cause structures that do not. Filled, labelled (H5): registered confirmation on a fresh stream; open, labelled (main grid): the main-grid cells, exploratory, where the margin was not registered, so the $Z$-basis $t=4$ cell appears once per stream.}
\label{fig:twirl}
\end{figure*}

\subsection{The intervention machinery}\label{sec:r-valid}

Before any cause set is read, the replay machinery has to be trusted. We checked it in three ways (Table~\ref{tab:checks}). First, it changes nothing when nothing should change: a null intervention, which splits a channel into two half-weight copies of its Kraus operators and replaces one copy by the identity, changed no record in either arm; with no intervention the engine hook reproduced the unmodified build bit for bit, the matrix-product-state and state-vector engines agreed bit for bit, every failing shot replayed identically, and no replay prefix diverged from the factual run. Second, where an independent answer exists, the engine gives it: on a Pauli arm its counterfactual agreed in decoded outcome with the Pauli-frame counterfactual computed from the ledger in all \val{H1.pairs} pairs of a failing shot and one of its sampled events (registered test H1, whose Wilson bound treats pairs from one shot as independent; counted by shot, all \val{A6.shots} failing shots agree). Third, the screening that finds cause sets cheaply agrees with an exhaustive scan: in an exploratory audit of \val{audit.n} failing shots in both arms of \val{audit.ncells} cells, including the two H5 cells, it missed \val{audit.missed} members and added none. Two consequences matter for what follows. Because the screening can miss a member but never adds one, the screened tails of Fig.~\ref{fig:sets} and Table~\ref{tab:sets} are lower bounds on the full ones, and misses in arm A can only widen the difference between the arms. And the misses are far too rare to erase that difference: erasing the H5 tail difference would need \val{A4.need.z} ($Z$) and \val{A4.need.x} ($X$) of the arm-B failures to move into the tail, whereas the audit found \val{A4.recl.B} of \val{A4.n.B}. Removing every site removed the failure in every failing shot of arm A, which checks the replay, not the completeness of the cause sets.

\subsection{The coupling-dependent layer: cause sets}\label{sec:r-dep}

\paragraph{Coherent noise spreads the cause sets.} Coherent noise makes failures depend on more sites at once. At $\fcoh=0$ most failing shots have a cause set of two sites (Fig.~\ref{fig:sets}a; an example is Fig.~\ref{fig:cases}a), as expected when two faults defeat distance three; $|E|$ counts sites to which the outcome is sensitive, not faults, and the two coincide in such shots. Moving part of the gate error into coherent $ZZ$ fattens the tail of $|E|$. The registered statistic is the tail mass $P(|E|\ge3\mid F)$, written $P(|E|\ge3)$ below; over the twelve cell pairs it rises from \val{H2a.f0} at $\fcoh=0$ to \val{H2a.f05} at $\fcoh=0.5$, a ratio of \val{H2a.R} (Table~\ref{tab:sets}, Fig.~\ref{fig:sets}b,c), and the dose response is monotone in all twelve. The rise does not depend on how it is tested or decoded. All twelve pairs reject with both decoders after Holm correction, in the registered test, which treats the two cells of a pair as independent, and in a paired re-analysis that respects their shared random numbers (exploratory; ratios \val{A1.t.R}, Table~\ref{tab:dep}); the dependence is weak, the failure indicators of the two cells correlating by $\phi=\val{A1.phi}$. A decoder fitted to the two-point detector correlations of the data gives the same ratio in the two cell pairs where it was run (Appendix~\ref{app:stat}). Unlike the mean cause count, the tail also depends on how the counterfactuals of different sites are coupled to one another; we use the same draws for all of them, a convention beyond the stepwise maximal pairing, and the registered tests H2a and H5 are statements under it (Section~\ref{sec:m-layers}).

\paragraph{The twirl does not produce a comparable tail.} The twirl, by contrast, leaves the tail small although it fails at an equivalent rate. In the registered comparison of the two arms at $\fcoh=0.5$, $\sigma_0$, $t=4$ on a fresh stream, the tail mass is \val{H5.z.tailA} in arm A against \val{H5.z.tailB} in arm B in the $Z$ basis, and \val{H5.x.tailA} against \val{H5.x.tailB} in the $X$ basis (Fig.~\ref{fig:twirl}a); the uniform decoder gives the same verdict. Because the arms share seeds, the main evidence is paired: resampling shot indices with both arms of an index kept together puts the difference at \val{T79.z.diff} and \val{T79.x.diff} percentage points (95\,\% intervals; chosen after the registration and therefore exploratory). The registered Fisher test ($p=\val{H5.z.p}$ and \val{H5.x.p}) treats the arms as independent samples; it is reported as registered and is not the basis of the conclusion. A bootstrap that reproduces the stopping rule and a permutation of the arm labels give the same picture (Table~\ref{tab:dep}; paired counts in Table~\ref{tab:pairs}). The two arms also fail largely in different shots: only \val{H5.z.shared} and \val{H5.x.shared} of the failing shots of arm A also fail in arm B with the same seed, an overlap that is itself a property of the pairing. And the interventions differ in kind: removing a twirled site deletes a sampled event or does nothing, whereas removing an exact site changes later Born probabilities in every shot; Section~\ref{sec:r-coup} gives a reference point with weakened interventions.

\paragraph{A second convention between sites.} A cause set is built from separate counterfactual replays, one per site. The stepwise maximal pairing fixes how each replay is paired with the factual run, but not how the replays are paired with one another. It therefore fixes, for each site, the probability that removing it removes the failure, and with it the mean cause count; it does not fix whether several sites are causes in the same shot, which is what the tail $P(|E|\ge3)$ counts. Knowing each coin's chance of heads does not tell how often three or more come up heads together; that depends on how the coins are correlated. Of the statistics above, only the tail depends on this choice (Table~\ref{tab:levels}). The results so far use the convention R0, the same seed for all replays (Section~\ref{sec:m-layers}), under which the replays depart from the factual run at the same draws and so move together. To check that the tail difference is not a product of this convention, we registered a second convention, R4, before running it. R4 keeps the probability $|p-p'|$ with which each replay disagrees with the factual run and scatters only where the disagreement falls, independently for each replay: where the intervention moves a measurement probability, the counterfactual draw is re-drawn uniformly inside the factual outcome's interval, $u'=v\,p$ or $p+v(1-p)$, with an auxiliary uniform $v$ independent for each intervention. Each single intervention keeps the stepwise maximal pairing, so only the dependence between replays changes. On the H5 stream (\val{N1.nA} failures of arm A per cell, every site tested individually) the difference between the arms remains: \val{N1.z.D.R4} ($Z$) and \val{N1.x.D.R4} ($X$) percentage points under R4 against \val{N1.z.D.R0} and \val{N1.x.D.R0} under R0 (95\,\% intervals; arm-A tails \val{N1.z.T.R4} and \val{N1.x.T.R4}), the outcome for which the registered reading keeps the headline. The mean cause count agreed between R0 and R4 (paired differences \val{N1.z.dE} and \val{N1.x.dE}), as it must if the mean does not depend on how the replays are coupled with one another. One of three engine gates failed through a flaw in its registered definition; Appendix~\ref{app:hyp} discloses it with a diagnostic.

\paragraph{Where the excess comes from.} The excess lies in the coherent rotations, and specifically in how many of them a failure needs. Splitting each cause set into coherent-$ZZ$ sites, whose set we also write $ZZ$, and the rest locates it: without the $ZZ$ sites the tail is small in both arms (\val{ZZ.z4.noZZ.A} against \val{ZZ.z4.noZZ.B} in the $Z$ basis at $t=4$). It is not that $ZZ$ sites are causes more often in arm A. At least one is in the cause set more often under the twirl ($P(E\cap ZZ\ne\emptyset\mid F)=\val{ZZ.z4.any.A}$ in arm A against \val{ZZ.z4.any.B} in arm B; the same holds in \val{ZZ.Bmore} cells with $\fcoh>0$). It is that when they are causes in arm A they come several at a time: \val{ZZ.perA} on average over the \val{ZZ.cells} cells with $\fcoh>0$, against \val{ZZ.perB} in arm B (per-cell values in the artifact). The mean number of $ZZ$ causes per failure needs only the pairing with each single intervention; it is \val{H7.z.kF.A} against \val{H7.z.kF.B} on the stream of Section~\ref{sec:r-free}. Put plainly, under the shared-seed pairing a failing shot of arm A whose cause set contains coherent rotations typically has several, each individually necessary for the failure; in arm B such a shot typically has one twirled event, and in the seed of Fig.~\ref{fig:cases}b, chosen for this contrast, no $ZZ$ site is a cause at all. We call this pattern, several individually necessary coherent rotations in one cause set, multiplicity; how much of it reaches quantities that do not depend on the pairing is what the next section asks.

\subsection{The coupling-free layer}\label{sec:r-free}

Section~\ref{sec:r-dep} showed that, under the shared-seed pairing, a failure of the exact model typically hinges on several coherent rotations at once, while a failure of the twirl hinges on one event. Another pairing of the same shots could change that. This section therefore asks whether the difference shows in quantities that every pairing must reproduce because they depend only on the marginals: the failure rate, the net effects of removing classes and groups of sites, and the marginal bounds of Eq.~(\ref{eq:sum}) on the cause count. The answer is that it barely does. The two models fail at equivalent rates, removing sites changes their failure rates by similar amounts, and neither registered test of joint necessity detected the multiplicity at the population level. Yet the exact model's cause sets lie well above the smallest count its marginals allow, while the twirl's lie close to it. The records show why: in the exact model, removing a coherent rotation rescues some failing shots and spoils about half as many succeeding ones, and the two cancel in every population quantity. Table~\ref{tab:free} collects the quantities side by side, and the paragraphs below take them in turn. Of these, only the failure rate was registered as an equivalence test; H6 and H7 are registered tests whose criteria were not met; the decomposition and the bounds were chosen after the data were seen and are exploratory.

\begin{table*}[t]
\centering\footnotesize
\begin{adjustbox}{max width=\linewidth}
\begin{tabular}{@{}llcccc@{}}
\toprule
& & \multicolumn{2}{c}{$Z$ basis} & \multicolumn{2}{c}{$X$ basis} \\
\cmidrule(lr){3-4}\cmidrule(lr){5-6}
Quantity & Layer, status & Arm A & Arm B & Arm A & Arm B \\
\midrule
\multicolumn{6}{@{}l}{\emph{Population quantities}} \\
Failure-rate ratio $p_A/p_B$ [90\,\%] & free; reg., met & \multicolumn{2}{c}{\val{H5.z.ratio}} & \multicolumn{2}{c}{\val{H5.x.ratio}} \\
\quad stopped-design bootstrap [90\,\%] & free; expl. & \multicolumn{2}{c}{\val{T79.z.rci}} & \multicolumn{2}{c}{\val{T79.x.rci}} \\
Class net effect, ratio B/A, median (IQR) & free; expl. & \multicolumn{4}{c}{cz \val{DG.cz} (\val{DG.cz.iqr}) \quad h \val{DG.h} (\val{DG.h.iqr}) \quad id \val{DG.id} (\val{DG.id.iqr})} \\
\quad cells with separated intervals & free; expl. & \multicolumn{4}{c}{cz \val{DG.cz.sep} \quad h none \quad id \val{DG.id.sep}} \\
Reduction from removing all $ZZ$, A over B & free; not tested & \multicolumn{2}{c}{\val{H6.z.zzratio}} & \multicolumn{2}{c}{\val{H6.x.zzratio}} \\
\midrule
\multicolumn{6}{@{}l}{\emph{Tests of joint necessity}} \\
Interaction $\eta$ of halves [95\,\%] & free; H6, not met & \val{H6.z.A} & \val{H6.z.B} & \val{H6.x.A} & \val{H6.x.B} \\
Redundancy $R_{\mathrm{red}}$ [95\,\%] & free; H7, not met & \val{H7.z.A} & \val{H7.z.B} & \val{H7.x.A} & \val{H7.x.B} \\
\quad $R_A-R_B$, index-paired [95\,\%] & free; expl. & \multicolumn{2}{c}{\val{A2.z.diff}} & \multicolumn{2}{c}{\val{A2.x.diff}} \\
\midrule
\multicolumn{6}{@{}l}{\emph{Decomposition of $\sum_i\Delta_i$ (expl.)}} \\
$ZZ$ causes per failure, $\mathbb E[k_F]$ & dep.\ (single) & \val{H7.z.kF.A} & \val{H7.z.kF.B} & \val{H7.x.kF.A} & \val{H7.x.kF.B} \\
Success-to-failure flips & dep.\ (single) & \val{H7.z.kS.A} & \val{H7.z.kS.B} & \val{H7.x.kS.A} & \val{H7.x.kS.B} \\
Cancelled share $\kappa_{ZZ}$ [95\,\%] & dep.\ (single) & \val{T71.z.kappa.A} & \val{T71.z.kappa.B} & \val{T71.x.kappa.A} & \val{T71.x.kappa.B} \\
\midrule
\multicolumn{6}{@{}l}{\emph{Marginal bounds on $\mathbb E[k_F]$ (expl.)}} \\
Signed bound $L_{ZZ}$ [95\,\%] & free & \val{T71.z.L.A} & \val{T71.z.L.B} & \val{T71.x.L.A} & \val{T71.x.L.B} \\
\quad $\mathbb E[k_F]/L_{ZZ}$ [95\,\%] & & \val{T71.z.ratio.A} & \val{T71.z.ratio.B} & \val{T71.x.ratio.A} & \val{T71.x.ratio.B} \\
Sharp bound $L^+_{ZZ}$, clipped & free & \val{T79.z.Lplus.A} & \val{T79.z.Lplus.B} & \val{T79.x.Lplus.A} & \val{T79.x.Lplus.B} \\
\quad cross-fitted [95\,\%] & free & \val{A3.z.cf.A} & \val{A3.z.cf.B} & \val{A3.x.cf.A} & \val{A3.x.cf.B} \\
$\mathbb E[k_F]/L^+_{ZZ}$, clipped & & \val{T79.z.ratio.A} & \val{T79.z.ratio.B} & \val{T79.x.ratio.A} & \val{T79.x.ratio.B} \\
\quad cross-fitted & & \val{A3.z.rcf.A} & \val{A3.z.rcf.B} & \val{A3.x.rcf.A} & \val{A3.x.rcf.B} \\
Excess $D=\mathbb E[k_F]-L^+_{ZZ}$, clipped [95\,\%] & & \val{A12.z.Dclip.A} & \val{A12.z.Dclip.B} & \val{A12.x.Dclip.A} & \val{A12.x.Dclip.B} \\
\quad cross-fitted [95\,\%] & & \val{A12.z.Dcf.A} & \val{A12.z.Dcf.B} & \val{A12.x.Dcf.A} & \val{A12.x.Dcf.B} \\
\bottomrule
\end{tabular}
\end{adjustbox}
\caption{Coupling-free quantities and the decomposition that relates them to the cause sets; $\fcoh=0.5$, $\sigma_0$, $t=4$, H7 stream (\val{H7.nF} failures and \val{H7.nS} successes per cell and arm) unless stated otherwise: failure-rate ratios from the H5 stream, class net effects from the main grid, $\eta$ from the H6 stream (\val{H6.n} shots per cell and arm). ``free'': coupling-free; ``dep.\ (single)'': coupling-dependent, but needs only the pairing with each single intervention; ``reg.'': registered test; ``expl.'': exploratory. Holm-adjusted $p$: H6 \val{H6.z.holm}, H7 \val{H7.z.holm} in both cells. Success-to-failure flips are summed over the \val{noise.nzz} $ZZ$ sites, so one success can count more than once. Class net effects: over \val{DG.cells} main-grid cells, the controlled-$Z$ ratio ranging over \val{DG.cz.range}; gate keys cz, controlled-$Z$; h, single-qubit gates; id, data idles. Clipped estimates of $L^+_{ZZ}$ are biased upward and their ratios and excesses $D$ downward, cross-fitted ones the other way; the intervals of the clipped excess are percentile intervals of this biased estimator, so its point estimate can lie at their upper end (Appendix~\ref{app:stat}).}
\label{tab:free}
\end{table*}

\paragraph{Failure rate and removal effects.} The failure rate is the one quantity shown equivalent: the ratio between the arms is \val{H5.z.ratio} ($Z$) and \val{H5.x.ratio} ($X$), with 90\,\% intervals inside the registered $\pm10\,\%$ margin (Fig.~\ref{fig:twirl}b), and they stay inside when a bootstrap reproduces the stopping rule (Table~\ref{tab:free}). The margin has a power rationale only, the sample size at which the planned test could show equivalence, not a physical one; the intervals themselves state which differences the data exclude. Removing a whole gate class changes the two failure rates by similar, but not identical, amounts: over \val{DG.cells} cells the ratio of arm B to arm A has median \val{DG.cz} for the controlled-$Z$ class, \val{DG.h} for single-qubit gates and \val{DG.id} for data idles, and the difference is resolved in some cells (Table~\ref{tab:free}; per arm, on the main-grid failures and the first 2,000 successes of the same index range, arm B measured in a later run; exploratory). Removing all coherent-$ZZ$ sites lowers the failure rate of arm A about as much as that of arm B (\val{H6.z.zzratio} to \val{H6.x.zzratio} times). The one qualitative difference is the expected one: the exact amplitude damping of the ancilla wait after reset leaves $\ket0$ unchanged and has no effect in arm A, while its twirl has an effect whose interval excludes zero in nearly every reset-wait cell, a check that the net effects see what the channels predict.

\paragraph{No joint necessity at the population level.} If several rotations were jointly necessary in the exact arm's failures, two population quantities should show it, and we registered a test of each. The first, H6, registered before its run, splits the $ZZ$ sites of each ancilla and round into the first and the second half of its controlled-$Z$ gates and measures the interaction $\eta$ between the halves; joint necessity of the halves would give $\eta$ near $-1/2$. The interactions are small and alike in the two arms, with overlapping intervals and lower limits of about \val{H6.Alow} or higher (Table~\ref{tab:free}); the registered criterion, $\eta_A<\eta_B$ with the interval of $\eta_A$ below zero, fails in both cells, and a control split by round parity gives the same picture. A split into halves cannot see joint necessity inside one half, so the second test, H7, registered after the first, works site by site: the redundancy $R_{\mathrm{red}}=\sum_i\Delta_i/\Delta_Z$, the sum of the single-site net effects of the \val{noise.nzz} $ZZ$ sites over the net effect of removing them all. Additive sites give $R_{\mathrm{red}}=1$. It equals $k$ if every failure that the $ZZ$ sites cause has exactly $k$ individually necessary $ZZ$ sites whose joint removal removes it, no removal turns a success into a failure, and no failure needs $ZZ$ sites without any one of them being individually necessary, as a failure that needs any two of three does. Each term is a difference of marginals, so $R_{\mathrm{red}}$ is coupling-free. It lies near one in both arms (Table~\ref{tab:free}), and the registered one-sided test of $R_A>R_B$ rejects in neither cell, also when the two arms of a shot index are kept together (Table~\ref{tab:dep}). The intervals of arm A are wide, so the test does not show the arms equal; but their upper limits, at most \val{H7.Aup}, exclude the value near three that the per-shot multiplicity of Section~\ref{sec:r-dep} would give under these idealized conditions. Both tests exclude an aggregate pattern; neither shows that rotations are never jointly necessary in a shot.

\paragraph{Where the multiplicity goes.} The two sides of $\sum_i\Delta_i$ reconcile the per-shot multiplicity with these population results. On the failure side, a failure of arm A has on average $\mathbb E[k_F]=\val{H7.z.kF.A}$ $ZZ$ sites in its cause set, $k_F=|E\cap ZZ|$, against \val{H7.z.kF.B} in arm B ($Z$ basis). On the success side, removing single $ZZ$ sites from the same number of successes creates far more failures in arm A, \val{H7.z.kS.A} against \val{H7.z.kS.B}. These success-to-failure flips cancel a share $\kappa_{ZZ}$ of the failure-to-success mass: about half in arm A (\val{T71.z.kappa.A}) and almost none under the twirl (\val{T71.z.kappa.B}); the $X$ basis is alike (Table~\ref{tab:free}). Since $\sum_i\Delta_i=\pfail\,\mathbb E[k_F](1-\kappa_{ZZ})$, a large cancellation hides a large cause count from the net effects, which is consistent with the non-rejection of H7. $\kappa\le1$ holds because the summed net effect of the $ZZ$ sites is positive, so an upper interval limit below one is a constraint of the data, not of the definition; because $\kappa$ rests on rare success-to-failure flips, it is estimated on the H7 stream with its \val{H7.nS} successes per cell and arm. Both terms are coupling-dependent; only their difference is not. The identity is algebraic, not a statement about the power of the test, and the redundancy test, $\kappa$ and the bound below are one decomposition seen three ways, not three independent pieces of evidence.

\paragraph{How far the cause sets are from the marginal bounds.} The same records say how far each arm's cause sets lie from the smallest count its marginals allow. The signed bound $L_{ZZ}$ of Eq.~(\ref{eq:sum}) does not separate the arms (differences \val{T71.z.dL} in the $Z$ basis and \val{T71.x.dL} in the $X$ basis), and the cause sets of arm A exceed it by $1/(1-\kappa_{ZZ})$, about twice. The sharp bound $L^+_{ZZ}=\sum_i\max(0,\Delta_i)/\pfail$ of Eq.~(\ref{eq:sum}), obtained by recording which $ZZ$ sites flip each shot, is the stricter reference; the comonotone coupling attains it. Under the shared-seed coupling the cause sets of the twirl lie close to it, at \val{T79.z.ratio.B} ($Z$) and \val{T79.x.ratio.B} ($X$) times, and those of the exact arm well above it, at \val{T79.z.ratio.A} and \val{T79.x.ratio.A} times (ratios computed before rounding). The estimate of $L^+_{ZZ}$ clips noisy site effects at zero and errs upward, so these ratios err downward; a cross-fitted estimate, taking each site's sign from one half of the shots and its size from the other, errs the other way and puts the exact arm's ratio at \val{A3.rcf.A}. Under both estimates the excess $D=\mathbb E[k_F]-L^+_{ZZ}$ of arm A excludes zero (Table~\ref{tab:free}; the bias and the calibration of the interval are in Appendix~\ref{app:stat}). The marginals do not force this excess: a coupling of the outcomes attains the bound, although not necessarily one that holds the trajectory's draws fixed. The range is wide at the other end too: with the same marginals the mean $ZZ$ cause count can rise to \val{T79.upper}, all $ZZ$ sites, and independent replays would put about \val{T79.z.indep.A} of them in an average cause set, so the shared-seed coupling lies near the bottom of the range in both arms. These re-analyses of the H7 shots were not registered.

\begin{figure*}[t]
\centering
\figin{\linewidth}{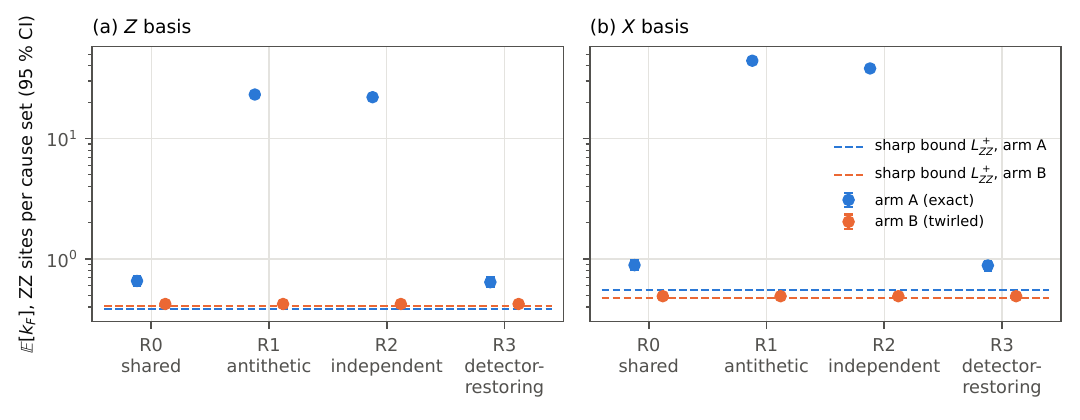}
\caption{Mean number of coherent-$ZZ$ sites in a cause set under four admissible pairings (log scale, 95\,\% intervals), with the sharp marginal bound of each arm (dashed), which is over all couplings of the outcomes and need not be attainable by an admissible pairing. (a) $Z$ basis, (b) $X$ basis; $\fcoh=0.5$, $\sigma_0$, $t=4$.}
\label{fig:coup}
\end{figure*}

\paragraph{What the layer shows.} Against these mostly small differences in coupling-free quantities, the cause sets of Section~\ref{sec:r-dep} differ by up to an order of magnitude. The difference in cause sets therefore cannot be read as a difference between the noise models alone. It is a property of the exact noise under the chosen pairing, in which coherent rotations both rescue and spoil shots and the two effects cancel in every population quantity we measured. Whether another admissible pairing would shrink it is the subject of Section~\ref{sec:r-coup} (Table~\ref{tab:coup}, Fig.~\ref{fig:coup}).

\begin{table*}[t]
\centering\small
\begin{adjustbox}{max width=\linewidth}
\begin{tabular}{@{}llccccc@{}}
\toprule
& & & \multicolumn{2}{c}{continuous} & \multicolumn{2}{c}{not continuous} \\
\cmidrule(lr){4-5}\cmidrule(lr){6-7}
Basis & Arm & $L^+_{ZZ}$ (clipped) & R0 shared & R3 detector-restoring & R1 antithetic & R2 independent \\
\midrule
$Z$ & A & \val{S13.z.Lplus.A} & \val{S13.z.kF.R0.A} & \val{S13.z.kF.R3.A} & \val{S13.z.kF.R1.A} & \val{S13.z.kF.R2.A} \\
$Z$ & B & \val{S13.z.Lplus.B} & \val{S13.z.kF.R0.B} & \val{S13.z.kF.R3.B} & \val{S13.z.kF.R1.B} & \val{S13.z.kF.R2.B} \\
$X$ & A & \val{S13.x.Lplus.A} & \val{S13.x.kF.R0.A} & \val{S13.x.kF.R3.A} & \val{S13.x.kF.R1.A} & \val{S13.x.kF.R2.A} \\
$X$ & B & \val{S13.x.Lplus.B} & \val{S13.x.kF.R0.B} & \val{S13.x.kF.R3.B} & \val{S13.x.kF.R1.B} & \val{S13.x.kF.R2.B} \\
\midrule
\multicolumn{7}{@{}l}{\emph{Ratio to the clipped $L^+_{ZZ}$}} \\
$Z$ & A & --- & \val{S13.z.r.R0.A} & \val{S13.z.r.R3.A} & \val{S13.z.r.R1.A} & \val{S13.z.r.R2.A} \\
$Z$ & B & --- & \val{S13.z.r.R0.B} & \val{S13.z.r.R3.B} & \val{S13.z.r.R1.B} & \val{S13.z.r.R2.B} \\
$X$ & A & --- & \val{S13.x.r.R0.A} & \val{S13.x.r.R3.A} & \val{S13.x.r.R1.A} & \val{S13.x.r.R2.A} \\
$X$ & B & --- & \val{S13.x.r.R0.B} & \val{S13.x.r.R3.B} & \val{S13.x.r.R1.B} & \val{S13.x.r.R2.B} \\
\midrule
\multicolumn{3}{@{}l}{R3 $-$ R0, arm A} & \multicolumn{2}{c}{$Z$: \val{S13.z.D.R3}; $X$: \val{S13.x.D.R3}} & & \\
\bottomrule
\end{tabular}
\end{adjustbox}
\caption{Mean number of coherent-$ZZ$ sites in a cause set, $\mathbb E[k_F]$ [95\,\%], under four admissible pairings of the same shots, with the sharp bound $L^+_{ZZ}$, and their ratio to the clipped bound (computed before rounding); $\fcoh=0.5$, $\sigma_0$, $t=4$, \val{S13.nF} failures per cell and arm. ``Coherent-$ZZ$ sites'' names locations: in arm B they carry the twirled channel. Continuity in the intervention strength is as measured in Section~\ref{sec:r-coup}. The sharp bound is over all couplings of the outcomes and need not be attainable by an admissible pairing; its clipped estimate is biased upward (Section~\ref{sec:r-free}).}
\label{tab:coup}
\end{table*}

\subsection{Other admissible pairings on the surface code}\label{sec:r-coup}

\begin{figure*}[t]
\centering
\figin{\linewidth}{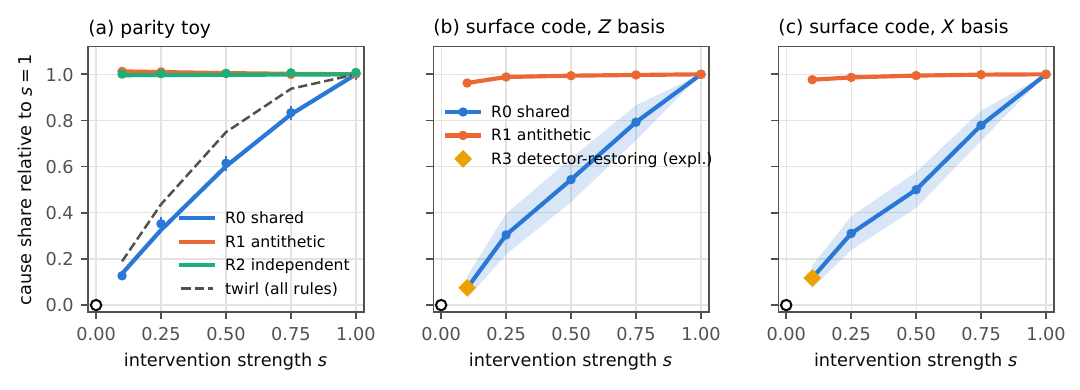}
\caption{Share of failing shots in which a weakened site is a cause, relative to full removal ($s=1$), against the strength $s$ of the intervention; $s=0$ is the factual run (open circle). (a) Parity toy, exact arm: closed form (lines) and engine (points, 95\,\% intervals), with the twirled arm (dashed), the same under every pairing. (b, c) Surface code, arm A, \val{S13.cont.nsites} sites per basis, shot-bootstrap 95\,\% bands; the diamond is R3 at $s=0.1$ (exploratory); on the surface code the registration covered R0 and R1 only, so R2 is not shown there. A pairing is continuous if the share falls to zero as the intervention weakens. The curves are at finite strengths; the limit $s\to0$ is the statement of Appendix~\ref{app:canon}. R0 and R3 fall towards zero, whereas under R1 (and R2 in the toy) a site weakened by a tenth is a cause almost as often as when it is removed. Continuity thus excludes the antithetic and independent pairings but does not single out R0.}
\label{fig:cont}
\end{figure*}

\paragraph{Four pairings.} The cause sets above were all computed under one pairing, the shared draw. To see how far the choice of pairing moves the cause sets of the surface code, we extended the engine with a ledger of measurement draws and their Born probabilities, and replayed the shots of the redundancy test (\val{S13.nF} failures and \val{S13.nS} successes per cell and arm, $\fcoh=0.5$, $\sigma_0$, $t=4$) under four pairings registered before the runs (Table~\ref{tab:coup}, Fig.~\ref{fig:coup}). All four hold fixed every measurement draw whose probability differs between the factual and the counterfactual run by at most \val{S13.eps}, a tolerance set from the floating-point spread of the Pauli arm (\val{S13.MB}); all four hold fixed the draws of noise sites and resets, so only measurement draws are varied. When the probabilities differ, they choose the counterfactual draw $u'$ from the factual draw $u$ in four ways: R0 keeps $u'=u$, the stepwise maximal pairing of Section~\ref{sec:m-layers} and the one used everywhere else in this paper; R1 takes $u'=1-u$; R2 draws $u'$ independently; R3 keeps $u$ if the previous measurement of the same qubit agrees between the two runs and takes $1-u$ otherwise, a greedy rule that pushes the detector of the round back towards its factual value and is the one candidate we expected could lower the cause count below R0. Each rule transforms the current draw using only the past, so the counterfactual draws remain independent and uniform and the counterfactual marginal is unchanged. Under R0 the extended engine reproduced the earlier records bit for bit on all \val{S13.Gb.n} shots, and on the Pauli arm the four pairings gave the same decoded outcome on every shot, as the lemma of Appendix~\ref{app:kraus} predicts.

\paragraph{The range.} The choice of pairing moves the cause sets of the exact arm a great deal and those of the twirl not at all (Table~\ref{tab:coup}, Fig.~\ref{fig:coup}). In arm A, R1 and R2, which are not continuous in the intervention strength (Fig.~\ref{fig:cont}), raise the mean $\mathbb E[k_F]$ to between \val{S13.r12.min} and \val{S13.r12.max} times the sharp bound, and in the $X$ basis a cause set of three or more $ZZ$ sites becomes the rule (\val{S13.x.tail.R1.A} under R1 against \val{S13.x.tail.R0.A} under R0). R3 is the only rule below R0, and only slightly: by \val{S13.z.Drel.R3} in the $Z$ basis, and not resolved in the $X$ basis. Arm B does not move: its four pairings give identical cause sets, at \val{S13.z.r.R0.B} and \val{S13.x.r.R0.B} times the bound. The smallest cause count over admissible pairings is therefore bracketed, $L^+_{ZZ}\le\inf_{\mathrm{adm}}\mathbb E[k_F]\le\min_R\mathbb E[k_F]$, between \val{S13.z.Lplus.A} and \val{S13.z.kF.R3.A} in the $Z$ basis and between \val{S13.x.Lplus.A} and \val{S13.x.kF.R3.A} in the $X$ basis; in ratio to the bound, the tested pairings span \val{S13.z.r.R3.A} to \val{S13.z.r.R1.A} ($Z$) and \val{S13.x.r.R3.A} to \val{S13.x.r.R1.A} ($X$), point estimates with the bias of Section~\ref{sec:r-free}. We registered a reading for each outcome: if some rule halved the excess over the bound in both cells, the excess would be largely a choice of pairing; if no rule went below R0, the shared draw would be the smallest of those tested; otherwise only the range is reported. The result is the third case. None of the tested admissible pairings removes a substantial part of the excess of the exact arm, and whether some other admissible pairing does is open; the analogue for classical dynamic models is the gap between bounds over all mechanisms and particular mechanisms~\cite{HaughSingal2023}.

\paragraph{A check on the coupling-free sum.} The sum $\sum_i\Delta_i$ over the $ZZ$ sites is coupling-free, so its estimates under the four pairings, from the same shots, should differ only by sampling variation. They agree closely: the paired difference from R0 just excludes zero under R1 in both cells and under R3 in the $Z$ basis, and a registered diagnostic on fresh shots found the R1 deviation consistent with sampling variation once the site that contributed most was removed (Appendix~\ref{app:stat}).

\paragraph{Continuity in the strength of the intervention.} Continuity rules out the extreme pairings but does not choose among the rest. It requires that weakening a site a little should change a shot's counterfactual a little. We replace a site's rotation angle $\theta$ by $(1-s)\theta$ for $s\in\{0.1,0.25,0.5,0.75,1\}$ and measure, relative to full removal, how often the weakened site is a cause of a failure. In the toy circuits, where the engine's weakened intervention matches the closed form (Appendix~\ref{app:canon}), the pairings split (Fig.~\ref{fig:cont}a). In the exact arm, a site weakened by a tenth is a cause only \val{S13.toy.C.R0} times as often as when it is removed under R0, as small as the intervention, but \val{S13.toy.C.R1} and \val{S13.toy.C.R2} times under R1 and R2, as often as under full removal. In the twirled arm the share is \val{S13.toy.C.twirl} under every pairing, since for Pauli noise the pairing does not change it; although the interventions of the two arms differ in kind (Section~\ref{sec:r-dep}), both give small shares when weakened under R0. The surface code shows the same split. Weakening in arm A the \val{S13.cont.nsites} $ZZ$ sites of each basis that are most often causes under R0, fixed from the R0 records of the H7 shots, gives at $s=0.1$ small shares under R0, \val{S13.z.C.R0} ($Z$) and \val{S13.x.C.R0} ($X$), and shares close to one under R1, \val{S13.z.C.R1} and \val{S13.x.C.R1} (Fig.~\ref{fig:cont}b,c). Under R1 a weakened site is a cause almost as often as a removed one not because of the site's effect but because of the rule itself, which flips every measurement draw whose probability changed at all. The registered criteria (upper limit below $0.5$ for R0, lower limit above $0.5$ for R1) are met in both cells. Continuity therefore excludes the idealized antithetic and independent rules; as implemented, the share barely falls over the strengths measured, for R1 on the surface code and for R1 and R2 in the toy circuits. It does not single out the shared draw: in an exploratory run R3 was continuous as well (\val{S13.z.C.R3} and \val{S13.x.C.R3}), yet it changes the cause count in the $Z$ basis. The discontinuity of R1 and R2 is a property of the idealized rules; the engine returns to $u'=u$ once a probability changes by less than \val{S13.eps}, which happens only at strengths far below those measured.

\begin{figure*}[t]
\centering
\figin{0.639\linewidth}{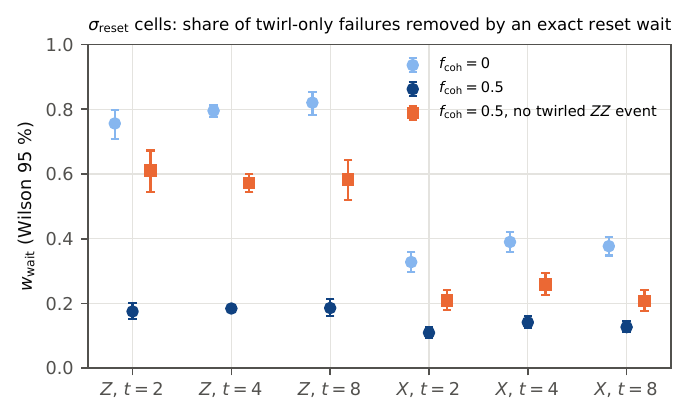}
\caption{Share $w_{\mathrm{wait}}$ of twirl-only failures removed by making the ancilla wait exact, for the six cell pairs with a reset wait, with Wilson 95\,\% intervals: $\fcoh=0$ (light circles), $\fcoh=0.5$ (dark circles), and $\fcoh=0.5$ restricted to failures without a twirled $ZZ$ event (squares). At $\fcoh=0$ the share is larger in the $Z$ basis than in the $X$ basis. The $Z$-basis $t=4$ pair predates the registered test H4$'$ and is shown for completeness.}
\label{fig:wait}
\end{figure*}

\paragraph{Cause frequency is not net effect.} Which sites would a practitioner remove first? The two natural rankings disagree, but acting on either made no detectable difference. Ranking the $ZZ$ sites by how often they are causes, $\mathrm{PN}_i$ under R0, and by their coupling-free net effect $\Delta_i$ gives different orders in the exact arm and nearly the same order under the twirl: Kendall's $\tau_b$ is \val{S13c.z.tau.A} in arm A against \val{S13c.z.tau.B} in arm B in the $Z$ basis (exploratory; both bases and the overlap of the leading sites in Appendix~\ref{app:stat}). We registered whether the difference matters: removing the eight sites that are most often causes lowers the failure rate of arm A less than removing the eight with the largest net effect. On \val{S13c.H8.n} fresh shots per cell the registered criterion is not met in either basis (Appendix~\ref{app:stat}). The orders differ, but the test detected no difference between the two reductions; it does not show that they are equal.

\subsection{What produces the multiplicity}\label{sec:r-mech}

\paragraph{A reading.} The records suggest a simple picture, offered as an interpretation that the two registered runs below examine. A failure of arm A typically hinges on one measurement whose draw $u$ lies just below its Born probability $p$. A removal that lowers the probability to $p'$ flips the outcome exactly when $u\in[p',p)$; these intervals are nested below $p$, so under the shared draw a draw just below $p$ lies in all of them, and removing any one rotation in the measurement's past removes the failure. A success just above the threshold becomes a failure the same way, which is the cancellation $\kappa$ of Section~\ref{sec:r-free}. The multiplicity then needs neither that the rotations add coherently nor any convention between sites, consistent with R4 leaving it in place (Section~\ref{sec:r-dep}).

\paragraph{Two checks.} If the multiplicity came from rotation amplitudes that add before a measurement, it should weaken when the rotations have random signs. In a registered run in which each $ZZ$ site rotates with a fixed random sign, it does not: on a fresh $Z$-basis stream the tail difference between the arms is \val{N2.D} percentage points against \val{N2.Dfix} with fixed signs, its interval above the registered threshold \val{N2.Dhalf}. The reading also predicts that the several causes of a failure flip one common measurement near its threshold. Replaying recorded shots with their Born draws bears this out: in arm A, \val{N3.z.A.F.accpct} ($Z$) and \val{N3.x.A.F.accpct} ($X$) of the failures with two or more $ZZ$ causes have all causes flipping a common measurement, at a median margin $|u-p|$ of \val{N3.z.A.F.margin} and \val{N3.x.A.F.margin}, whereas in arm B the causes mostly flip disjoint measurements, as separate Pauli events do. The design of both runs, their net-effect quantities and the counts are in Appendix~\ref{app:extra}.

\begin{figure*}[t]
\centering
\figin{\linewidth}{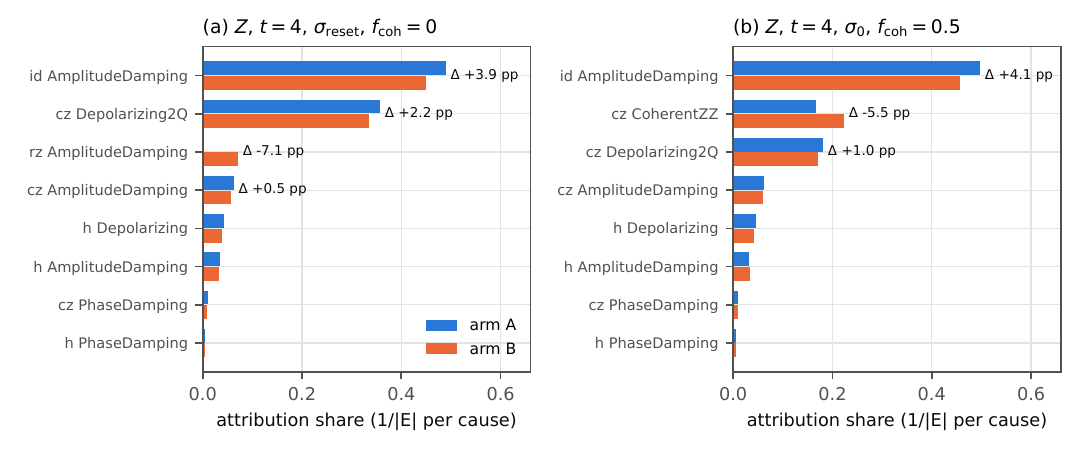}
\caption{Attribution mass by channel, arm A against arm B, for two cells; each member of a cause set receives $1/|E|$ of its shot (equal-share rule); differences, in percentage points of attribution share, are labelled where their 95\,\% bootstrap interval excludes zero. Labels name the gate key and the channel: cz, controlled-$Z$; h, single-qubit gates; id, data idles; rz, the ancilla wait after reset.}
\label{fig:map}
\end{figure*}

\subsection{A case: the ancilla wait after reset}\label{sec:r-wait}

The ancilla wait after reset (Fig.~\ref{fig:wait}) is where the exact and twirled channels differ by construction: the exact amplitude damping leaves the reset state $\ket0$ unchanged, and its twirl does not. It is therefore a positive control, a place where attribution should find a difference between the models that is known in advance. Consider, in the cells with a reset wait, the twirl-only failures, shots that fail in arm B and succeed in arm A under the same seed. The share of them whose arm-B cause set contains a reset-wait site, $w_{\mathrm{wait}}$, is a point-level quantity: sites are removed one at a time, and because the exact channel acts as the identity on $\ket0$, removing such a site is the same as making it exact. At $\fcoh=0$ the reset wait accounts for a large part of the twirl-only failures, as the channels predict. The share falls from \val{H4p.f0} at $\fcoh=0$ to \val{H4p.f05} at $\fcoh=0.5$ (circles), and the registered one-sided test (H4$'$, Appendix~\ref{app:hyp}), which covers \val{H4p.nreg} of these cell pairs, rejects in all of them: as coherent noise grows, twirled $ZZ$ events take over as causes of the twirl-only failures. Part of the fall is a change in the set of such shots, but not all of it, since the share is lower than at $\fcoh=0$ also among failures without a twirled $ZZ$ event (Fig.~\ref{fig:wait}, squares). A registered test at the class level, which makes the whole wait class exact at once, asked whether the effect differs with the state in which the ancilla waits (H4, Appendix~\ref{app:hyp}); its registered criterion, separated intervals at $t=2$, was not met.

\subsection{The attribution map}\label{sec:r-map}

Summed over failing shots, the cause sets say to which channels the failures are attributed, and the two models agree on the large picture while differing where the channels predict. Aggregating cause sets by channel (each member receiving $1/|E|$ of its shot, an equal-share rule, and shares averaged over failing shots with a non-empty cause set) gives an attribution map. Data-idle amplitude damping and two-qubit depolarizing noise carry most of the mass in both arms. The largest differences between the arms are where the channels predict --- the reset wait, which only arm B attributes --- and in the share of coherent $ZZ$, which is smaller in arm A because its $ZZ$ sites share their shot with other members; smaller differences whose intervals exclude zero appear in other channels as well (Fig.~\ref{fig:map}). Over the cells of the main grid measured together with the class net effects (all but the four $Z$-basis $t=4$ cells with $\fcoh\in\{0,0.5\}$, with and without reset wait, which were measured earlier), the gate classes that receive map mass are exactly those with a positive coupling-free class net effect in \val{M2.sign} cells of arm A and \val{M2B.sign} of arm B, and the classes are ranked alike in \val{M2.rank} and \val{M2B.rank} cells, the rank changes being swaps of two adjacent classes (in arm A always the two leading ones). A gate class collects the sites of one gate key (cz, h, id or rz) over all their channels. The map is coupling-dependent; this agreement is a consistency check, not a validation.

\section{Discussion}\label{sec:discussion}

\paragraph{What the two layers mean.} The two arms fail at equivalent rates and differ little in the coupling-free quantities we measured, yet under the shared-seed pairing a failing shot of the exact model has a cause set of several coherent sites, each individually necessary. Monotonicity under the chosen coupling reconciles the two pictures. A twirled $ZZ$ site acts only when it fires an event, so removing it rarely creates a failure, and its cause sets lie close to the sharp marginal bound. A coherent rotation acts in every shot; removing it shifts later Born probabilities, which moves some failing shots below a measurement threshold and some succeeding shots above one, and when one measurement lies close to its threshold several rotations do so at once (Section~\ref{sec:r-mech}). The first kind enters cause sets and the second cancels it in the failure rate, so the cause sets grow above the bound without a matching change at the population level; the marginals do not force this growth, and none of the admissible pairings we tested removes much of it (Section~\ref{sec:r-coup}). Whether a cause set is ``large'' under coherent noise is therefore a statement about the coupling as much as about the noise, and a comparison of cause sets between noise models cannot be read as a comparison of the noise models alone.

\paragraph{Per-trajectory error labels.} Trajectory samplers record the error branches of each trajectory as labels for supervised learning, for example of decoders~\cite{Patti2025}. For unitary-mixture noise these labels specify the canonical event-removal interventions of Section~\ref{sec:m-layers}. Extended to amplitude damping or coherent rotations by counterfactual replay, they may inherit the coupling dependence measured here, and a model trained on them may learn part of the pairing; we have not tested this. Labels derived from coupling-free quantities do not have this problem. What cause sets add to the coupling-free quantities is diagnostic: they show multiplicity that the net effects cancel, as $\kappa$ and the excess over $L^+$ do on the surface code, and they point to the measurement on which a failure hinges (Section~\ref{sec:r-mech}). Whether training on them helps a decoder is a separate study we have not done.

\paragraph{Relation to known mechanisms.} Coherent errors matter for decoders because rotation amplitudes can add before a measurement projects them~\cite{Bravyi2018,TakouBrown2026}. The multiplicity of the exact arm's cause sets does not need this addition: it survives random signs, and the records show several rotations each tipping the same measurement that lies near its threshold (Section~\ref{sec:r-mech}). We do not claim a new physical effect; the reading is about attribution under a pairing. Failure rates that agree within a margin do not by themselves imply equal sensitivities. An exploratory sweep that made all excess CZ error coherent and scaled the noise by four factors ($Z$ basis, $t=4$, $\sigma_0$, each scale run until arm A had 1,500 failures), over failure rates of \val{T72.pfail}, kept the rate ratio at \val{T72.ratio} with every 95\,\% interval inside \val{T72.ci}; a regime where the twirl misses the rate may exist for other bases, round counts or idle placements.

\paragraph{Which pairing to report.} Shared random numbers are one coupling among many with the same marginals. For Pauli noise every admissible one gives the decoded outcome of event removal, so the question does not arise; for non-Pauli noise admissibility leaves much of the measurement randomness free (Section~\ref{sec:r-coup}). Requiring the pairing to be maximal at every random step removes this freedom for the channels used here and gives the shared draw (Appendix~\ref{app:canon}); at their two-branch steps this is counterfactual stability~\cite{OberstSontag2019} applied step by step, as it has been applied to the transitions of decision processes~\cite{Tsirtsis2021}. It is a principle we adopt, not a fact about the physics, and it settles less than it may seem. It fixes the pairing of the factual run with each intervention and so the mean cause count, but not how different interventions are coupled with one another; the tail depends on that convention, although on this circuit the one alternative we tested, R4, left the tail difference in place (Section~\ref{sec:r-dep}). It is not an optimum: the detector-restoring pairing gives slightly smaller cause sets in the $Z$ basis, and the sharp marginal bound lies lower still. Continuity excludes the idealized antithetic and independent rules but does not single out the maximal one. And the branch correspondence that maximality presupposes is fixed by the unravelling: another unravelling of amplitude damping or dephasing could give different cause sets, and for a channel with three or more state-dependent branches the maximal coupling is not unique; we varied neither. Whether some pairing is physically preferred --- for example one tied to a measurement record of the environment --- is open. None of these choices changes the coupling-free results or the marginal bounds.

\paragraph{A reporting checklist.} A shot-level attribution under non-Pauli noise is reproducible only if it states the following: (1) the pairing of factual and counterfactual runs, whether it is the stepwise maximal one, and, for any statistic beyond the mean cause count, how the counterfactuals of different interventions are coupled; (2) the unravelling of each non-Pauli channel and the order of its Kraus operators when a channel has three or more state-dependent branches; (3) the decoder, since cause sets depend on it; (4) the site resolution, since splitting or merging channels changes the cause sets; and (5) next to every cause-set statistic, its coupling-free counterparts, the net effects $\Delta_i$ and the sharp bound $L^+$, together with the coupling-dependent cancelled share $\kappa$.

\paragraph{Diffuse failures.} Diffuse failures ($E=\emptyset$) are excluded from the map; their share in arm A is \val{E4.rho} over the grid. Appendix~\ref{app:extra} relates the map's equal share to a Shapley value.

\paragraph{Reweighting a decoder by the attribution map.} To see whether cause sets are of use beyond diagnosis, we reweighted the decoder's DEM by an attribution map built from the arm-A cause sets with unit mass per member, rather than the equal shares of Section~\ref{sec:r-map} (four $Z$-basis $t=4$ cells, 100,000 training and 300,000 test shots per cell). Its logical error rate differs from those of two attribution-free baselines, a state-aware DEM (the twirled DEM with the reset wait set to its exact effect, the identity) and a DEM fitted to detector correlations, by at most \val{E7.maxrel} (relative), and no paired McNemar test was significant. Choosing which sites to remove fares the same way: cause frequency and net effect rank the sites differently, but no difference was detected between removing the leading eight of either ranking (Section~\ref{sec:r-coup}). The map is diagnostic.

\paragraph{Limitations.} The theoretical statements --- the Pauli lemma, where the coupling dependence enters, and the uniqueness of the stepwise maximal pairing for each intervention --- hold for the stated circuit class and channels and do not depend on the distance or the device. The numbers do: they are for distance three with readout noise off (except the null-intervention check and H4), one hand-authored device model, two bases and, for the pairings and the mechanism, two cells at $t=4$ with $\fcoh=0.5$. The registered distance-five cell pair (H2c) was not run because it failed a preregistered cost gate (Appendix~\ref{app:hyp}); a meaningful distance comparison would need a distance-aware statistic and a new registration, since a fixed threshold such as $|E|\ge3$ means something different when a failure needs more faults. Of the choices we did not vary, we expect the sizes to be most sensitive to the unravelling, which changes cause sets at fixed net effects already in a one-qubit toy (Section~\ref{sec:m-layers}); then to the decoder and the distance, which set how many measurements a failure hinges on; and least to readout noise, which adds independent flips at the end. Of the five coupling-free quantities compared, only the failure rate was registered as an equivalence test, and class net effects differ resolvably in some cells; the site-level test bounds redundancy from above (exact-arm intervals up to \val{H7.Aup}) rather than showing the arms equal, and the decomposition, the marginal bounds, the paired re-analyses and the toy circuits are exploratory. The smallest cause count over admissible couplings is bracketed, not determined, by four tested pairings that vary only the measurement draws. We varied the sign of the rotations and one convention between sites, both only at measurement draws, and not the unravelling of the surface code, the draws of noise sites with state-dependent branch probabilities or readout noise. The completeness audit covers four cells, cause sets depend on the decoder, and all results are from simulation, where trajectories can be replayed.

\section{Conclusion}\label{sec:conclusion}

Which error caused a quantum error-correction failure has a canonical answer under Pauli noise and depends on the pairing of factual and counterfactual trajectories under non-Pauli noise. We claim three things. The first is where the pairing enters: for Pauli noise in a stabilizer memory every admissible pairing gives the Pauli-frame counterfactual, while for non-Pauli noise the dependence enters where a fixed draw is compared with a Born probability and where later draws are reused. The second is what fixes it: maximality at every random step fixes the pairing of the factual run with each intervention, to the shared draw; continuity narrows the choice but does not fix it; and neither fixes how interventions are coupled with one another. The third is how much it matters: on a distance-three surface code the exact model and its Pauli twirl fail at rates equivalent within a registered margin, yet under the shared draw \val{H5.z.tailA} of exact-model failures need three or more sites, against \val{H5.z.tailB} under the twirl. Other admissible pairings move the exact model's coherent cause count from \val{S13.x.r.R3.A} to \val{S13.r12.max} times the smallest count its marginals allow, while the coupling-free quantities we measured barely separate the models.

The first two claims are proved. The Pauli lemma follows by induction on the Pauli difference between the replay and the frame counterfactual, and the uniqueness of the stepwise maximal pairing and its gap from continuity follow from explicit couplings; toy circuits exhibit each case (Section~\ref{sec:m-toy}, Appendices~\ref{app:kraus} and~\ref{app:canon}). The third is measured on matrix-product-state trajectories of QPlayer, checked against a Pauli-frame simulator (H1). Hypotheses, test statistics and stopping rules were registered in a versioned protocol, for some versions committed before the first shot, and all registered tests are reported, including the four whose criteria were not met. The tail difference remains under a second convention between sites and when the rotations have random signs. Replaying the recorded draws locates its source: several rotations each tipping the same measurement near its threshold.

The value of this work is a reporting discipline for shot-level attribution under realistic noise. Agreeing failure rates do not imply agreeing explanations, and an explanation obtained by replaying trajectories is a property of the noise model and the pairing together. A shot-level attribution should therefore state its pairing and report coupling-free quantities, the net effects and the sharp marginal bounds, alongside. These do not depend on any choice made here and are the basis on which noise models can be compared. The checklist of Section~\ref{sec:discussion} makes such attributions reproducible, and the artifact regenerates every number. How the cause-set tail scales with distance, how low the cause count can go over admissible pairings, and whether the measurement record of the environment singles out a physically preferred pairing are the next questions.

\section*{Code and data availability}
The artifact of this paper is archived at Zenodo~\cite{JinPaper2Artifact}. It contains the versioned registration protocol, the experiment drivers, the raw results, the scripts that produce every figure and table from the frozen values, the source of the QPlayer release used here, v6.0.1-Lynx, and the research patches to it with a script that builds them. The QPlayer emulator~\cite{Jin2023QPlayer} is open source at \url{https://github.com/eQuantumOS/QPlayer}. The artifact records a check that this release, with the patches, reproduces the builds used for the measurements bit for bit, and that the frozen values, figures and tables are regenerated exactly.

\section*{Author contributions}
K.J. proposed the core idea of the study, developed the theoretical formalism, and carried out the step-by-step experiments and the detailed analysis. G.C. contributed to the computation of concrete examples and provided relevant considerations throughout the research process. Both authors verified the results presented and contributed to the writing of the manuscript. Large language models were used only for writing numerical computation scripts and polishing the language.

\section*{Acknowledgments}
This work was supported by the Institute of Information \& Communications Technology Planning \& Evaluation (IITP) grant funded by the Korea government (MSIT) (No.~2020-0-00014, A Technology Development of Quantum OS for Fault-tolerant Logical Qubit Computing Environment).

\clearpage
\onecolumn
\appendix

\section{Noise parameters}\label{app:noise}

Table~\ref{tab:noise} lists the noise of the main grid and where each parameter comes from.

\begin{table}[H]
\centering\footnotesize
\begin{tabularx}{\linewidth}{>{\raggedright\arraybackslash}p{0.17\linewidth}>{\raggedright\arraybackslash}p{0.19\linewidth}LL}
\toprule
Location & Channel & Parameter & Source \\
\midrule
all qubits & relaxation, dephasing & $T_1=\val{nz.T1}$, $T_2=\val{nz.T2}$ & hand-authored device model from representative values~\cite{Willow2025}, in the artifact \\
single-qubit gate (\val{nz.1q.ns}~ns) & amplitude damping, dephasing & $\gamma=\val{nz.1q.gamma}$, $\lambda=\val{nz.1q.lambda}$ & from $T_1,T_2$ and duration \\
 & depolarizing & $p=\val{nz.1q.depol}$ & average error \val{nz.1q.err} minus the average error of relaxation over the gate \\
controlled-$Z$ (\val{nz.cz.ns}~ns) & amplitude damping, dephasing & $\gamma=\val{nz.cz.gamma}$, $\lambda=\val{nz.cz.lambda}$ & from $T_1,T_2$ and duration \\
 & excess error & \val{nz.cz.excess} & average error \val{nz.cz.err} minus the average error of relaxation over the gate \\
 & two-qubit depolarizing & $p=\val{nz.cz.depol.f0}$ ($\fcoh=0$), \val{nz.cz.depol.f05} ($\fcoh=0.5$) & excess error, share $1-\fcoh$ \\
 & coherent $\exp(-i\theta ZZ/2)$ & average error \val{nz.cz.rcoh}, $\theta=\val{noise.theta}$ ($\fcoh=0.5$) & \textbf{set by this study}: share $\fcoh$, fixed sign \\
data idle (\val{nz.idle.ns}~ns) & amplitude damping, dephasing & $\gamma=\val{nz.idle.gamma}$, $\lambda=\val{nz.idle.lambda}$ & from $T_1,T_2$ and duration, the part of the syndrome cycle not covered by gates \\
ancilla reset wait (\val{nz.anc.ns}~ns) & amplitude damping, dephasing & $\gamma=\val{nz.anc.gamma}$, $\lambda=\val{nz.anc.lambda}$ & \textbf{set by this study}: optional cell axis \\
readout & --- & noiseless & \textbf{set by this study} \\
\bottomrule
\end{tabularx}
\caption{Noise of the main grid. Arm B replaces every channel by its Pauli twirl. For a duration $t_{\mathrm d}$, $\gamma=1-e^{-t_{\mathrm d}/T_1}$ and $\lambda=1-e^{-2t_{\mathrm d}/T_\phi}$ with $1/T_\phi=1/T_2-1/(2T_1)$, the amplitude- and phase-damping parameters of~\cite{NielsenChuang}. Dephasing is registered in both arms in its unitary-mixture form, $Z$ with probability $(1-\sqrt{1-\lambda})/2$, which is its own twirl. A depolarizing $p$ is the total probability of a non-identity Pauli on a Hilbert space of dimension $N$, with average error $Np/(N+1)$; a coherent rotation has average error $4\sin^2(\theta/2)/5$. Errors are combined to first order.}
\label{tab:noise}
\end{table}

\clearpage
\section{Registered hypotheses}\label{app:hyp}

Table~\ref{tab:hyp} lists every registered confirmatory hypothesis with its verdict.

\begin{table}[H]
\centering\footnotesize
\begin{tabularx}{\linewidth}{p{0.05\linewidth}L>{\raggedright\arraybackslash}p{0.16\linewidth}>{\raggedright\arraybackslash}p{0.22\linewidth}}
\toprule
& Hypothesis and statistic & Registered & Result \\
\midrule
H1 & Engine counterfactual equals Pauli-frame counterfactual; Wilson lower bound of pair agreement $\ge0.99$ & main family & criterion met \\
H2a & Tail mass $P(|E|\ge3)$ larger at $\fcoh=0.5$ than at $0$; one-sided Fisher, twelve cell pairs & main family & criterion met in all twelve pairs, both decoders \\
H4 & Reset-wait attribution separates idle states (class level), $Z$ and $X$ & main family & criterion not met (intervals overlap at $t=2$; readout noise on) \\
H4$'$ & Share of twirl-only failures removed by an exact wait falls with $\fcoh$ & outside family & criterion met in all five registered pairs \\
H5 & Tail mass of arm A larger than arm B, and failure-rate ratio equivalent within $\pm10\,\%$ & outside family, after earlier data & criterion met in both cells \\
H6 & Interaction $\eta_A<\eta_B$ with $\eta_A$ below zero (two halves of the $ZZ$ sites) & outside family & criterion not met in either cell \\
H7 & Site-level redundancy $R_A>R_B$ & outside family & criterion not met in either cell \\
H2c & Distance five, tail mass rises with $\fcoh$ & conditional on a cost gate & not run (gate failed: pilot \val{d5.f0.h}~h $>$ 3~h) \\
H8 & Removing the eight sites most often causes lowers the failure rate less than removing the eight with the largest net effect & outside family & criterion not met in either cell \\
--- & Four admissible pairings; prespecified reading (halved excess / none below R0 / range only) & registered comparison & range only (Section~\ref{sec:r-coup}) \\
--- & Continuity: shared pairing tends to zero, antithetic does not & registered comparison & both criteria met in both cells \\
N1 & Tail difference under the convention R4 between sites; prespecified reading (kept / convention-dependent / range) & exploratory, registered before the run & kept in both cells; one engine gate failed (see text) \\
N2 & Tail difference with random signs of the rotations; prespecified reading (sign-free / same-sign / range) & exploratory, registered before the run & sign-free \\
N3 & Measurements flipped by coherent causes (accumulation / parity) & exploratory, registered before the run, no verdict & reported (Section~\ref{sec:r-mech}) \\
\bottomrule
\end{tabularx}
\caption{Every registered confirmatory hypothesis. ``Criterion not met'' means that the registered test did not reject; it does not show the opposite. The Wilson bound of H1 treats pairs from one failing shot as independent (Section~\ref{sec:r-valid}). The main family was Holm-corrected together; the others were registered individually with their own correction. H2b (joint causes concentrate at $\fcoh=0.5$) and H3 (the attribution maps of the two arms differ) were registered as exploratory; H3 is the arm difference of the map in Section~\ref{sec:r-map}, and H2b is reported in the artifact. Analyses that respect the dependence of the registered tests are in Table~\ref{tab:dep}.}
\label{tab:hyp}
\end{table}

\paragraph{Deviations.} Two deviations concern N1. Its third engine gate tested whether the re-drawn $u'$ of R4 are uniform by one Kolmogorov--Smirnov test on all re-drawn values of 20 failing shots per cell. Within one shot and measurement, every intervention re-draws inside the same factual interval, so the \val{T87.z.nv} values of the $Z$ basis fall into \val{T87.z.ncl} clusters (median size \val{T87.z.clmed}) and are not independent; the test gave $p=\val{N1.z.ghp}$ ($X$: \val{N1.x.ghp}). A diagnostic written after the run replays the same ledger: every re-drawn value lies inside its factual interval, and the tests whose independence holds pass --- the position inside the interval ($p=\val{T87.z.conditional}$ and \val{T87.x.conditional}), one value per measurement (\val{T87.z.one}, \val{T87.x.one}) and the factual draws (\val{T87.z.factual}, \val{T87.x.factual}). We report N1 with this disclosure. The gate was also registered to be passed before the main run and was evaluated on the main run's first failing shots instead.

\FloatBarrier
\section{Statistics and the screening algorithm}\label{app:stat}

\paragraph{Stopping and intervals.} Each cell of the main grid and the registered comparison H5 ran shots in index order until arm A reached its failure target with the twirled decoder (main grid 10,000 failures in the four $Z$-basis $t=4$ cells with $\fcoh\in\{0,0.5\}$ and 2,000 in the others; H5 2,000); arm B was evaluated on the same indices. H6 ran \val{H6.n} shots per cell and arm, and H7 collected \val{H7.nF} failures and \val{H7.nS} successes per cell and arm. Proportions carry Wilson intervals. Tail masses are compared with one-sided Fisher exact tests and Holm correction within each registered family, as registered; the test treats the compared samples as independent, although the two cells of an H2a pair run on the same random numbers and the two arms of H5 share seeds; for H5, Section~\ref{sec:r-dep} adds a paired bootstrap over shot indices and a within-index permutation of the arm labels. The paired bootstrap resamples shot indices given the realized stopping point and does not reproduce the stopping rule; the permutation assumes the two arms of an index exchangeable under the null and serves as a sensitivity check. The interval of the H5 rate ratio is also recomputed by a parametric bootstrap that reproduces the stopping rule: shots are drawn until arm A has its fixed number of failures, with the joint outcomes of an index drawn from their observed rates; the same construction, with each arm's outcome split into success, failure with $|E|<3$ and failure with $|E|\ge3$, gives the stopped-design interval of the tail difference; the full $3\times3$ joint counts are in the artifact. Table~\ref{tab:pairs} gives the paired counts and Table~\ref{tab:streams} the streams; the steps that reproduce them are in the artifact~\cite{JinPaper2Artifact}. H7 ran each cell and arm from the same first index until the arm had its failure target, and took the first \val{H7.nS} successes of that range; $\pfail$ in H7 is the arm's failure count over the shots run. Ratios of failure rates, $R_{\mathrm{red}}$, $\kappa$, the bound ratios and the attribution-map differences carry percentile bootstrap intervals over shots, resampling failing and succeeding shots separately and drawing $\pfail$ binomially where it enters; ratios of the two arms resample shot indices with both arms of an index kept together, because both arms share seeds. Intervals for the sharp bound are not given, because the clipped estimator is biased (Section~\ref{sec:r-free}).
The clipped estimate of the sharp bound $L^+_{ZZ}$ (Section~\ref{sec:r-free}) is biased upward, by \val{T79.bias.A} in arm A and \val{T79.bias.B} in arm B in synthetic data that take the measured per-site rates as the truth, and the cross-fitted estimate errs downward. A nominal 95\,\% lower limit of the clipped ratio, \val{T79.z.lo.A} ($Z$) and \val{T79.x.lo.A} ($X$) in arm A, lay at or below the truth in every replicate of the calibration below; that the clipped bound is biased upward and the ratio downward is a tendency observed there, not a guarantee. The lower limit of the bound ratio is the lower end of a nominal 95\,\% percentile interval from the joint bootstrap. It was calibrated in \val{T80.nscen} scenarios that draw sites independently and in two that resample whole recorded shots, with \val{T80.R} replicates each: each replicate draws a synthetic data set from the stated truth, computes the lower limit from \val{T81.Bcov} resamples, and counts whether it lies at or below the true ratio. It did in every replicate, and the mean ratio estimate lay below the truth, by \val{T80.under.A} and \val{T81.under.A} in arm A.

\begin{table}[H]
\centering\small
\begin{tabularx}{\linewidth}{Lrrrrrrrr}
\toprule
Basis & Shots & Both fail & A only & B only & Neither & Tail A & Tail B & Both tail \\
\midrule
$Z$ & \val{H5.z.N} & \val{H5.z.both} & \val{H5.z.Aonly} & \val{H5.z.Bonly} & \val{H5.z.neither} & \val{H5.z.tailA.n} & \val{H5.z.tailB.n} & \val{H5.z.tailboth} \\
$X$ & \val{H5.x.N} & \val{H5.x.both} & \val{H5.x.Aonly} & \val{H5.x.Bonly} & \val{H5.x.neither} & \val{H5.x.tailA.n} & \val{H5.x.tailB.n} & \val{H5.x.tailboth} \\
\bottomrule
\end{tabularx}
\caption{Paired outcomes of the H5 comparison, $\fcoh=0.5$, $\sigma_0$, $t=4$, twirled decoder: shots run until arm A reached its failure target, failures by arm on the same index, failing shots of each arm with a screened cause set of three or more sites, and indices at which both arms have one. These shots come from the fresh stream of the registered comparison, not from the main grid of Table~\ref{tab:sets}.}
\label{tab:pairs}
\end{table}

\begin{table}[H]
\centering\footnotesize
\begin{tabularx}{\linewidth}{lLL}
\toprule
Stream & Size & Used in \\
\midrule
Main grid & \val{E4.cells} cells, each until arm A has 10,000 failures (four $Z$-basis $t=4$ cells) or 2,000 (the others); class net effects add the first 2,000 successes per cell and arm & H2a; H4$'$; class net effects; Figs.~\ref{fig:sets}, \ref{fig:cases} and~\ref{fig:wait} and the open points of Fig.~\ref{fig:twirl}; Table~\ref{tab:sets}; the attribution map (Section~\ref{sec:r-map}); Shapley values and the reweighting map (Section~\ref{sec:discussion}) \\
Fitted-correlation decoder & four $Z$-basis $t=4$ cells; 100,000 fitting shots per cell on a separate stream, then the first 2,000 failures it decodes in the main-grid range & Section~\ref{sec:r-dep} \\
Engine checks & null intervention: \val{null.n} shots, $Z$, $t=4$, $\fcoh=0$, reset wait, readout noise on; bit identity: 400 shots, $Z$, $t=2$, $\fcoh=0$, reset wait; H1: 10,000 failing shots of arm B, $Z$, $t=4$, $\fcoh=0$, $\sigma_0$ & Section~\ref{sec:r-valid} \\
H4 & $Z$ and $X$, $t\in\{2,4\}$, three states of the waiting ancilla, 15,000 to 50,000 shots per cell, readout noise on & H4 (Section~\ref{sec:r-wait}) \\
H5 & two cells, \val{H5.z.N} and \val{H5.x.N} shots, until arm A has 2,000 failures & H5; Fig.~\ref{fig:twirl}; Table~\ref{tab:pairs} \\
H6 & two cells, \val{H6.n} shots per cell and arm & H6 \\
H7 & two cells, \val{H7.nF} failures and \val{H7.nS} successes per cell and arm & H7; $\kappa$, $L$, $L^+$ and the bound ratios (Section~\ref{sec:r-free}) \\
Four pairings & replays of H7 shots, \val{S13.nF} failures and \val{S13.nS} successes per cell and arm & Table~\ref{tab:coup}; Fig.~\ref{fig:coup}; continuity and rank comparison (Section~\ref{sec:r-coup}) \\
Ledger check & two cells, 20 shots per cell and arm, every fourth noise site & share of moved draws (Section~\ref{sec:m-layers}) \\
Toy circuits & closed form; engine check with 20,000 shots per combination & Section~\ref{sec:m-toy}; Fig.~\ref{fig:cont}a \\
H8 & two cells, \val{S13c.H8.n} shots per cell & H8 (Section~\ref{sec:r-coup}) \\
Diagnostic & two cells, \val{S13.diag.n} shots per cell & check on the coupling-free sum (Section~\ref{sec:r-coup}) \\
N1 & H5 stream, two cells, \val{N1.nA} failures of arm A per cell and the arm-B failures of the same range & convention R4 (Section~\ref{sec:r-dep}) \\
N2 & $Z$ basis, fresh stream, \val{N2.nA} failures of arm A, \val{N2.nS} successes per arm & random signs (Section~\ref{sec:r-mech}) \\
N3 & replays of H7 shots with a $ZZ$ cause or failure-creating site, \val{N3.n} shots & flipped measurements (Section~\ref{sec:r-mech}) \\
Audit & \val{audit.ncells} cells, each until arm A has 300 failures, \val{audit.n} failing shots & completeness of the screening (Section~\ref{sec:r-valid}) \\
Regime sweep & one cell, four noise scales, each until arm A has 1,500 failures & Section~\ref{sec:discussion} \\
Decoder reweighting & four cells, 300,000 test and 100,000 training shots per cell & Section~\ref{sec:discussion} \\
\bottomrule
\end{tabularx}
\caption{Data streams. The audit, regime sweep and decoder reweighting are exploratory; the ledger check is an engine gate on shots not used elsewhere. Each confirmatory stream starts from its own offset of the confirmatory seed base; the two cells of a main-grid pair, and the two arms of every stream, share random numbers.}
\label{tab:streams}
\end{table}

\paragraph{Other intervals and tests.} The tail ratio of an H2a cell pair (Table~\ref{tab:sets}) and that of the fitted-correlation decoder (Section~\ref{sec:r-dep}) carry an interval from a parametric bootstrap that draws the two tail counts as independent binomials, ignoring the random numbers that the two cells share. The share $w_{\mathrm{wait}}$ of H4$'$ is compared between $\fcoh=0$ and $\fcoh=0.5$ with a one-sided Fisher exact test, Holm-corrected over the five registered pairs. Class net effects resample failing and succeeding shots with $\pfail$ held fixed. The interaction $\eta$ resamples the joint outcome patterns of its two groups, and H6 takes the share of resamples on the wrong side as its $p$ value; H7 does the same with failing and succeeding shots resampled separately. In both, and for the signed bound $L$ and its arm difference, the two arms are resampled independently although they share seeds. The cause counts under the four pairings and their differences, the per-site net effects, Kendall's $\tau_b$ and the comparison H8 resample shots, keeping the records of one shot under all pairings or removals together. The diagnostic of Section~\ref{sec:r-coup} uses a Wald interval of the paired difference of failure rates. Bootstrap intervals are percentile intervals from 4,000 to 20,000 resamples.

\begin{table}[H]
\centering\footnotesize
\begin{tabularx}{\linewidth}{p{0.07\linewidth}>{\raggedright\arraybackslash}p{0.2\linewidth}>{\raggedright\arraybackslash}p{0.22\linewidth}LL}
\toprule
& Registered test & Dependence & Analysis that respects it (exploratory) & Conclusion \\
\midrule
H1 & Wilson bound over pairs & pairs from one shot share it & count by shot: \val{A6.shots}/\val{A6.shots}, Wilson lower \val{A6.lo} & unchanged \\
H2a & Fisher, cells independent & two cells of a pair share random numbers & common index range, stopping rule reproduced; tail ratio \val{A1.t.R} (smallest lower 95\,\% limit \val{A1.t.lomin}); Holm $p\le\val{A1.t.holm}$ in all \val{A1.npairs} pairs, $\phi=\val{A1.phi}$ & unchanged \\
H5 & Fisher, arms independent & the arms share seeds & paired bootstrap: \val{T79.z.diff} ($Z$), \val{T79.x.diff} ($X$) points; stopped-design bootstrap: \val{T81.z.dci}, \val{T81.x.dci}; label permutation, assuming the arms of an index exchangeable under the null: no difference above \val{T79.nullmax} points in \val{T79.B} permutations, Monte Carlo $p=1/\val{T79.Bp1}$ & unchanged \\
H6 & bootstrap, arms independent & the arms share seeds & not possible: per-index patterns not recorded & registered only \\
H7 & bootstrap, arms independent & the arms share seeds & index-paired: $p=\val{A2.z.p}$, \val{A2.x.p} & unchanged (not met) \\
\bottomrule
\end{tabularx}
\caption{Registered tests whose samples share random numbers, and the analyses that respect the dependence. Class net effects are stored as per-arm aggregates, so their arm difference cannot be paired; non-overlap of their intervals is exploratory.}
\label{tab:dep}
\end{table}

\begin{table}[H]
\centering\footnotesize
\begin{tabular}{llrrrrrrr}
\toprule
Cell & Arm & Failing shots & Sets differ & Missed & Added & Tail, screened & Tail, full & Upper 95\,\% \\
\midrule
z t4 $\sigma_0$ $f_{\mathrm{coh}}$ 0.5 & A & 300 & 1 & 2 & 0 & 64 & 64 & 1.6\,\% \\
z t4 $\sigma_0$ $f_{\mathrm{coh}}$ 0.5 & B & 322 & 0 & 0 & 0 & 7 & 7 & 0.9\,\% \\
x t4 $\sigma_0$ $f_{\mathrm{coh}}$ 0.5 & A & 300 & 0 & 0 & 0 & 75 & 75 & 1.0\,\% \\
x t4 $\sigma_0$ $f_{\mathrm{coh}}$ 0.5 & B & 297 & 1 & 1 & 0 & 5 & 5 & 1.6\,\% \\
z t8 $\sigma_0$ $f_{\mathrm{coh}}$ 0.5 & A & 300 & 1 & 1 & 0 & 68 & 69 & 1.6\,\% \\
z t8 $\sigma_0$ $f_{\mathrm{coh}}$ 0.5 & B & 270 & 0 & 0 & 0 & 3 & 3 & 1.1\,\% \\
z t4 $\sigma_0$ $f_{\mathrm{coh}}$ 0 & A & 300 & 0 & 0 & 0 & 5 & 5 & 1.0\,\% \\
z t4 $\sigma_0$ $f_{\mathrm{coh}}$ 0 & B & 300 & 0 & 0 & 0 & 5 & 5 & 1.0\,\% \\
\midrule
pooled & A & 1,200 & 2 & & & & & \\
pooled & B & 1,189 & 1 & & & & & \\
\bottomrule
\end{tabular}

\caption{Screening audit by cell and arm: failing shots scanned exhaustively, shots whose cause sets differ, members missed and added by the screening, failing shots in the tail by each method, and the one-sided 95\,\% upper bound on the share that differ. Exploratory. A full scan replays a failing shot \val{A9.runs} times.}
\label{tab:audit}
\end{table}

\begin{table}[H]
\centering\footnotesize
\begin{tabularx}{\linewidth}{>{\raggedright\arraybackslash}p{0.25\linewidth}>{\raggedright\arraybackslash}p{0.25\linewidth}L}
\toprule
Check & Sample & Result \\
\midrule
Null intervention & \val{null.n} shots in each arm & no record changed \\
No intervention, engine hook & 400 shots & bit-identical to the unmodified build \\
Two engines & --- & matrix-product-state and state-vector bit-identical; every failing shot replays identically \\
Replay prefix & \val{H1.runs} intervention runs & no prefix diverged from the factual run \\
H1, engine against Pauli frame (registered) & \val{H1.pairs} pairs & \val{H1.pairs}/\val{H1.pairs} agree; Wilson lower bound \val{H1.lo}; binomial test against $0.99$, $p=\val{H1.p}$ \\
\quad counted by shot (exploratory) & \val{A6.shots} failing shots & \val{A6.shots}/\val{A6.shots} agree; Wilson lower bound \val{A6.lo} \\
All sites removed & \val{E4.allundo} failing shots of arm A & failure removed in all \\
Screening audit (exploratory) & \val{audit.n} failing shots, \val{audit.ncells} cells, \val{audit.sites.min} to \val{audit.sites.max} sites each & \val{audit.same} identical (one-sided 95\,\% upper bound on the share that differ \val{audit.up}); \val{audit.missed} members missed, none added; \val{audit.tailchg} shot, in arm A, crossed the tail threshold (upper bound \val{audit.tailup}) \\
Robustness of H5 (exploratory) & arm-B failures of the audit & erasing the tail difference needs \val{A4.need.z} ($Z$) and \val{A4.need.x} ($X$) reclassified; found \val{A4.recl.B} of \val{A4.n.B} (upper bound \val{A4.upB}) \\
\bottomrule
\end{tabularx}
\caption{Checks of the intervention machinery (Section~\ref{sec:r-valid}); the cells and streams are in Table~\ref{tab:streams}, per-cell audit counts in Table~\ref{tab:audit}. The registered Wilson bound and binomial test of H1 treat pairs from one shot as independent and are reported as registered.}
\label{tab:checks}
\end{table}

\paragraph{Fitted-correlation decoder.} Fitted on 100,000 separate shots per cell and evaluated on the first 2,000 failures it decodes in the main-grid range, the decoder fitted to the two-point detector correlations gives tail ratios of \val{M1.s0} at $\sigma_0$ and \val{M1.sr} with reset wait in the two $Z$-basis $t=4$ cell pairs, against \val{M1.s0.tw} and \val{M1.sr.tw} with the twirled decoder (the $Z$-basis $t=4$ rows of Table~\ref{tab:sets}, there from 10,000 failures per cell against 2,000 here). There the spread does not depend on the decoder; the uniform decoder gives the same verdict as the twirled one in all twelve pairs.

\paragraph{The coupling-free sum across pairings.} Estimated under each pairing from the same shots, the coupling-free sum $\sum_i\Delta_i$ over the $ZZ$ sites need not agree exactly. Its paired difference from R0 has 95\,\% intervals \val{A5.z.R1} ($Z$) and \val{A5.x.R1} ($X$) under R1, \val{A5.z.R2} and \val{A5.x.R2} under R2, and \val{A5.z.R3} and \val{A5.x.R3} under R3. A registered diagnostic on \val{S13.diag.n} fresh shots per cell found R1 minus R0 of \val{S13.diag.z.T2} ($Z$) and \val{S13.diag.x.T2} ($X$) after removing the site that contributed most, against \val{S13.diag.z.pred} and \val{S13.diag.x.pred} predicted from the deviation, consistent with sampling variation; that site was chosen from the records it was then tested on, which the registration disclosed.

\paragraph{Rank comparison and H8.} Kendall's $\tau_b$ between the rankings of the $ZZ$ sites by $\mathrm{PN}_i$ and by $\Delta_i$ is \val{S13c.z.tau.A} ($Z$) and \val{S13c.x.tau.A} ($X$) in arm A against \val{S13c.z.tau.B} and \val{S13c.x.tau.B} in arm B, and the eight leading sites coincide in \val{S13c.z.ov8.A} and \val{S13c.x.ov8.A} in arm A against \val{S13c.z.ov8.B} and \val{S13c.x.ov8.B} in arm B (exploratory). In H8, removing the eight most frequent causes and the eight largest net effects lowered the failure rate of arm A by \val{S13c.H8.z.dC} and \val{S13c.H8.z.dN} in the $Z$ basis (one-sided $p=\val{S13c.H8.z.p}$) and by \val{S13c.H8.x.dC} and \val{S13c.H8.x.dN} in the $X$ basis ($p=\val{S13c.H8.x.p}$, Holm-adjusted \val{S13c.H8.x.holm}).

\paragraph{Screening.} For a failing shot with sites $\{1,\dots,n\}$, the $ZZ$ sites are removed from the list and each is tested alone. The remaining sites are split into $24$ contiguous blocks. Every block whose removal turns the failure into a success is halved, and the halves that do so are halved again, down to single sites, which are members of $E$. A block that removes the failure while neither of its halves does is recorded as an unresolved block; its sites are not tested individually, which is how the screening can miss a member whose effect is masked inside the block. If $E$ is empty, every site is tested alone before the shot is classified as diffuse.

\FloatBarrier

\section{Kraus order and the coupling}\label{app:kraus}

A site with branch probabilities $(w_1,\dots,w_m)$ chooses branch $j$ when the draw $r$ falls in the $j$-th interval of the cumulative sums. For $m=2$ reversing the order maps $r$ to $1-r$ in both runs, a measure-preserving map that does not depend on the weights, so the joint distribution of factual and counterfactual branches is unchanged. If the weights do not depend on the state, any reordering is one fixed piecewise translation of $r$, again the same in both runs. If $m\ge3$ and the weights differ between the two runs, the reordering map depends on them. With factual weights $(1/2,1/4,1/4)$ and counterfactual weights $(1/4,1/4,1/2)$, the order $(1,2,3)$ gives probability $0$ that both runs take branch $2$, while the order $(2,1,3)$ gives $1/4$. The channels of this paper are all of the first two kinds.

\paragraph{Admissible replay and the Pauli frame.} \emph{Lemma.} Assume (i) a fixed Clifford circuit with stabilizer-state preparation, Pauli measurements and resets and no feed-forward; (ii) Pauli noise, whose branch probabilities do not depend on the state; (iii) detectors and a logical observable that are parities of the measurement record and deterministic in the noiseless circuit; and (iv) a decoder that reads only the detectors. Then, for the removal of one sampled Pauli event, every admissible coupling gives the same detectors, observable and decoded outcome as the Pauli-frame counterfactual. The raw measurement records can differ.

\emph{Proof.} By (ii) every site's branch probabilities are the same in both runs, so an admissible coupling holds every site draw fixed, and the only freedom is at measurements. Up to the first measurement whose outcome is random and that anticommutes with the current frame difference, the admissible replay and the frame counterfactual coincide, because outcomes of probability $0$ or $1$ are forced and those of probability $1/2$ that commute with the difference are the same in both. At such a measurement $M$ with outcome $o$, the frame counterfactual records $o\oplus1$ and the admissible replay $o$. If $\psi$ is the counterfactual state before $M$, a random outcome means that some stabilizer $V$ of $\psi$ anticommutes with $M$, and $\Pi_o\psi=V\,\Pi_{o\oplus1}\psi$ for the projectors $\Pi$ onto the outcomes. In the frame simulation the admissible replay is therefore the frame counterfactual with $V$ added as an extra Pauli error before $M$: the extra error flips the record of $M$, while on the state itself $V\psi=\psi$. Since $\psi$ differs from the noiseless state by a Pauli operator, $V$ is, up to sign, a stabilizer of the noiseless state at that point (that state depends on earlier random outcomes, but its stabilizer group up to signs does not), so as an error it flips no parity that is deterministic in the noiseless circuit, and by (iii) no detector and not the observable. The argument concerns these deterministic parities only, not the distribution of later raw outcomes on either branch. Repeating it at each such measurement, the two counterfactuals agree on every detector and on the observable, and by (iv) the decoder returns the same correction.

\emph{The induction, stated as an invariant.} Before every step, the admissible replay and the frame counterfactual differ by a Pauli operator $Q$ on the state and by a record difference $\delta$ on the measurements so far, such that $\delta$ flips no parity that is deterministic in the noiseless circuit. A site step keeps $Q$, a Clifford gate conjugates it, a measurement either leaves $\delta$ unchanged or, at a random outcome that anticommutes with $Q$, adds the flip of a stabilizer as above. A reset to $\ket0$ after a $Z$ measurement removes the part of $Q$ on that qubit, because both runs leave it in the same state; for example, if the two runs differ by $X$ on an ancilla before its measurement and reset, the records may differ in that outcome but not afterwards. The lemma therefore distinguishes three levels: raw records, which can differ; detectors and the observable, which agree; and the decoded outcome, which agrees by (iv). Admissibility (A1 of Appendix~\ref{app:canon}) is a condition on draws, holding fixed every draw whose probability does not change; matching branches by name is the convention that says which branch of the counterfactual run a draw selects. For Pauli noise the two coincide, because every probability that does not change belongs to a branch that exists in both runs.

\section{The stepwise maximal pairing}\label{app:canon}

\paragraph{Setting.} Fix the unravelling. A shot is a sequence of random steps $k=1,\dots,K$ --- the branch choice of a noise site, or the outcome of a measurement or reset --- in an order that is the same in both runs, since the circuit has no feed-forward. Step $k$ chooses a branch $b_k$ with probability $\pi_k(\cdot\mid h_{<k})$ given the history; write $\pi'_k$ for the run in which site $i$ is weakened with strength $s$, so that $s=1$ is the removal $\doI{i}$ and $s=0$ the factual run. Branches of the two runs correspond by name: the same measurement outcome, or the same position in the list of Kraus operators. A pairing is co-adapted if, given the two histories $(h,h')=(h_{<k},h'_{<k})$, it chooses a joint distribution of $(b_k,b'_k)$ with marginals $\pi_k(\cdot\mid h)$ and $\pi'_k(\cdot\mid h')$ without looking at later steps; the four pairings of Section~\ref{sec:r-coup} are of this kind. Before the first split $h=h'$; after it the two runs carry different histories, and every requirement below compares the factual and counterfactual conditional distributions given their own histories. We consider three requirements: (A1) admissibility, $b_k=b'_k$ whenever $\pi_k(\cdot\mid h)=\pi'_k(\cdot\mid h')$; (A2) continuity, $P(h\ne h'_s)\to0$ as $s\to0$, where $h'_s$ is the counterfactual history at strength $s$, so that $\mathrm{PN}_i(s)\to0$; and (A3) stepwise maximality, $P(b_k\ne b'_k\mid h,h')=d_{\mathrm{TV}}\big(\pi_k(\cdot\mid h),\pi'_k(\cdot\mid h')\big)$ at every step, before and after a split.

\emph{Lemma.} For two branches with probabilities $p$ and $p'$ of branch $1$, the only coupling with $P(b\ne b')=|p-p'|$ puts $\min(p,p')$ on $(1,1)$, $\min(1-p,1-p')$ on $(0,0)$ and the remainder on the one off-diagonal cell in the direction of $p-p'$~\cite{LevinPeresWilmer}. \emph{Proof.} The two off-diagonal cells differ by $p-p'$, so their sum is at least $|p-p'|$ with equality only if one of them is empty; the marginals then fix the rest. The shared threshold draw, $b=\mathbf 1[u<p]$ and $b'=\mathbf 1[u<p']$, is this coupling; the antithetic draw disagrees with probability $1-|p+p'-1|$ and an independent draw with $p(1-p')+p'(1-p)$.

\emph{Proposition 1.} Suppose every step at which the two runs can have different branch probabilities either has two branches or has a deterministic counterfactual branch, and every other step has the same probabilities in both runs. Then, for the given unravelling and correspondence of branches, the co-adapted pairing that satisfies (A3) is unique in law and is the shared draw. Uniqueness is of the joint law of the factual run and one intervention, not of a map of uniform draws. \emph{Proof.} By induction on $k$: given the histories $(h,h')$, before or after a split, the lemma fixes the joint law of a two-branch step, and the shared draw realizes it; a step with a deterministic counterfactual branch admits only one coupling, which the shared draw also gives; a step with the same probabilities in both runs has $d_{\mathrm{TV}}=0$, so (A3) forces $b_k=b'_k$, which the shared draw also gives. The channels and interventions of this paper satisfy the assumption: measurements and resets have two outcomes, amplitude and phase damping two Kraus operators, and a coherent rotation one branch; depolarizing and twirled channels have state-independent probabilities, which therefore agree between the runs except at the removed site itself, whose counterfactual branch is the identity; and only coherent rotations are weakened. The proposition fixes the joint law of the factual run and one intervention; it does not fix how the counterfactuals of different interventions are coupled with one another, which matters for the full law of the cause set but not for its mean (Section~\ref{sec:m-layers}).

\emph{Proposition 2.} For a finite number $K$ of random steps in an order fixed in advance (no feed-forward), (A3) implies (A1) and (A2). (A1) is the case $d_{\mathrm{TV}}=0$. Write $d_k(h)=d_{\mathrm{TV}}(\pi_k(\cdot\mid h),\pi'_k(\cdot\mid h))$ for the distance at step $k$ given the history $h$. Under (A3) the two runs share their history until the first split, so
\[
P(\text{split})=\sum_k\mathbb E\big[\mathbf 1\{\text{no split before }k\}\,d_k(h)\big]\le\sum_k\mathbb E_h\big[d_k(h)\big],
\]
the last expectation over the factual history $h=h_{<k}$, because a common history is reached without a split with at most its factual probability. For every history of positive factual probability the counterfactual branch probabilities are continuous in $s$ and equal the factual ones at $s=0$, so each of the $K$ terms, and with them $P(\text{split})$, tends to zero with $s$; since a flip of the decoded outcome requires a split, $\mathrm{PN}_i(s)\le P(\text{split})/\pfail\to0$.

\emph{Proposition 3.} (A1) and (A2) do not imply (A3), and do not fix the cause count. The antithetic and independent pairings violate (A2): once a probability moves away from its factual value, their probability of splitting tends to $2\min(p,1-p)$ and $2p(1-p)$, not to zero for $0<p<1$ (Section~\ref{sec:r-coup}). The detector-restoring pairing R3 satisfies (A1) and (A2) --- it coincides with the shared draw until the first split, which therefore has the same vanishing probability --- but not (A3) after a split: when the previous measurement of the same qubit differs between the runs it takes $u'=1-u$, which at a two-branch step with branch probabilities strictly between $0$ and $1$ disagrees with probability above $d_{\mathrm{TV}}$ (lemma above). It changes the cause count on the surface code (Section~\ref{sec:r-coup}). Already for a single measurement with $p'<p$, the pairings with off-diagonal mass $(1+\beta)(p-p')$ on (factual $1$, counterfactual $0$) and $\beta(p-p')$ on the other cell, for $0\le\beta\le\min(p',1-p)/(p-p')$ so that the diagonal stays non-negative, have the right marginals, are admissible (the probabilities differ, so (A1) does not constrain the step), are continuous, since both off-diagonal masses vanish as $p'\to p$, and give $\mathrm{PN}=(1+\beta)(p-p')/p$; $\beta=0$ is the shared draw.

\emph{Proposition 4.} (A3) is a stepwise principle of minimal disturbance and does not minimize the cause count. The detector-restoring pairing gives fewer $ZZ$ sites per cause set in the $Z$ basis, and the sharp bound $L^+$ is lower still; it is attained by a coupling of the final outcomes that in general must look at later steps and is not co-adapted. The smallest cause count over co-adapted admissible pairings is open.

\paragraph{The rules keep the counterfactual marginal.} Each rule of Section~\ref{sec:r-coup} maps the factual draw to $u'_k=g_k(u_k;h_{<k},h'_{<k})$ with $g_k(\cdot\,;h,h')$ measure-preserving on $[0,1]$ for every pair of histories --- the identity, $u\mapsto1-u$, an independent draw, or the identity or reflection chosen from the past. Since $u_k$ is uniform and independent of the past, so is $u'_k$, and the counterfactual run has its own law. R4 re-draws uniformly inside the factual outcome's interval, which also preserves the uniform law, since the interval is chosen with its length as probability. The engine's weakened intervention in the toy circuits of Section~\ref{sec:r-coup}, run for 20,000 shots per combination, agrees with the closed form within four standard errors in all \val{S13.toy.ok} of \val{S13.toy.cells} combinations of configuration, arm, pairing and strength, which tests these marginals on small circuits.

\paragraph{Relation to known principles.} For a binary step, (A3) is monotonicity, and counterfactual stability reduces to monotonicity for binary variables~\cite{OberstSontag2019}; applying it transition by transition has precedents in sequential decision processes~\cite{Tsirtsis2021}. We therefore present Proposition 1 as an application of these principles to quantum trajectories. What is specific to trajectories is that removing a single site moves a large share of the later Born probabilities (at least \val{S13.moved.A} of the measurement draws in the check of Section~\ref{sec:m-layers}), so that (A1), which is enough for Pauli noise, leaves them free and a further principle is required; and that the correspondence of branches on which (A3) rests is fixed by the unravelling.

\FloatBarrier

\section{Example failing shots}\label{app:extra}

This appendix shows the three example failing shots referred to in the main text, relates the map's equal share to a Shapley value, and gives the detail of the random-sign run and the flipped-measurement records of Section~\ref{sec:r-mech}.

\begin{figure}[H]
\centering
\figin{\linewidth}{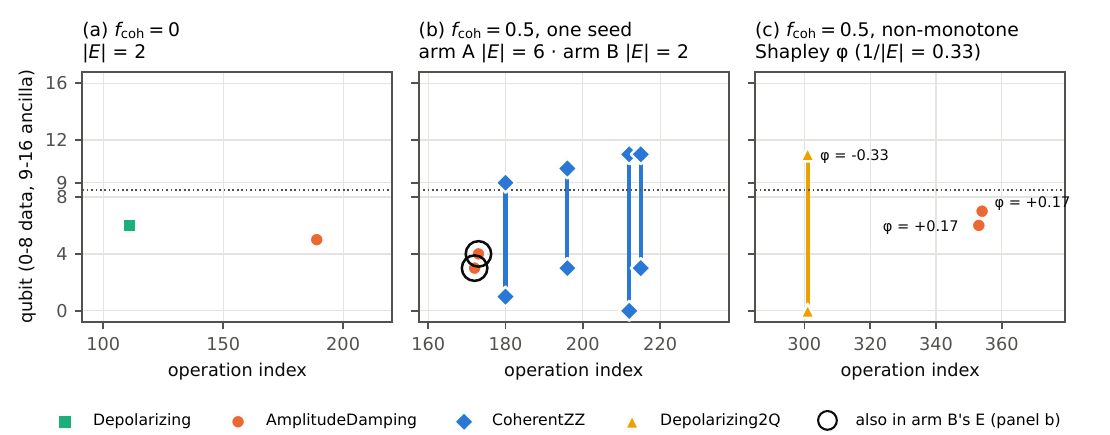}
\caption{Three failing shots on the qubit $\times$ operation plane. Each shot is the first that meets a fixed rule, $Z$ basis, $t=4$, $\sigma_0$. (a) The first failing shot with a cause set of two sites at $\fcoh=0$, the most common size. (b) The first seed at $\fcoh=0.5$ at which both arms fail, arm A with at least six and arm B with at most two sites, chosen as an extreme contrast: arm A has two amplitude-damping sites and four coherent-$ZZ$ sites in its cause set; arm B fails on the same seed with the two amplitude-damping sites only (circled); comparing the arms on one seed is itself a property of the shared-seed pairing. (c) The first shot at $\fcoh=0.5$ with a negative Shapley value and at most six sites, a non-monotone game: removing any one site removes the failure, but removing the two-qubit depolarizing site together with either other site brings it back. The game is restricted to $E$, a coalition having value $1$ if its removal removes the failure; removing the two sites other than the two-qubit depolarizing site together removes the failure (value $1$), and removing all three sites of $E$ together, with every other site of the shot kept, leaves it, so the Shapley values shown sum to zero. Shapley values and the restricted game are defined above.}
\label{fig:cases}
\end{figure}

\paragraph{Shapley values.} The map's equal share $1/|E|$ is the Shapley value of the game restricted to $E$, in which a coalition has value $1$ if removing it removes the failure, whenever every nonempty coalition of $E$ removes the failure. A site outside $E$ can still belong to a removing coalition, so $1/|E|$ is an equal-share rule, not a Shapley value over all sites. It equals the restricted Shapley value on \val{S10.f05.min} or more of the first 1,000 failing shots of arm A with a non-empty cause set in each of the four $Z$-basis $t=4$ cells. Cause sets of more than eight sites, estimated from 2,000 random orders, are counted among the exceptions, so the share is a lower bound; the other exceptions are games in which removing some coalition brings the failure back (Fig.~\ref{fig:cases}c).

\paragraph{Random signs.} In the run of Section~\ref{sec:r-mech} each of the \val{noise.nzz} $ZZ$ sites rotates by $\exp(-i\varepsilon_i\theta ZZ/2)$ with a fixed random sign $\varepsilon_i=\pm1$; arm B, whose twirl does not see the sign, is unchanged. Its fresh $Z$-basis stream has \val{N2.nA} failures of arm A, with tails of \val{N2.TA} in arm A against \val{N2.TB} in arm B. The signs are one fixed pattern from the seed \val{N2.seed}, and the engine gates passed before the run (all signs $+1$ reproduce the fixed-sign circuit bit for bit; \val{N2.Gi} shots). With \val{N2.nS} successes per arm, a failure of arm A in the random-sign run has \val{N2.kF.A} $ZZ$ causes on average against \val{N2.kF.B} in arm B, and \val{N2.mult.A} at a time when it has any (arm B \val{N2.mult.B}); the ratio to the sharp bound is \val{N2.rclip.A} (clipped) to \val{N2.rcf.A} (cross-fitted) in arm A and \val{N2.rclip.B} to \val{N2.rcf.B} in arm B, and $\kappa$ is \val{N2.kappa.A} against \val{N2.kappa.B}. These are close to the fixed-sign values of Section~\ref{sec:r-free}; the stream differs, so the comparison is not paired. The failure rate ratio of the arms is \val{N2.ratio} (90\,\%).

\begin{figure}[H]
\centering
\figin{\linewidth}{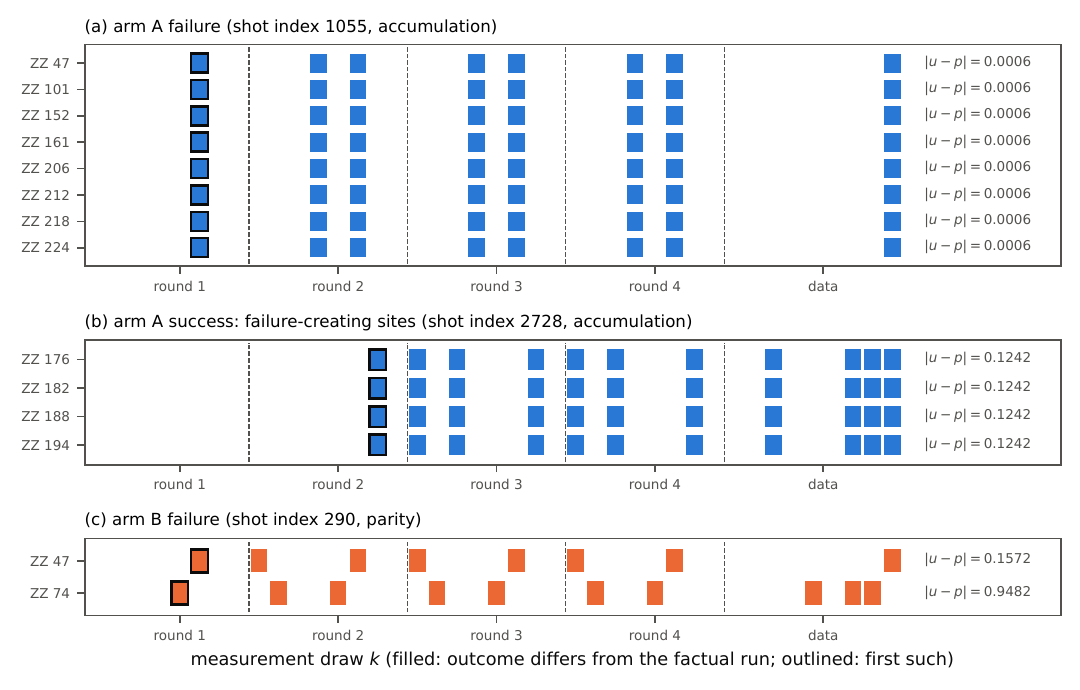}
\caption{Measurements whose outcome differs between the factual run and the removal of one coherent-$ZZ$ site, for three recorded shots of the $Z$-basis H7 stream: the first of each kind by shot index among (a) arm-A failures of accumulation type with at least three $ZZ$ causes, (b) arm-A successes with at least three failure-creating $ZZ$ sites, and (c) arm-B failures of parity type with at least two (no arm-B parity failure has three). Columns are measurement draws, eight ancilla measurements per round and then the nine data measurements; the first flip of each site is outlined and its factual margin $|u-p|$ is given at the right. In (a) and (b) every site flips the same measurements; in (c) the two sites flip disjoint ones. This rule of selection was fixed after the summary counts were seen.}
\label{fig:flips}
\end{figure}

\paragraph{Flipped measurements.} We replayed \val{N3.n} H7 shots with the ledger of Born draws and listed, for every coherent-$ZZ$ cause, the measurements whose outcome differs between the factual run and the removal. A shot with several causes is of accumulation type if all causes flip a common measurement and of parity type if their flipped sets are disjoint. Every replay reproduced its recorded decision. Of the failures of arm A with two or more $ZZ$ causes, \val{N3.z.A.F.acc} of \val{N3.z.A.F.multi} ($Z$) and \val{N3.x.A.F.acc} of \val{N3.x.A.F.multi} ($X$) are of accumulation type, \val{N3.z.A.F.par} and \val{N3.x.A.F.par} of parity type and the rest mixed. The median factual margin of the first flipped measurement is \val{N3.z.A.F.margin} and \val{N3.x.A.F.margin}, and removing the site moves its probability by a median of \val{N3.z.A.F.shift} and \val{N3.x.A.F.shift}; in \val{N3.z.A.F.same} and \val{N3.x.A.F.same} of these failures all causes flip the same first measurement (a summary added after the registration). The successes with failure-creating $ZZ$ sites are of accumulation type in \val{N3.z.A.S.acc} of \val{N3.z.A.S.multi} and \val{N3.x.A.S.acc} of \val{N3.x.A.S.multi}. In arm B, \val{N3.z.B.F.par} of \val{N3.z.B.F.multi} and \val{N3.x.B.F.par} of \val{N3.x.B.F.multi} multi-cause failures are of parity type, with a median shift of \val{N3.z.B.F.shift}. Figure~\ref{fig:flips} shows one shot of each kind.

\FloatBarrier


\begin{thebibliography}{10}

\bibitem{Gidney2021Stim}
Craig Gidney.
\newblock ``Stim: a fast stabilizer circuit simulator''.
\newblock \href{https://dx.doi.org/10.22331/q-2021-07-06-497}{Quantum {\bf 5},
  497}~(2021).

\bibitem{BravyiVargo2013}
Sergey Bravyi and Alexander Vargo.
\newblock ``Simulation of rare events in quantum error correction''.
\newblock \href{https://dx.doi.org/10.1103/PhysRevA.88.062308}{Phys. Rev. A
  {\bf 88}, 062308}~(2013).
\newblock  \href{http://arxiv.org/abs/1308.6270}{arXiv:1308.6270}.

\bibitem{Beverland2025}
Michael~E. Beverland, Malcolm Carroll, Andrew~W. Cross, and Theodore~J. Yoder.
\newblock ``Fail fast: techniques to probe rare events in quantum error
  correction''~(2025).
\newblock  \href{http://arxiv.org/abs/2511.15177}{arXiv:2511.15177}.

\bibitem{Ghosh2012}
Joydip Ghosh, Austin~G. Fowler, and Michael~R. Geller.
\newblock ``Surface code with decoherence: {A}n analysis of three
  superconducting architectures''.
\newblock \href{https://dx.doi.org/10.1103/PhysRevA.86.062318}{Phys. Rev. A
  {\bf 86}, 062318}~(2012).
\newblock  \href{http://arxiv.org/abs/1210.5799}{arXiv:1210.5799}.

\bibitem{Willow2025}
{Google Quantum AI and Collaborators}.
\newblock ``Quantum error correction below the surface code threshold''.
\newblock \href{https://dx.doi.org/10.1038/s41586-024-08449-y}{Nature {\bf
  638}, 920--926}~(2025).
\newblock  \href{http://arxiv.org/abs/2408.13687}{arXiv:2408.13687}.

\bibitem{Darmawan2017}
Andrew~S. Darmawan and David Poulin.
\newblock ``Tensor-network simulations of the surface code under realistic
  noise''.
\newblock \href{https://dx.doi.org/10.1103/PhysRevLett.119.040502}{Phys. Rev.
  Lett. {\bf 119}, 040502}~(2017).
\newblock  \href{http://arxiv.org/abs/1607.06460}{arXiv:1607.06460}.

\bibitem{Bravyi2018}
Sergey Bravyi, Matthias Englbrecht, Robert K{\"o}nig, and Nolan Peard.
\newblock ``Correcting coherent errors with surface codes''.
\newblock \href{https://dx.doi.org/10.1038/s41534-018-0106-y}{npj Quantum Inf.
  {\bf 4}, 55}~(2018).
\newblock  \href{http://arxiv.org/abs/1710.02270}{arXiv:1710.02270}.

\bibitem{Tuloup2026}
Thomas Tuloup and Thomas Ayral.
\newblock ``Computing logical error thresholds with the {P}auli frame sparse
  representation''~(2026).
\newblock  \href{http://arxiv.org/abs/2603.14670}{arXiv:2603.14670}.

\bibitem{Barone2026}
Francesco~Pio Barone, Daniel Jaschke, Ilaria Siloi, and Simone Montangero.
\newblock ``Color code thresholds under circuit-level noise beyond the {P}auli
  framework''.
\newblock \href{https://dx.doi.org/10.1088/2058-9565/ae8df5}{Quantum Sci.
  Technol. {\bf 11}, 035065}~(2026).
\newblock  \href{http://arxiv.org/abs/2511.05719}{arXiv:2511.05719}.

\bibitem{Hines2026}
Jordan Hines, Corey Ostrove, Kenneth Rudinger, Stefan Seritan, Kevin Young,
  Robin Blume-Kohout, and Timothy Proctor.
\newblock ``Simulating quantum error correction beyond {P}auli stochastic
  errors''~(2026).
\newblock  \href{http://arxiv.org/abs/2603.18457}{arXiv:2603.18457}.

\bibitem{Garner2025}
Sean~R. Garner, Nathan~M. Myers, Meng Wang, Samuel Stein, Chenxu Liu, and Ang
  Li.
\newblock ``Simulation of thermal-relaxation noise for quantum error
  correction''.
\newblock \href{https://dx.doi.org/10.1103/hgnj-v4j7}{Phys. Rev. Research {\bf
  8}, 023134}~(2026).
\newblock  \href{http://arxiv.org/abs/2512.09189}{arXiv:2512.09189}.

\bibitem{HouRavi2026}
Yuewen Hou and Gokul~Subramanian Ravi.
\newblock ``Scalable exact non-{P}auli noise simulation for fault-tolerant
  quantum circuits''~(2026).
\newblock  \href{http://arxiv.org/abs/2610.15975}{arXiv:2610.15975}.

\bibitem{TianPearl2000}
Jin Tian and Judea Pearl.
\newblock ``Probabilities of causation: bounds and identification''.
\newblock \href{https://dx.doi.org/10.1023/A:1018912507879}{Ann. Math. Artif.
  Intell. {\bf 28}, 287--313}~(2000).
\newblock  \href{http://arxiv.org/abs/1301.3898}{arXiv:1301.3898}.

\bibitem{OberstSontag2019}
Michael Oberst and David Sontag.
\newblock ``Counterfactual off-policy evaluation with {Gumbel}-max structural
  causal models''.
\newblock In Proceedings of the 36th International Conference on Machine
  Learning.
\newblock Volume~97 of PMLR, pages 4881--4890.
\newblock ~(2019).
\newblock  \href{http://arxiv.org/abs/1905.05824}{arXiv:1905.05824}.

\bibitem{ShaoLiu2026}
Yuguo Shao and Zi-Wen Liu.
\newblock ``Optimizing quantum error correction through error
  attribution''~(2026).
\newblock  \href{http://arxiv.org/abs/2610.03399}{arXiv:2610.03399}.

\bibitem{FuStaplesThompson2026}
Winston Fu, J.~Wilson Staples, and Jeff~D. Thompson.
\newblock ``Efficient estimation of logical sensitivities through
  fault-counting''~(2026).
\newblock  \href{http://arxiv.org/abs/2610.10531}{arXiv:2610.10531}.

\bibitem{Pearl2009}
Judea Pearl.
\newblock ``{Causality: Models, Reasoning, and Inference}''.
\newblock \href{https://dx.doi.org/10.1017/CBO9780511803161}{Cambridge
  University Press}. ~(2009).
\newblock 2nd edition.

\bibitem{Wallman2016}
Joel~J. Wallman and Joseph Emerson.
\newblock ``Noise tailoring for scalable quantum computation via randomized
  compiling''.
\newblock \href{https://dx.doi.org/10.1103/PhysRevA.94.052325}{Phys. Rev. A
  {\bf 94}, 052325}~(2016).

\bibitem{Tsirtsis2021}
Stratis Tsirtsis, Abir De, and Manuel Gomez-Rodriguez.
\newblock ``Counterfactual explanations in sequential decision making under
  uncertainty''.
\newblock In Advances in Neural Information Processing Systems.
\newblock Volume~34.
\newblock ~(2021).
\newblock  \href{http://arxiv.org/abs/2107.02776}{arXiv:2107.02776}.

\bibitem{Higgott2023}
Oscar Higgott and Craig Gidney.
\newblock ``Sparse blossom: correcting a million errors per core second with
  minimum-weight matching''.
\newblock \href{https://dx.doi.org/10.22331/q-2025-01-20-1600}{Quantum {\bf 9},
  1600}~(2025).
\newblock  \href{http://arxiv.org/abs/2303.15933}{arXiv:2303.15933}.

\bibitem{Dalibard1992}
Jean Dalibard, Yvan Castin, and Klaus M{\o}lmer.
\newblock ``Wave-function approach to dissipative processes in quantum
  optics''.
\newblock \href{https://dx.doi.org/10.1103/PhysRevLett.68.580}{Phys. Rev. Lett.
  {\bf 68}, 580}~(1992).

\bibitem{TakouBrown2026}
Evangelia Takou and Kenneth~R. Brown.
\newblock ``Estimating and decoding coherent errors of {QEC} experiments with
  detector error models''~(2025).
\newblock  \href{http://arxiv.org/abs/2510.23797}{arXiv:2510.23797}.

\bibitem{Patti2025}
Taylor~L. Patti, Thien Nguyen, Justin~G. Lietz, Alexander~J. McCaskey, and
  Brucek Khailany.
\newblock ``Augmenting simulated noisy quantum data collection by orders of
  magnitude using pre-trajectory sampling with batched execution''.
\newblock In Proceedings of the International Conference for High Performance
  Computing, Networking, Storage and Analysis (SC '25).
\newblock \href{https://dx.doi.org/10.1145/3712285.3759871}{Pages 762--773}.
\newblock ACM~(2025).
\newblock  \href{http://arxiv.org/abs/2504.16297}{arXiv:2504.16297}.

\bibitem{Shah2026}
Jaineet Shah.
\newblock ``Causal agent replay: Counterfactual attribution for {LLM}-agent
  failures''~(2026).
\newblock  \href{http://arxiv.org/abs/2606.08275}{arXiv:2606.08275}.

\bibitem{HaughSingal2023}
Martin~B. Haugh and Raghav Singal.
\newblock ``Counterfactual analysis in dynamic latent state models''.
\newblock In Proceedings of the 40th International Conference on Machine
  Learning.
\newblock Volume 202 of PMLR, pages 12647--12677.
\newblock ~(2023).
\newblock  \href{http://arxiv.org/abs/2205.13832}{arXiv:2205.13832}.

\bibitem{LevinPeresWilmer}
David~A. Levin, Yuval Peres, and Elizabeth~L. Wilmer.
\newblock ``{Markov Chains and Mixing Times}''.
\newblock \href{https://dx.doi.org/10.1090/mbk/058}{American Mathematical
  Society}. ~(2009).

\bibitem{Allen2017QCM}
John-Mark~A. Allen, Jonathan Barrett, Dominic~C. Horsman, Ciar{\'a}n~M. Lee,
  and Robert~W. Spekkens.
\newblock ``Quantum common causes and quantum causal models''.
\newblock \href{https://dx.doi.org/10.1103/PhysRevX.7.031021}{Phys. Rev. X {\bf
  7}, 031021}~(2017).
\newblock  \href{http://arxiv.org/abs/1609.09487}{arXiv:1609.09487}.

\bibitem{Barrett2019QCM}
Jonathan Barrett, Robin Lorenz, and Ognyan Oreshkov.
\newblock ``Quantum causal models''~(2019).
\newblock  \href{http://arxiv.org/abs/1906.10726}{arXiv:1906.10726}.

\bibitem{Suresh2023QCF}
Ardra~Kooderi Suresh, Markus Frembs, and Eric~G. Cavalcanti.
\newblock ``A semantics for counterfactuals in quantum causal models''~(2023).
\newblock  \href{http://arxiv.org/abs/2302.11783}{arXiv:2302.11783}.

\bibitem{WisemanMilburn2009}
Howard~M. Wiseman and Gerard~J. Milburn.
\newblock ``{Quantum Measurement and Control}''.
\newblock \href{https://dx.doi.org/10.1017/CBO9780511813948}{Cambridge
  University Press}. ~(2009).

\bibitem{Jin2023QPlayer}
Ki-Sung Jin and Gyu-Il Cha.
\newblock ``{QPlayer}: Lightweight, scalable, and fast quantum simulator''.
\newblock \href{https://dx.doi.org/10.4218/etrij.2021-0442}{ETRI J. {\bf 45},
  304--317}~(2023).

\bibitem{Jin2023Multilayered}
Ki-Sung Jin and Gyu-Il Cha.
\newblock ``Multilayered logical qubits and synthesized quantum bits''.
\newblock \href{https://dx.doi.org/10.1088/2058-9565/accec5}{Quantum Sci.
  Technol. {\bf 8}, 035008}~(2023).

\bibitem{Jin2026Affine}
Kisung Jin, Jinho On, and Gyuil Cha.
\newblock ``Tracking affine subspace with {G}aussian elimination for adaptive
  quantum circuit simulation''.
\newblock \href{https://dx.doi.org/10.1145/3815191}{ACM Trans. Quantum Comput.
  {\bf 7}, 1--22}~(2026).

\bibitem{JinPaper2Artifact}
Kisung Jin and Gyuil Cha.
\newblock ``Counterfactual attribution of surface-code failures under
  non-{Pauli} noise: dependence on trajectory pairing --- code and data,
  version 1.0.0''.
\newblock Zenodo~(2026).
\newblock
  \href{https://doi.org/10.5281/zenodo.23940106}{doi:10.5281/zenodo.23940106}.

\bibitem{NielsenChuang}
Michael~A. Nielsen and Isaac~L. Chuang.
\newblock ``{Quantum Computation and Quantum Information}''.
\newblock \href{https://dx.doi.org/10.1017/CBO9780511976667}{Cambridge
  University Press}. ~(2010).

\end{thebibliography}
\end{document}